\pdfoutput=1
\documentclass[bibyear]{aa} 

\usepackage{graphicx}
\usepackage{txfonts}
\begin{document}

   \title{(Sub)millimetre interferometric imaging of a sample of COSMOS/AzTEC submillimetre galaxies -- 
I. Multiwavelength identifications and redshift distribution\thanks{Based on observations carried out with the IRAM 
Plateau de Bure Interferometer. IRAM is supported by INSU/CNRS (France), MPG (Germany), and IGN (Spain).}}


   \author{O.~Miettinen\inst{1}, V.~Smol\v{c}i\'{c}\inst{1}, M.~Novak\inst{1}, M.~Aravena\inst{2}, A.~Karim\inst{3}, D.~Masters\inst{4}, D.~A.~Riechers\inst{5}, R.~S.~Bussmann\inst{5}, H.~J.~McCracken\inst{6}, O.~Ilbert\inst{7}, F.~Bertoldi\inst{3}, P.~Capak\inst{4}, C.~Feruglio\inst{8,9}, C.~Halliday\inst{10}, J.~S.~Kartaltepe\inst{11}, F.~Navarrete\inst{3}, M.~Salvato\inst{12}, D.~Sanders\inst{13}, E.~Schinnerer\inst{14}, and K.~Sheth\inst{15}}

   \institute{Department of Physics, University of Zagreb, Bijeni\v{c}ka cesta 32, HR-10000 Zagreb, Croatia\\ \email{oskari@phy.hr} \and N\'ucleo de Astronom\'{\i}a, Facultad de Ingenier\'{\i}a, Universidad Diego Portales, Av. Ej\'ercito 441, Santiago, Chile \and Argelander-Institut f\"{u}r Astronomie, Universit\"{a}t Bonn, Auf dem H\"{u}gel 71, D-53121 Bonn, Germany \and Spitzer Science Center, 314-6 Caltech, Pasadena, CA 91125, USA \and Astronomy Department, Cornell University, 220 Space Sciences Building, Ithaca, NY 14853, USA \and Institute d'Astrophysique de Paris, UMR7095 CNRS, Universit\'e Pierre et Marie Curie, 98 bis Boulevard Arago, 75014 Paris, France \and Aix Marseille Universit\'e, CNRS, LAM (Laboratoire d'Astrophysique de Marseille) UMR 7326, 13388, Marseille, France \and Institut de RadioAstronomie Millim\'etrique, 300 rue de la Piscine, Domaine Universitaire, 38406 Saint Martin d'H\`eres, France \and Scuola Normale Superiore, Piazza dei Cavalieri 7, I-56126 Pisa, Italy \and 23, rue d'Yerres, 91230 Montgeron, France \and National Optical Astronomy Observatory, 950 North Cherry Avenue, Tucson, AZ 85719, USA \and Max-Planck-Institut f\"{u}r extraterrestrische Physik, Giessenbachstrasse 1, D-85748 Garching bei M\"{u}nchen, Germany \and Institute for Astronomy, University of Hawaii, 2680 Woddlawn Drive, Honolulu, HI, 96822 \and Max-Planck-Institut f\"{u}r Astronomie, K\"{o}nigstuhl 17, 69117 Heidelberg, Germany \and National Radio Astronomy Observatory/NAASC, 520 Edgemont Road, Charlottesville, VA 22903, USA}

   \date{Received ; accepted}

\authorrunning{Miettinen et al.}
\titlerunning{Plateau de Bure Interferometer 1.3 mm imaging of SMGs}

  \abstract{
We used the Plateau de Bure Interferometer (PdBI) to map a sample of 15 submillimetre galaxies (SMGs) 
in the COSMOS field at the wavelength of 1.3 mm. 
The target SMGs were originally discovered in the James Clerk Maxwell Telescope (JCMT)/AzTEC 1.1 mm continuum survey at 
${\rm S/N}_{\rm 1.1\, mm}=4-4.5$. This paper presents, for the first time, interferometric millimetre-wavelength observations of these sources.
The angular resolution of our observations, $\sim 1\farcs8$, allowed us to accurately determine the positions of the target SMGs. 
Using a detection threshold of ${\rm S/N}_{\rm 1.3\, mm}>4.5$ regardless of multiwavelength counterpart association, and 
$4<{\rm S/N}_{\rm 1.3\, mm}\leq4.5$ if a multiwavelength counterpart within $1\farcs5$ is also present, the total number of detections in our survey is 
22. The most significant PdBI detection of ${\rm S/N}_{\rm 1.3\, mm}=10.3$ is towards AzTEC19.
Three of our detected SMGs (AzTEC21, 27, and 28; which corresponds to $20\%$) are marginally resolved at our 
angular resolution, and these sources are found to have elongated or clumpy morphologies and/or multiple components. 
Using optical to near-infrared photometric redshifts, available spectroscopic redshifts, and redshifts estimated from the
radio-to-submm spectral index we infer a median redshift of $\tilde{z}=3.20\pm0.25$ for our sample. To study the overall multiplicity and redshift 
distribution of flux-limited samples of SMGs we combined these sources with the 15 brightest JCMT/AzTEC SMGs detected at 1.1 mm, AzTEC1--15, 
and studied previously. This constitutes a complete, flux- and S/N-limited 1.1-mm selected sample. We find that the median
redshift for the 15 brightest JCMT/AzTEC SMGs ($\tilde{z}=3.05\pm0.44$) is consistent with that for AzTEC16--30. 
This conforms to recent observational findings that SMGs do not exhibit any significant trend between the redshift and (sub)mm flux density.
For the combined AzTEC1--30 sample we derive a median redshift of $\tilde{z}=3.17\pm0.27$, consistent with previous results based on mm-selected samples.
We further infer that within the combined AzTEC1--30 sample $\sim25\pm9\%$ of sources separate into multiple components.}


   \maketitle
%

\section{Introduction}

When the first extragalactic submillimetre continuum surveys were 
carried out at the end of the 1990s, a population of heavily dust-obscured (i.e.
optically faint) galaxies at high redshift was discovered (\cite{smail1997}; 
\cite{hughes1998}; \cite{barger1998}). These sources are generally referred to
as submillimetre galaxies or SMGs (see \cite{blain2002}; \cite{casey2014} for reviews).

The bulk of SMGs are observed at redshifts $z \simeq 2-3$ (e.g. 
\cite{chapman2005}; \cite{wardlow2011}; \cite{lindner2011}; \cite{casey2013}; 
\cite{simpson2014}; \cite{umehata2014}). However, the number of known high-redshift 
($z > 4$) SMGs has increased considerably in
the past few years (e.g. \cite{schinnerer2008}; \cite{daddi2009a},b; 
\cite{coppin2009}; \cite{riechers2010}; \cite{capak2011}; \cite{smolcic2011}; 
\cite{cox2011}; \cite{combes2012}; \cite{walter2012}; \cite{swinbank2012}; 
\cite{weiss2013}; \cite{smolcic2015}). 
The SMG with the highest spectroscopically confirmed redshift currently 
known is HFLS3 at $z=6.34$ (\cite{riechers2013}), which demonstrates 
that these sources were already present when the universe 
was only $\lesssim 890$ Myr old (see the end of this section for our adopted cosmology).   

Submillimetre galaxies have parent dark matter haloes -- i.e. the sites of galaxy 
formation originating in the dark-matter-dominated density perturbations in 
the early universe (e.g. \cite{benson2010}) -- with characteristic masses of 
$\sim 10^{12}-10^{13}$ M$_{\sun}$ (\cite{blain2004}; \cite{swinbank2008}; \cite{hickox2012}). 
The physical properties of SMGs are found to be extreme. In particular, their very high 
infrared (IR; 8--1\,000 $\mu$m) luminosities of $L_{\rm IR} \sim 10^{12}-10^{13}$~L$_{\sun}$ 
are indicative of extreme star formation rates (SFRs) of $\sim100-1\,000$~M$_{\sun}$~yr$^{-1}$, 
making SMGs the most intense known starbursts in the universe. 
Observations of CO rotational transitions with upper rotational-energy levels of
$J_{\rm u}=2-7$ suggest H$_2$ gas masses of $M_{\rm H_2}\sim 10^{10}-10^{11}$~M$_{\sun}$ 
in SMGs (e.g. \cite{greve2005}; \cite{tacconi2006}; \cite{bothwell2013}), 
while CO($J=1-0$) observations yield gas masses up to several times $10^{11}$~M$_{\sun}$ 
(\cite{ivison2011}; \cite{riechers2011}). Submillimetre galaxies are therefore among 
the most gas-rich systems in the universe. For instance, the median 
$M_{\rm H_2}$ value of $3.0\pm1.6 \times 10^{10}$~M$_{\sun}$ (within a $\sim 2$ 
kpc radius) derived for SMGs studied by Greve et al. (2005) is about four times higher than in 
the most luminous local ultraluminous IR galaxies or ULIRGs (\cite{solomon1997}). 
These authors also estimated that the typical gas-consumption timescale in SMGs is 
$\gtrsim 40$~Myr, but they noted that if feedback processes slow down the star formation activity 
(i.e. negative feedback such as radiation pressure acting on dust, stellar winds, outflows, supernovae, 
and the associated turbulence), the above timescale can be significantly longer. 
The derived stellar masses in SMGs are typically in the range 
$M_{\star}\sim 10^{11}-10^{12}$ M$_{\sun}$ (e.g. \cite{borys2005}; \cite{dye2008}; 
\cite{wardlow2011}; \cite{hainline2011}; \cite{michalowski2012}; \cite{simpson2014}). 
While some authors suggest that SMGs might predominantly constitute the high-mass end of the star-forming galaxies' main
sequence (the $M_{\star}$--SFR relationship) at $z \geq 2$ (\cite{michalowski2012}), a fair fraction certainly lies above it 
(e.g. \cite{daddi2009a}).

Since SMGs are found to have very high SFRs, the 
question then arises as to which physical process(es) are responsible for these rates. 
It has been suggested that galaxy mergers can trigger a significant burst of star 
formation (e.g. \cite{barnes1991}). The basic idea behind this is that, when 
dynamical friction within a parent halo causes galaxies to collide, 
the dissipation of angular momentum during the process allows the gas to 
be funneled to the central region of the system. Numerical simulations have also 
demonstrated how gas inflows associated with gas-rich or ``wet'' mer\-gers can feed vigorous 
star formation (\cite{mihos1996}; \cite{chakrabarti2008}). 
More recently, hydrodynamic simulations by Narayanan et al. (2010) 
uggested that SMGs can naturally form via galaxy mergers. 
From an observational point of view, this is supported by the
clumpy or disturbed morphologies of SMGs and their complex kinematic signatures (e.g. \cite{smail1998}; 
\cite{tacconi2008}; \cite{engel2010}; \cite{swinbank2011}; 
\cite{sharon2013}; \cite{riechers2013}; \cite{hung2013}; \cite{toft2014}; 
\cite{neri2014}; \cite{riechers2014}; \cite{chen2015}). 
Engel et al. (2010) concluded that most of the SMGs with IR
luminosities of $L_{\rm IR} \gtrsim 5 \times 10^{12}$ L$_{\sun}$ are probably 
major-merger systems [i.e. systems where the interacting galaxies have a mass ratio of 
$>1/3$ (e.g. \cite{casey2014})]. An alternative mechanism behind galaxy formation and the
fuelling of their star formation is the accretion of gas from the 
intergalactic medium through filamentary structures (the so-called cold-mode 
accretion; \cite{keres2005}, 2009; \cite{dekel2009}). 
Cosmological hydrodynamic simulations performed by Dav{\'e} et al. (2010) 
suggest that SMGs could generally obtain their gas reservoirs via such 
accretion processes (rather than through mer\-gers). In these 
simulations, the galaxies often had complex morphologies and gas kinematics -- 
signatures often interpreted as evidence of an ongoing merger. However,
as a result of cold-mode accretion, an extended disk-like gas
structure undergoing rotation is also expected, and some SMGs are indeed found to 
show such signatures (e.g. \cite{carilli2010}; \cite{hodge2012}; \cite{debreuck2014}).
Finally, we note that simulations suggest that during the course of their evolution, 
SMGs can exhibit properties that are reminiscent of both normal star-forming galaxies and 
vigorous starbursts (see \cite{hayward2013a},b)\footnote{When a galaxy is on the 
main sequence, it is often said to be a ``normal'' star-forming galaxy. In 
contrast, if the galaxy has a clearly enhanced SFR with respect to its stellar 
mass (i.e. outlier above the main sequence), it is defined to be a starburst 
galaxy (e.g. \cite{magdis2011}).}. For example, numerical simulations 
(\cite{springel2005}) have demonstrated that a disk-like structure can form soon 
after the merging of gas-rich galaxies because of the rapid cooling (see also 
\cite{hopkins2009}). This conforms to the idea that SMGs are a heterogeneous galaxy population, 
probably caught at different stages of evolution.

As an SMG increases its gas reservoir (through whatever mechanism), its
central supermassive black hole (SMBH) can accrete increasing amounts of gas, which
is driven to the nuclear region (e.g. \cite{granato2006}). Some SMGs can therefore host 
an active galactic nucleus (AGN) as revealed by deep X-ray observations 
(\cite{alexander2003}, 2005; \cite{wang2013}). For example, Chapman et al. 
(2005) concluded that about 20--30\% of radio-identified SMGs display AGN activity, and
radio-detected SMGs indeed appear to have a higher AGN fraction than the general SMG population 
(\cite{wang2013}). However, most of the bolometric IR luminosity of SMGs is found 
to originate in star formation activity (dust-reprocessed radiation) and not 
gas accretion onto the SMBH of an AGN. An important characteristic 
of these accreting central black holes is that they can influence the properties of their host 
galaxies through radiative and mechanical feedback. In particular, besides 
the exhaustion of the gas reservoir, AGN feedback can lead to the 
``quenching'' or shut down of the star formation 
(e.g. \cite{springeletal2005}; \cite{dimatteo2005}; \cite{hopkins2006}).

Another intriguing question is the role played by SMGs 
in galaxy evolution over cosmic time. It has been found that SMGs are promising 
candidates for the progenitors of the most massive, passive (i.e. with little or 
no ongoing star formation) elliptical galaxies seen in the 
present-day universe (e.g. \cite{lilly1999}; \cite{swinbank2006}; 
\cite{fu2013}; \cite{toft2014}; \cite{simpson2014}). The existence of quiescent, 
red massive galaxies already at $z \sim 2-3$ with old stellar populations 
indicates that these galaxies have experienced a short-lived starburst phase 
in their past (e.g. \cite{renzini2006}; \cite{capak2008}; \cite{coppin2010}). 
High-redshift ($z \sim4-5$) SMGs could well represent these galaxy precursors. 
Besides their physical characteristics, the strong clustering 
of SMGs is consistent with this evolutionary picture (\cite{blain2004}; 
\cite{aravena2010a}). Toft et al. (2014) found compelling evidence 
that the evolution of the giant red-and-dead ellipticals observed in the nearby 
universe, starting from $z > 3$ SMGs, goes through a transition stage manifested as compact quiescent 
galaxies at $z \sim 2-3$. 

Before the physical properties of SMGs can be studied in detail, 
the position of the source giving rise to the (sub)mm continuum emission 
must be accurately determined. The source counterparts at other wavelengths can only be correctly identified 
if the exact location of the FIR/(sub)mm emission is known, which in practice requires the analysis of 
FIR or (sub)mm interferometric observations to achieve this goal (e.g. \cite{frayer2000}; 
\cite{younger2007}, 2008, 2009; \cite{dannerbauer2008}; \cite{aravena2010b}; \cite{smolcic2012a},b; 
\cite{karim2013}; \cite{hodge2013}). To date, however, only a few flux-limited SMG samples have been followed up with 
interferometers (\cite{younger2007}, 2009; \cite{smolcic2012b}; \cite{barger2012}; \cite{karim2013}; \cite{hodge2013}).

In this paper, we present the results of our intermediate-resolution ($1\farcs8$) Plateau de Bure 
Interferometer (PdBI; \cite{guilloteau1992}) 1.3 mm continuum imaging of 
a sample of 15 SMGs discovered by Scott et al. (2008) in the Cosmic Evolution Survey 
(COSMOS; \cite{scoville2007a}) field. The good angular resolution of the present data allows us to 
accurately pinpoint source positions and match them with correct multiwavelength counterparts.
Accurate SMG positions are needed for their targeted spectroscopic redshift measurements, 
and knowing the proper multiwavelength counterparts allows us to determine the photometric redshifts of the sources 
-- a prerequisite for a rigorous analysis of the physical properties. After describing the source sample, 
observations, data reduction, and ancillary data in Sect.~2, the direct 
observational results and analysis are presented in Sect.~3. In Sect.~4, we 
present our ana\-lysis of the redshifts of our SMGs. We then discuss our results in Sect.~5, 
and a summary is given in Sect.~6.

In the present paper, we adopt a 
concordance $\Lambda$ cold dark matter ($\Lambda$CDM) cosmology, with 
the Hubble constant $H_0~=~71$ km~s$^{-1}$~Mpc$^{-1}$ [i.e. the reduced Hubble constant 
$h \equiv H_0/(100\,{\rm km~s^{-1}~Mpc^{-1}})=0.71$], 
total (dark+luminous baryonic) matter density $\Omega_{\rm m}=0.27$, 
and dark energy density $\Omega_{\Lambda}=0.73$ (\cite{spergel2007}; \cite{larson2011}). 
In this spatially flat universe, $1\arcsec$ corresponds to a physical spatial scale of 
8.04, 8.48, and 7.83 kpc at redshifts of $z=1$, 2, and 3, respectively. 
The corresponding cosmic times are 5.94, 3.34, and 2.19~Gyr. 
Magnitudes in the present paper refer to the AB magnitude system (see \cite{oke1974}).

\section{Observations, data, and data reduction}

\subsection{Source sample}

Our new PdBI 1.3~mm observations, described in the next subsection, were made towards 
the SMGs listed in Table~\ref{table:sample}. These SMGs were originally 
discovered in the 1.1~mm continuum survey of a north-west 
subfield (0.15 deg$^2$ in size) of the 2 deg$^2$ COSMOS field by Scott et al. (2008). 
The survey was carried out with the Aztronomical Thermal Emission Camera, 
or the AzTEC bolometer array (\cite{wilson2008}), on the 15 m James Clerk Maxwell Telescope (JCMT), 
and the target field was centred on a prominent large-scale structure 
traced by the galaxy overdensity (\cite{scoville2007b}) that includes a 
massive ($\sim10^{15}$ M$_{\sun}$) galaxy cluster at a redshift 
of $z \simeq 0.73$ (\cite{guzzo2007}). In total, Scott et al. (2008) 
reported 50 candidate SMGs with a detection signal-to-noise ratio 
S/N$_{\rm 1.1\, mm} \geq 3.5$ (see their Table~1). 

While our PdBI observations targeted the COSMOS/AzTEC SMGs AzTEC16--30, the
15 brightest SMGs detected by Scott et al. (2008), i.e. AzTEC1--15 (S/N$_{\rm 1.1\, mm} \geq 4.6$), 
had previously been imaged (and detected) with the Submillimetre Array (SMA) at 890 $\mu$m 
($2\arcsec$ angular resolution) by Younger et al. (2007, 2009). Spectroscopic and/or photometric 
redshifts have been assigned to them by Younger et al. (2007, 2009), Riechers et al. (2010), 
Capak et al. (2011), Smol{\v c}i{\'c} et al. (2011, 2012b), and M.~S.~Yun et al. (in prep.). 
Combining these data with the present observations towards AzTEC16--30 provides 
us with a flux-limited sample of 30 SMGs (S/N$_{\rm 1.1\, mm} \geq 4$), which have all been
followed up with intermediate-resolution interferometric observations. This allows us to carry out 
a statistically meaningful study of their redshift distribution.

\begin{table}
\renewcommand{\footnoterule}{}
\caption{Source list.}
{\small
\begin{minipage}{1\columnwidth}
\centering
\label{table:sample}
\begin{tabular}{c c c c}
\hline\hline 
Source & JCMT/AzTEC ID & $S_{\rm 1.1\, mm}^{\rm db}$ & S/N$_{\rm 1.1\, mm}$\\
       & & [mJy] & \\
\hline
AzTEC16 & AzTEC$_{-}$J095950.29+024416.1 & $3.9\pm1.3$ & 4.5 \\
AzTEC17 & AzTEC$_{-}$J095939.30+023408.0 & $3.8\pm1.4$ & 4.4 \\
AzTEC18 & AzTEC$_{-}$J095943.04+023540.2 & $3.8_{-1.5}^{+1.3}$ & 4.3 \\
AzTEC19 & AzTEC$_{-}$J100028.94+023200.3 & $3.8_{-1.6}^{+1.3}$ & 4.3 \\
AzTEC20 & AzTEC$_{-}$J100020.14+024116.0 & $3.8_{-1.6}^{+1.3}$ & 4.3 \\
AzTEC21 & AzTEC$_{-}$J100002.74+024645.0 & $3.4_{-1.4}^{+1.3}$ & 4.2 \\
AzTEC22 & AzTEC$_{-}$J095950.69+022829.5 & $3.6_{-1.6}^{+1.5}$ & 4.2 \\
AzTEC23 & AzTEC$_{-}$J095931.57+023601.5 & $3.4_{-1.5}^{+1.4}$ & 4.1 \\
AzTEC24 & AzTEC$_{-}$J100038.72+023843.8 & $3.3_{-1.5}^{+1.4}$ & 4.1 \\
AzTEC25 & AzTEC$_{-}$J095950.41+024758.3 & $3.3\pm1.4$ & 4.1 \\
AzTEC26 & AzTEC$_{-}$J095959.59+023818.5 & $3.3_{-1.5}^{+1.4}$ & 4.0 \\
AzTEC27 & AzTEC$_{-}$J100039.12+024052.5 & $3.3_{-1.6}^{+1.4}$ & 4.0 \\
AzTEC28 & AzTEC$_{-}$J100004.54+023040.1 & $3.3_{-1.6}^{+1.5}$ & 4.0 \\
AzTEC29 & AzTEC$_{-}$J100026.68+023753.7 & $3.3_{-1.6}^{+1.4}$ & 4.0 \\
AzTEC30 & AzTEC$_{-}$J100003.95+023253.8 & $3.3_{-1.6}^{+1.4}$ & 4.0 \\
\hline 
\end{tabular} 
\tablefoot{The 1.1 mm flux densities listed in column~(3) are deboosted values 
from Scott et al. (2008; their Table~1) and the quoted errors represent the 68\% confidence interval.}
\end{minipage} 
}
\end{table}

\subsection{Intermediate-resolution 1.3 mm imaging}

The PdBI 1.3 mm (230.5 GHz) continuum observations towards AzTEC16--30 
(project {\tt W0AE}) were carried out 
between January and November 2013. The array of six antennas
was mostly in its C configuration, i.e. the second most-compact configuration 
with 15 baselines ranging from 24 to 176 m (which corresponds to 18.5--135.4 k$\lambda$). 
On 16 April, however, antenna station E04 was not available, resulting in only ten baselines. 
On 12 October, when part of the observations towards AzTEC16--22 were performed, 
the array was in its most compact D configuration (baselines in the range 
24--97 m or 18.5--74.6 k$\lambda$). The lower-sideband (LSB) system temperature was typically
$\sim200$~K, except on 16 April and 3 May when it was 300--350~K. The 
atmospheric precipitable water vapour (PWV) was typically in the range 2--4~mm,
except on 16 April when it was 6 mm. The best weather conditions were on 30 October,
when the PWV value was only about 1 mm. The phase centres used were the AzTEC 
1.1~mm peak positions of the sources detected by Scott et al. (2008), and 
the on-source observing time per source was $\sim1.5$~hr.

The backend used was the WideX correlator, 
which is composed of four units [two for both orthogonal linear 
(horizontal and vertical) polarisation modes], 
each being 2 GHz wide and split into 1\,024 channels 
(corresponding to a channel width of about 1.95 MHz). 
The total effective bandwidth is about 1.8~GHz for each unit or about 3.6 GHz 
for both polarisations. The correlator bandpass calibration was achieved by 
observing 3C84 (NGC 1275), 3C279, and B0923+392. Gain phases and amplitudes 
were calibrated by observing B0906+015 and 1005+066. 
The standard source 3C84 was used the most often as a flux calibrator, with the 
adopted 230 GHz flux density of $S_{\rm 230\, GHz}=9.85-12.36$ Jy depending on 
the observing day. The other flux calibrators used were 3C279 (10.68 Jy), 
0851+202 (3.04 Jy), and 0923+392 (2.5--3.16 Jy). The absolute flux-calibration 
uncertainty was estimated to be about 20\%, which is based on the observed scatter 
in the calibrators' flux densities. The primary beam (PB) of the PdBI at 
the observing frequency is $21\farcs3$ (Half-Power Beam Width or HPBW). 
At $z=2$, this corresponds to about 180.5 physical kpc. 

Calibration and imaging were performed using the CLIC (Continuum and Line 
Interferometer Calibration) and MAPPING programs of the {\tt GILDAS} software 
package\footnote{Grenoble Image and Line Data Analysis Software is provided and 
actively developed by IRAM, and is available at 
{\tt http://www.iram.fr/IRAMFR/GILDAS}}, respectively. 
When creating the maps, natural weighting was applied to the calibrated 
visibilities (i.e. weighting according to the number of measurements within a given region of 
the $uv$-plane). The CLEAN algorithm was used for deconvolution, and applied in 
regions centred on the strongest emission features. The typical resulting  
synthesised beam size (Full Width at Half Maximum or FWHM) is $1\farcs8$, and the 
restored continuum maps (dual polarisation) have $1\sigma$ root mean square (rms) noise 
values of $\sim0.2$ mJy~beam$^{-1}$. Hence, the statistical positional error, 
$\Delta \theta_{\rm stat}\simeq \sqrt{\theta_{\rm maj}\theta_{\rm min}}/(2{\rm S/N})$ 
(where $\theta_{\rm maj}$ and $\theta_{\rm min}$ are the major and minor axes 
of the beam; \cite{reid1988}; \cite{younger2007}), can be estimated to be about 
$0\farcs18/S_{\rm 1.3\, mm}[{\rm mJy~beam^{-1}}]$, which is about $0\farcs23$ for 
a typical $4\sigma$ source. We note that merging the C and D configuration visibilities together 
for AzTEC16--22 improved the $uv$ coverage and produced maps of higher spatial dynamic-range than
those of AzTEC23--30. The phase centres, both the synthesised beam sizes and position angles, 
and the rms noises of the final cleaned, continuum maps are listed in Table~\ref{table:obs}.

\subsection{The COSMOS field -- ancillary data}

Since our target sources lie within the COSMOS field, they have been observed 
with several ground- and space-based telescopes at wavelengths spanning from 
the X-rays to the radio regime. 

Observations at X-ray wavelengths were performed with the \textit{XMM-Newton} 
(\cite{hasinger2007}), and \textit{Chandra} satellites (C-COSMOS; \cite{elvis2009}; \cite{puccetti2009}). 
The \textit{Galaxy Evolution Explorer}, or \textit{GALEX}, imaged the COSMOS field 
in the near-UV (NUV) and far-UV (FUV) (\cite{zamojski2007}). \textit{Hubble Space Telescope} 
(\textit{HST}) Advanced Camera for Surveys (ACS) Wide Field Channel (WFC) observations of the COSMOS 
field were done in the $I$ band (the F814W filter) (\cite{scoville2007a}; 
\cite{koekemoer2007}). Observations at optical/near-IR (NIR)
wavelengths have been carried out with the 8.2 m Subaru telescope, 
the 3.6 m Canada France Hawaii Telescope (CFHT), the 3.8 m United Kingdom Infrared
Telescope (UKIRT), the 2.2 m University of Hawaii telescope called UH88 (or UH2.2),
and the 4 m telescopes of the Kitt-Peak National Observatory 
(KPNO), the Cerro Tololo Inter-American Observatory (CTIO), and the National 
Optical Astronomy Observatory (NOAO) [see Capak et al. (2007), Taniguchi et al. 
(2007), and McCracken et al. (2010) for details]. 

New NIR imaging of the COSMOS field in the $Y$ (1.02~$\mu$m), $J$ (1.25~$\mu$m), 
$H$ (1.65~$\mu$m), and $K_{\rm s}$ (2.15~$\mu$m) bands is being collected by the UltraVISTA survey 
(\cite{mccracken2012}; \cite{ilbert2013})\footnote{The data 
products are produced by TERAPIX; see {\tt http://terapix.iap.fr}}. 
Mid-infrared (MIR) observations were obtained with the Infrared Array Camera 
(IRAC; 3.6--8.0 $\mu$m) and the Multiband Imaging Photometer for Spitzer 
(MIPS; 24--160~$\mu$m) on board the \textit{Spitzer} Space Telescope as 
part of the COSMOS Spitzer survey (S-COSMOS; \cite{sanders2007}). 
Far-infrared (70, 160, and 250~$\mu$m) to submm (350 and 500~$\mu$m) \textit{Herschel} 
continuum observations were performed as part of the 
Photodetector Array Camera and Spectrometer (PACS) Evolutionary Probe (PEP; 
\cite{lutz2011}) and the Herschel Multi-tiered Extragalactic Survey (HerMES\footnote{{\tt http://hermes.sussex.ac.uk}}; 
\cite{oliver2012}) programmes. Radio-continuum imaging was done at 20 cm (1.4~GHz) with the Very 
Large Array or VLA (\cite{schinnerer2007}, 2010), and at 10 cm (3 GHz) with 
the Karl G. Jansky Very Large Array (VLA-COSMOS 3~GHz Large Project; PI: V.~Smol\v{c}i\'{c}). 
In addition to the imaging observations, a large spectroscopic redshift survey of galaxies in 
the COSMOS field has been carried out with the Very Large Telescope (VLT), a survey known 
as zCOSMOS (\cite{lilly2007}, 2009), and with the Keck telescope (M.~Salvato et al., in prep.). 
Photometric redshifts towards sources in the COSMOS field have been computed using 30 wavelength 
bands spanning from UV to MIR (\cite{ilbert2009}, 2013).

Most of the extensive multiwavelength datasets are publicly available 
from the NASA/IPAC Infrared Science Archive\footnote{\tt http://irsa.ipac.caltech.edu/Missions/cosmos.html}.

\begin{table}
\renewcommand{\footnoterule}{}
\caption{Observational parameters.}
{\scriptsize
\begin{minipage}{1\columnwidth}
\centering
\label{table:obs}
\begin{tabular}{c c c c c c}
\hline\hline 
Source & $\alpha_{2000.0}$ & $\delta_{2000.0}$ & $\theta_{\rm syn}$ & P.A. & $\sigma_{\rm rms}$\\
       & [h:m:s] & [$\degr$:$\arcmin$:$\arcsec$] & [$\arcsec \times \arcsec$] & [$\degr$] & [mJy~beam$^{-1}$] \\
\hline
AzTEC16 & 09 59 50.29 & +02 44 16.1 & $1.87 \times 1.18$ & 12.82 & 0.247\\
AzTEC17 & 09 59 39.30 & +02 34 08.0 & $1.85 \times 1.15$ & 13.49 & 0.239\\
AzTEC18 & 09 59 43.04 & +02 35 40.2 & $1.85 \times 1.14$ & 13.75 & 0.256\\
AzTEC19 & 10 00 28.94 & +02 32 00.3 & $1.85 \times 1.14$ & 13.67 & 0.302\\
AzTEC20 & 10 00 20.14 & +02 41 16.0 & $1.85 \times 1.14$ & 14.42 & 0.252\\
AzTEC21 & 10 00 02.74 & +02 46 45.0 & $1.86 \times 1.14$ & 14.82 & 0.256\\
AzTEC22 & 09 59 50.69 & +02 28 29.5 & $1.80 \times 1.21$ & 0.00 & 0.227\\
AzTEC23 & 09 59 31.57 & +02 36 01.5 & $1.76 \times 1.03$ & 4.19 & 0.205\\
AzTEC24 & 10 00 38.72 & +02 38 43.8 & $1.76 \times 1.03$ & 4.53 & 0.188\\
AzTEC25 & 09 59 50.41 & +02 47 58.3 & $1.75 \times 1.03$ & 5.00 & 0.191\\
AzTEC26 & 09 59 59.59 & +02 38 18.5 & $1.76 \times 1.03$ & 5.29 & 0.178\\
AzTEC27 & 10 00 39.12 & +02 40 52.5 & $1.76 \times 1.02$ & 5.19 & 0.215\\
AzTEC28 & 10 00 04.54 & +02 30 40.1 & $1.76 \times 1.02$ & 5.28 & 0.225\\
AzTEC29 & 10 00 26.68 & +02 37 53.7 & $1.75 \times 1.02$ & 5.57 & 0.212\\
AzTEC30 & 10 00 03.95 & +02 32 53.8 & $1.77 \times 1.01$ & 6.09 & 0.205\\
\hline 
\end{tabular} 
\tablefoot{The equatorial coordinates refer to the JCMT/AzTEC 1.1 mm peak positions 
(\cite {scott2008}), and they represent the phase centres of the PdBI observations 
presented here.}
\end{minipage} 
}
\end{table}

\section{Source catalogue, multiwavelength counterparts, and multiplicity}

\subsection{Source extraction and multiwavelength counterparts}

\begin{figure*}
\begin{center}
\includegraphics[width=0.315\textwidth]{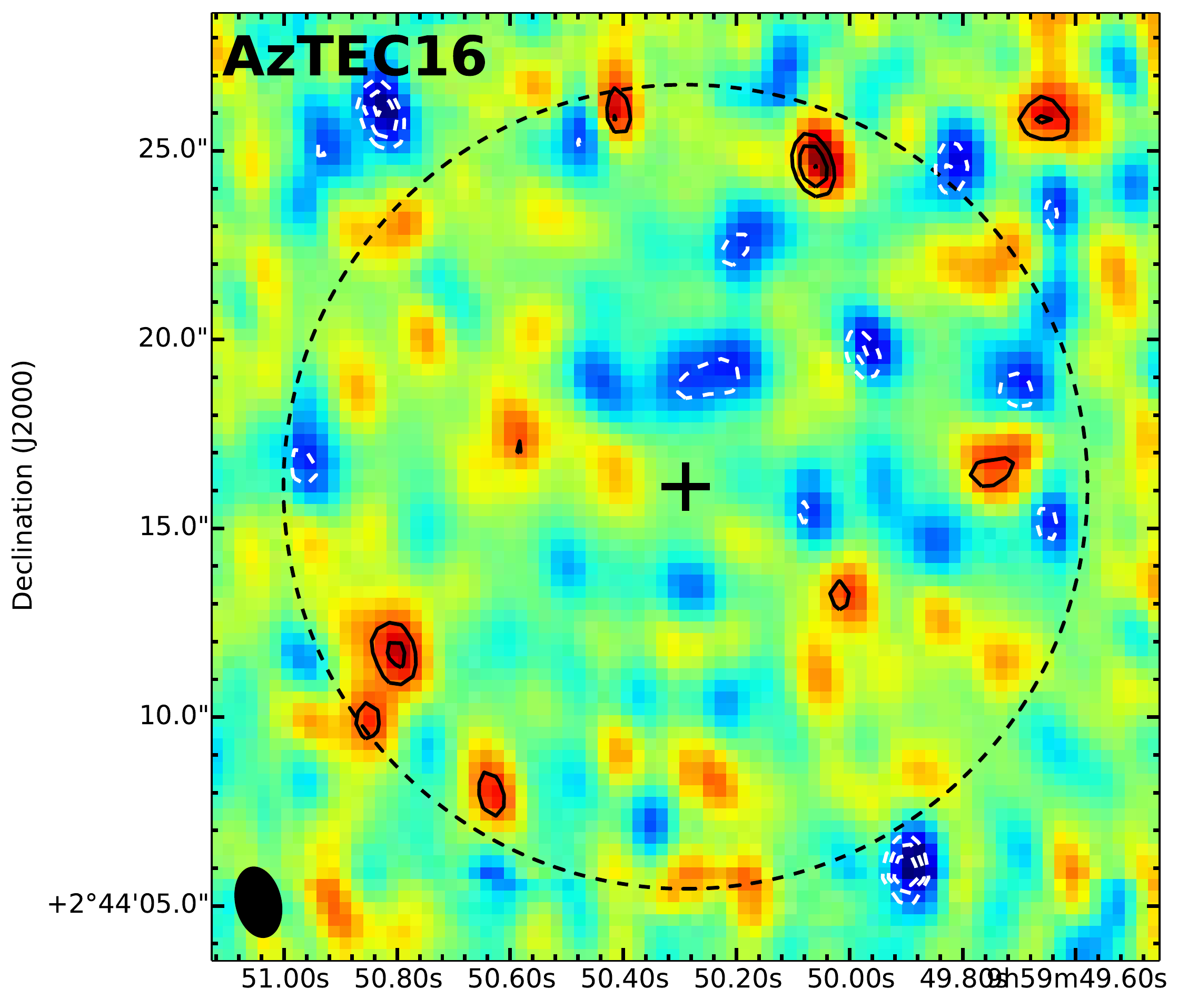}
\includegraphics[width=0.31\textwidth]{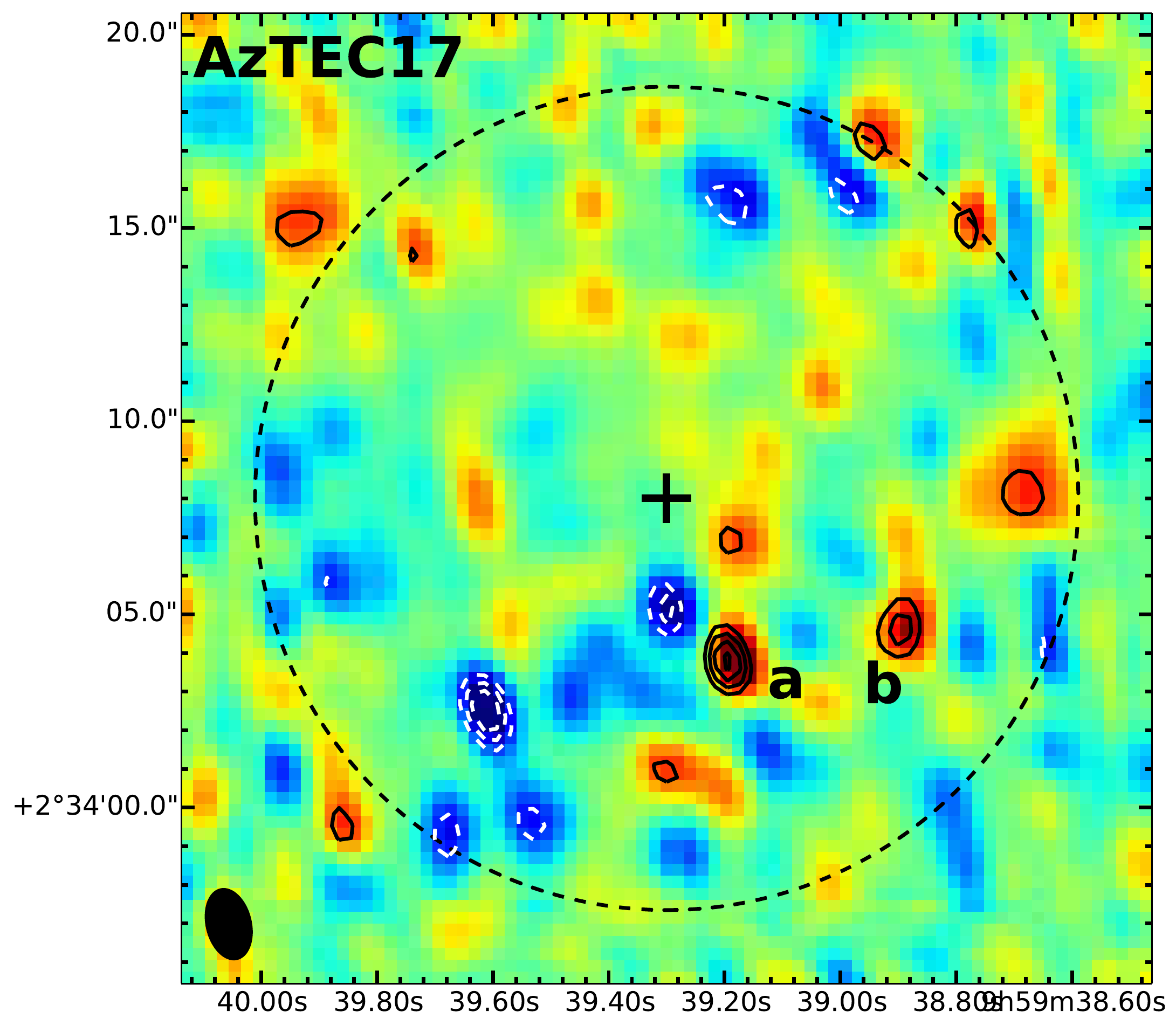}
\includegraphics[width=0.31\textwidth]{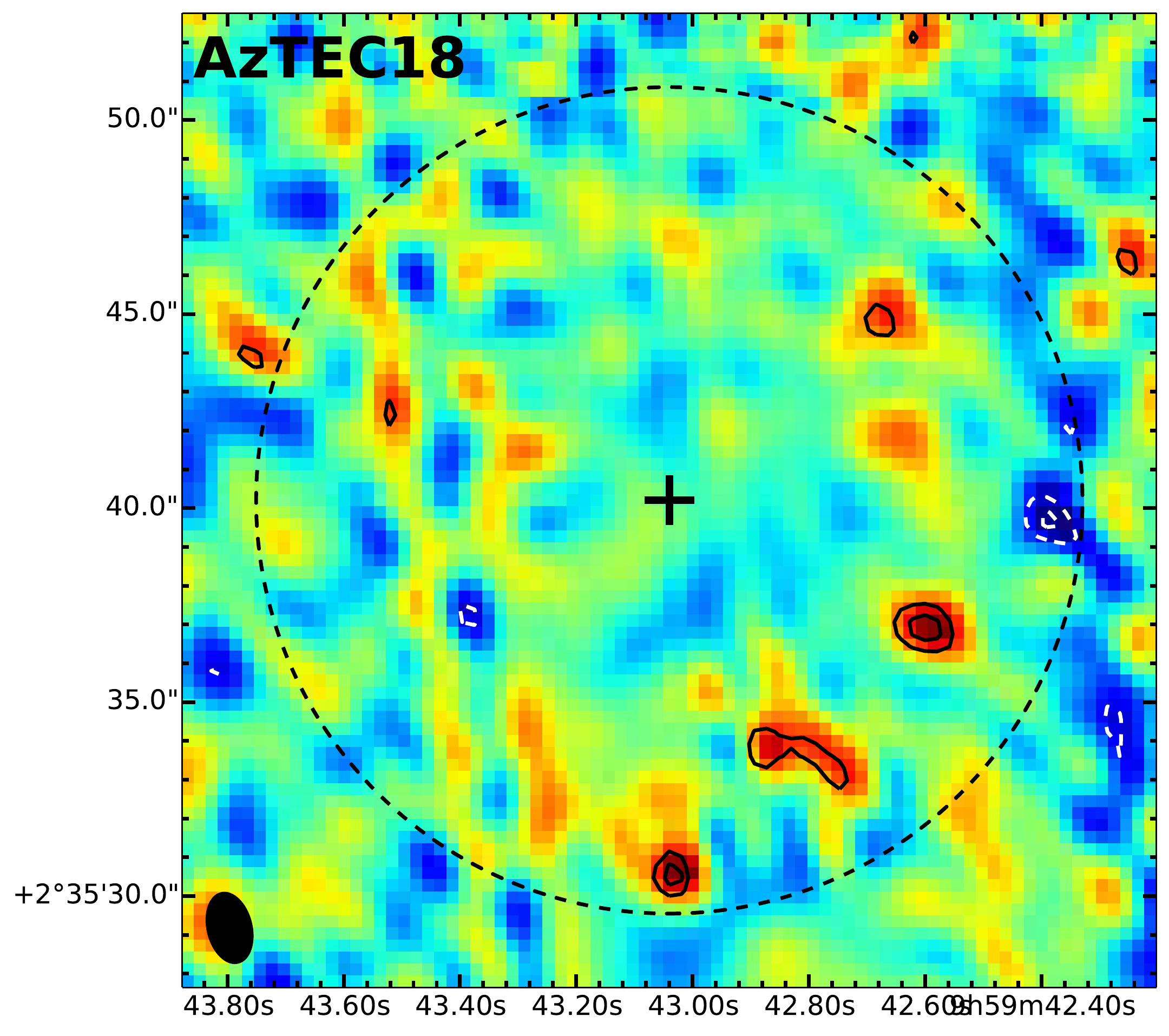}
\includegraphics[width=0.317\textwidth]{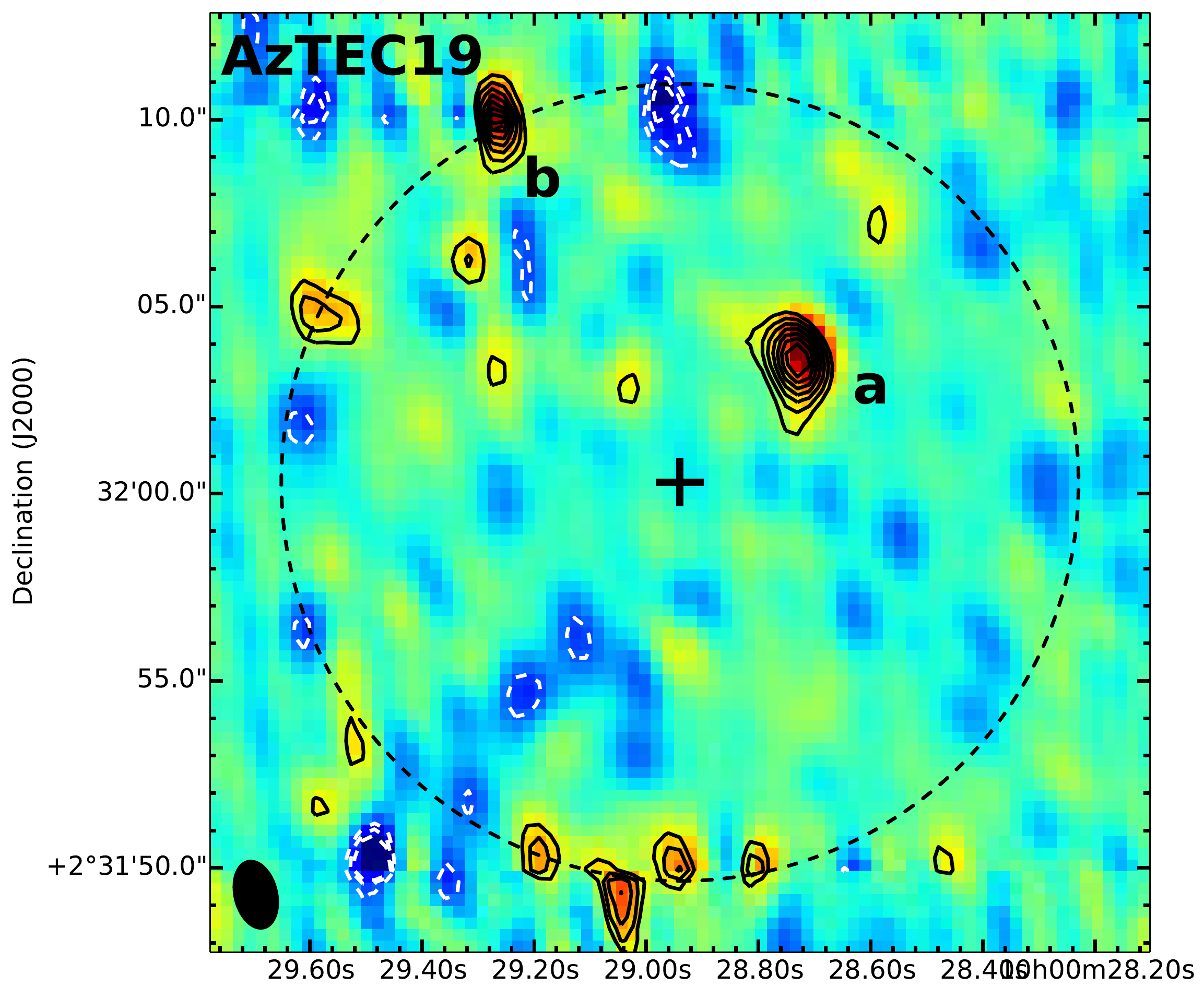}
\includegraphics[width=0.31\textwidth]{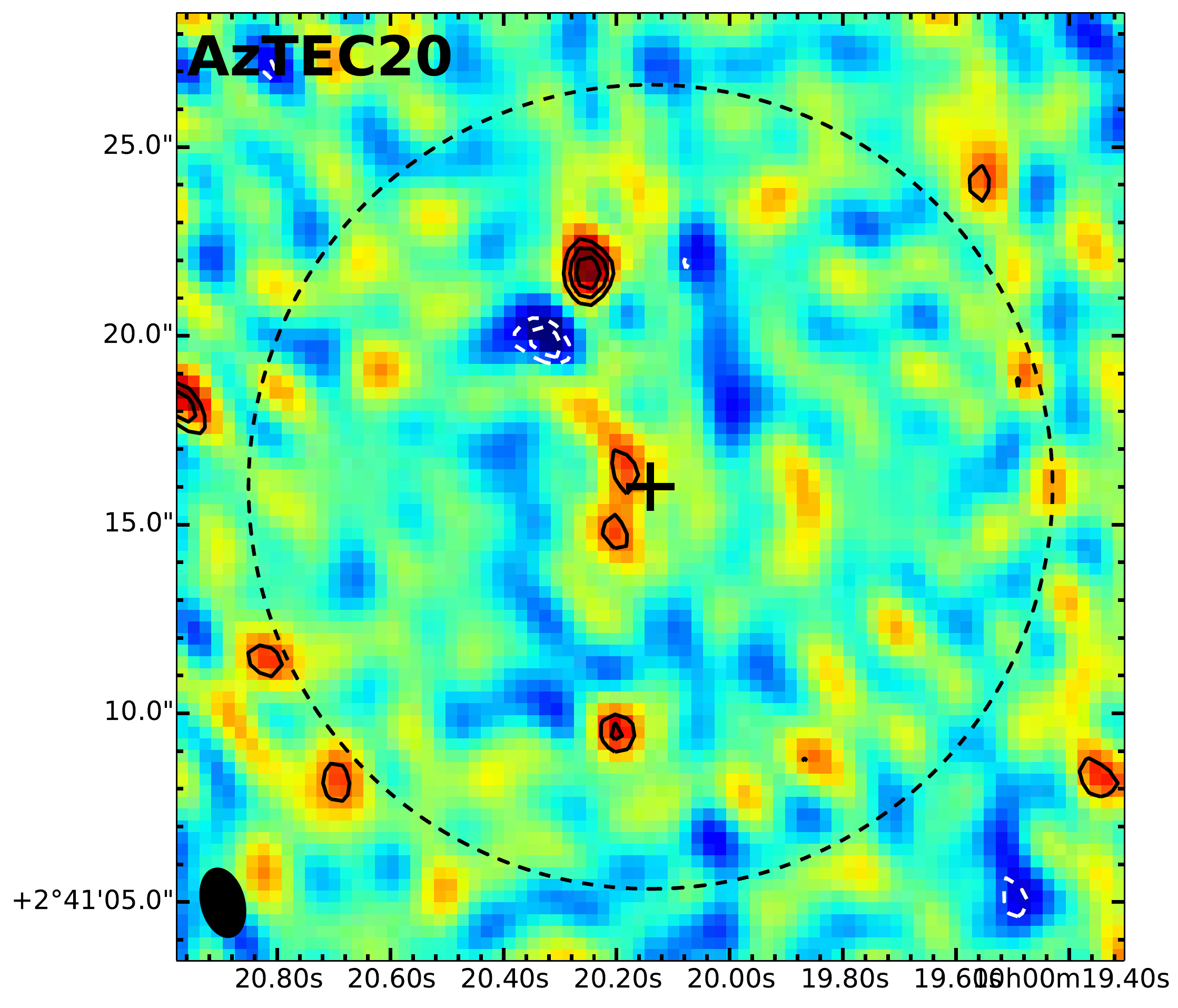}
\includegraphics[width=0.31\textwidth]{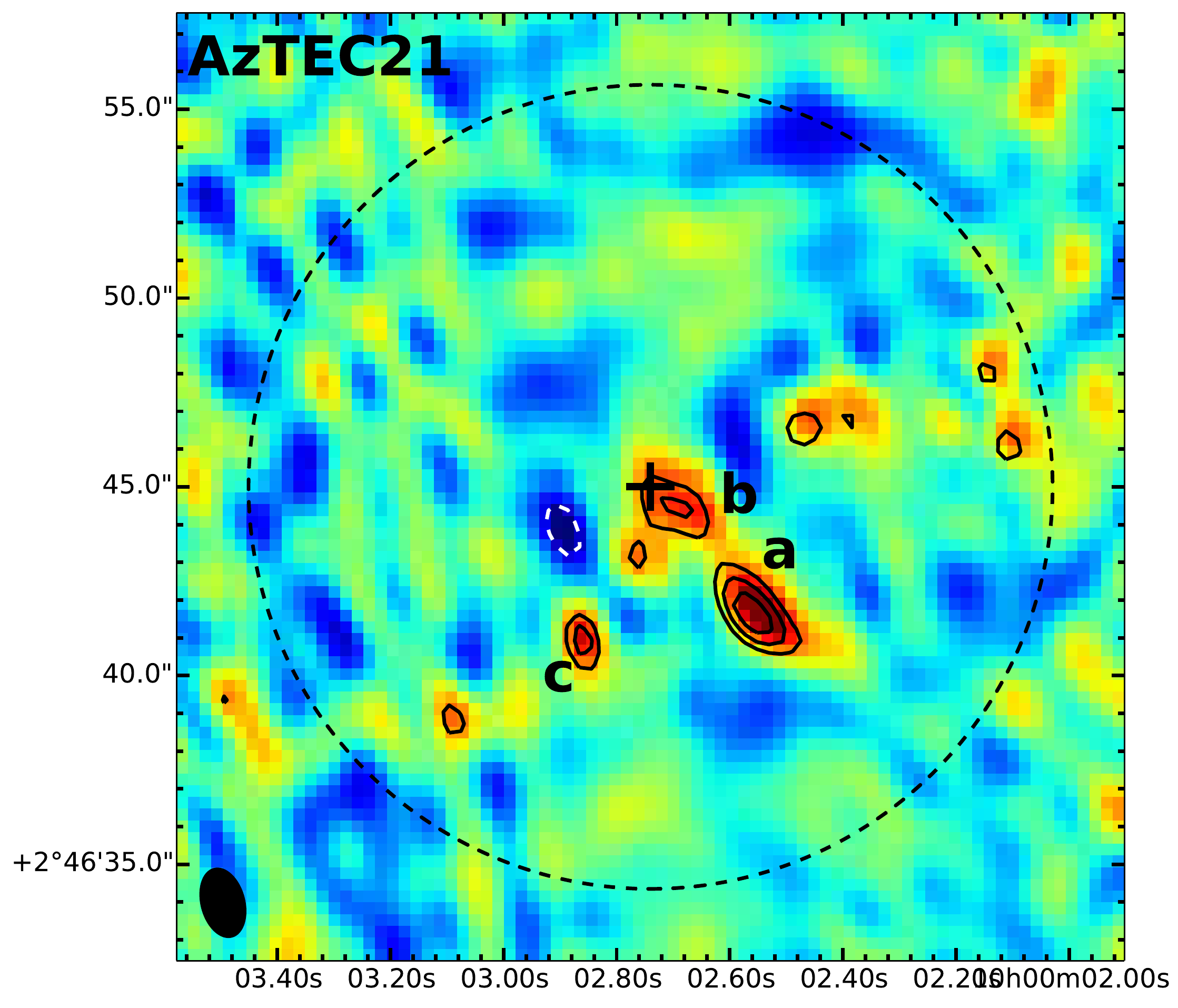}
\includegraphics[width=0.31\textwidth]{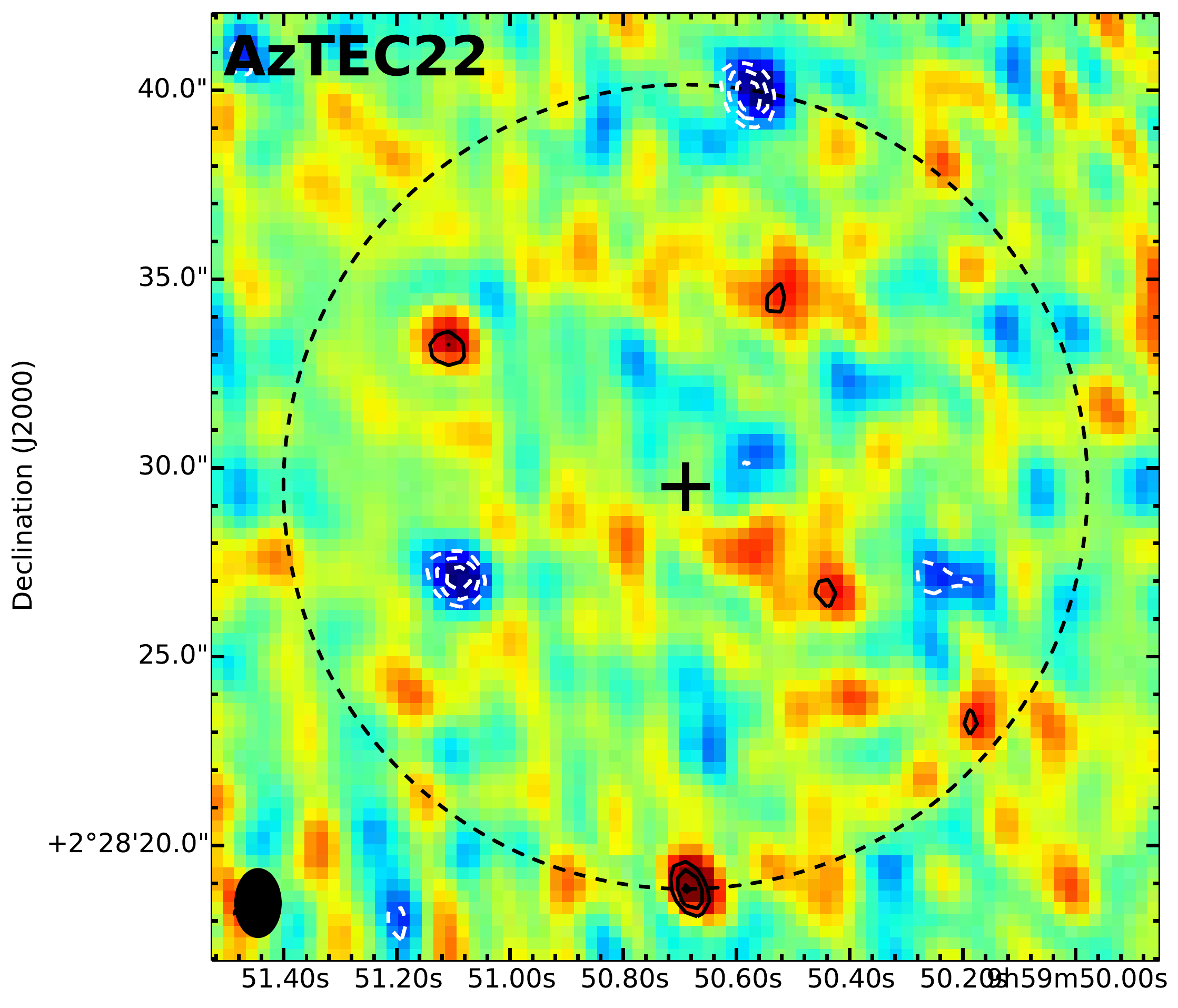}
\includegraphics[width=0.317\textwidth]{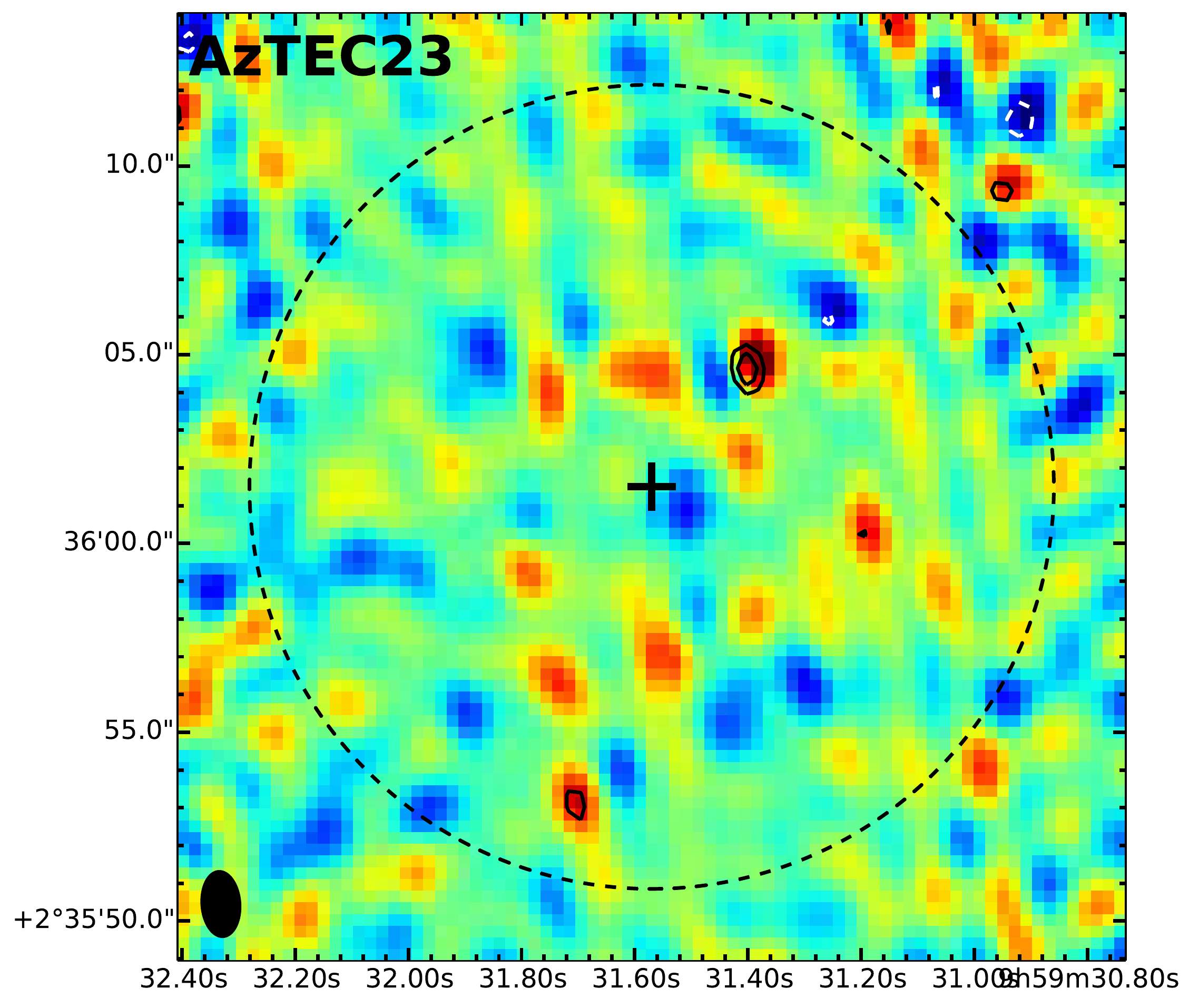}
\includegraphics[width=0.31\textwidth]{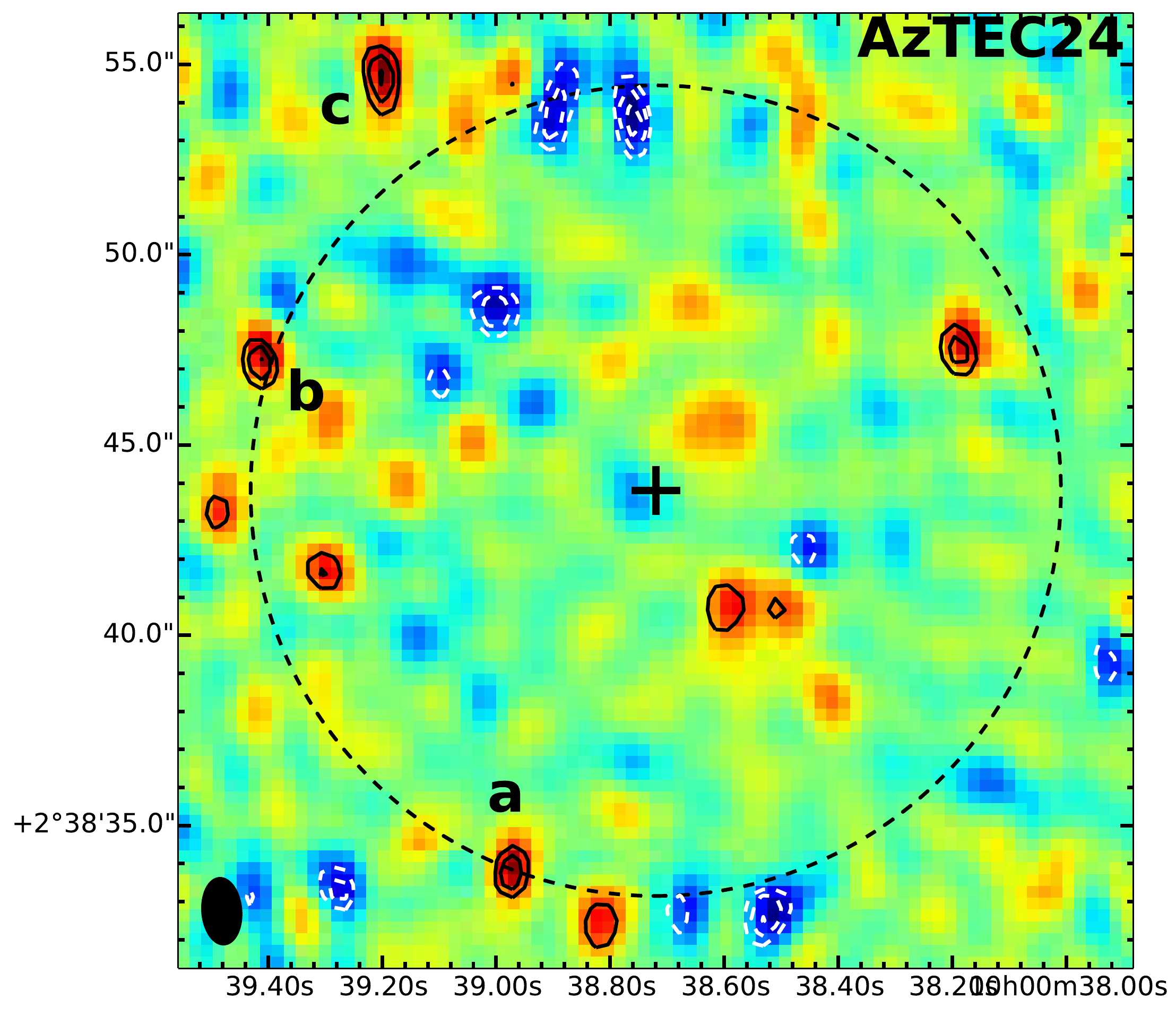}
\includegraphics[width=0.319\textwidth]{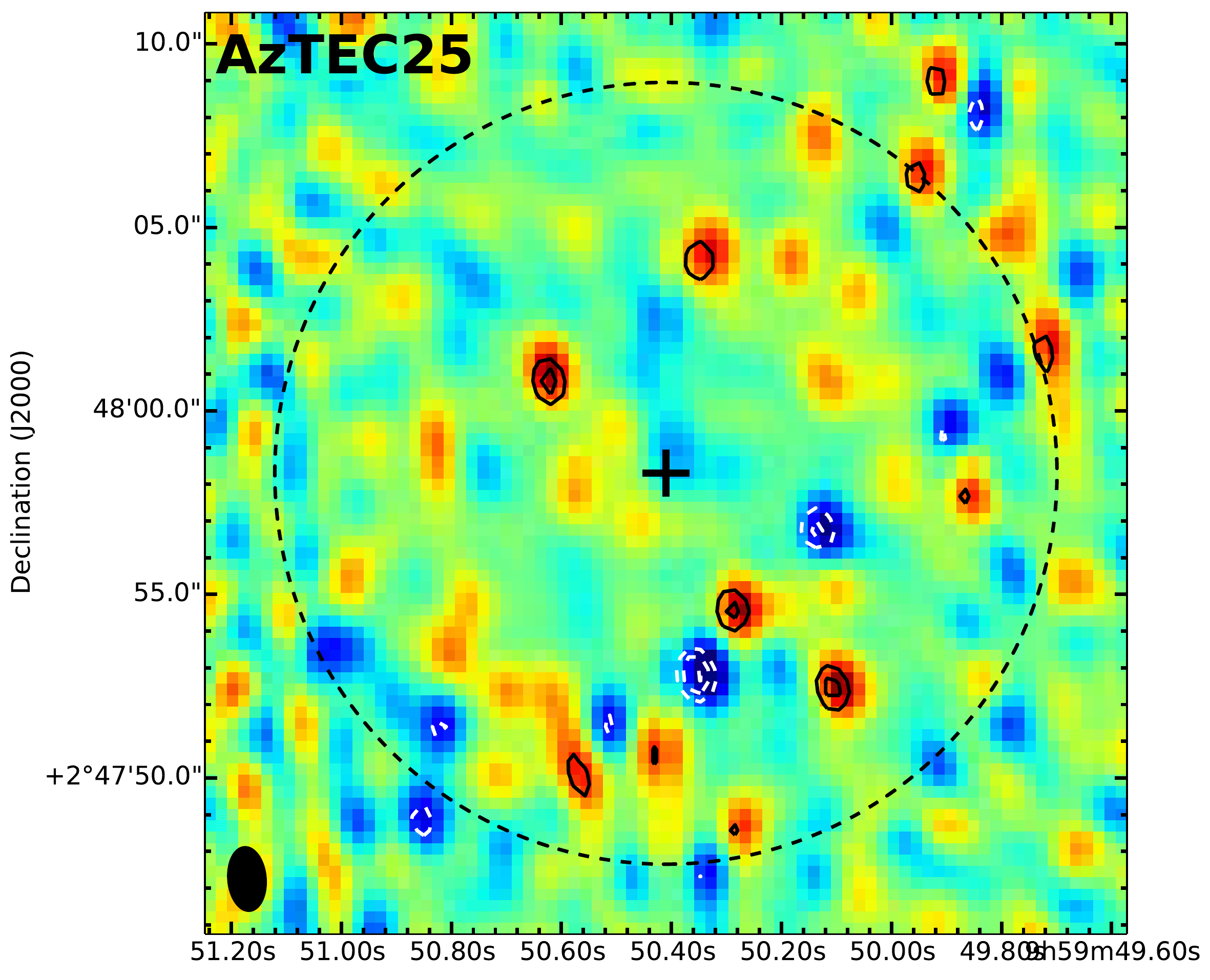}
\includegraphics[width=0.31\textwidth]{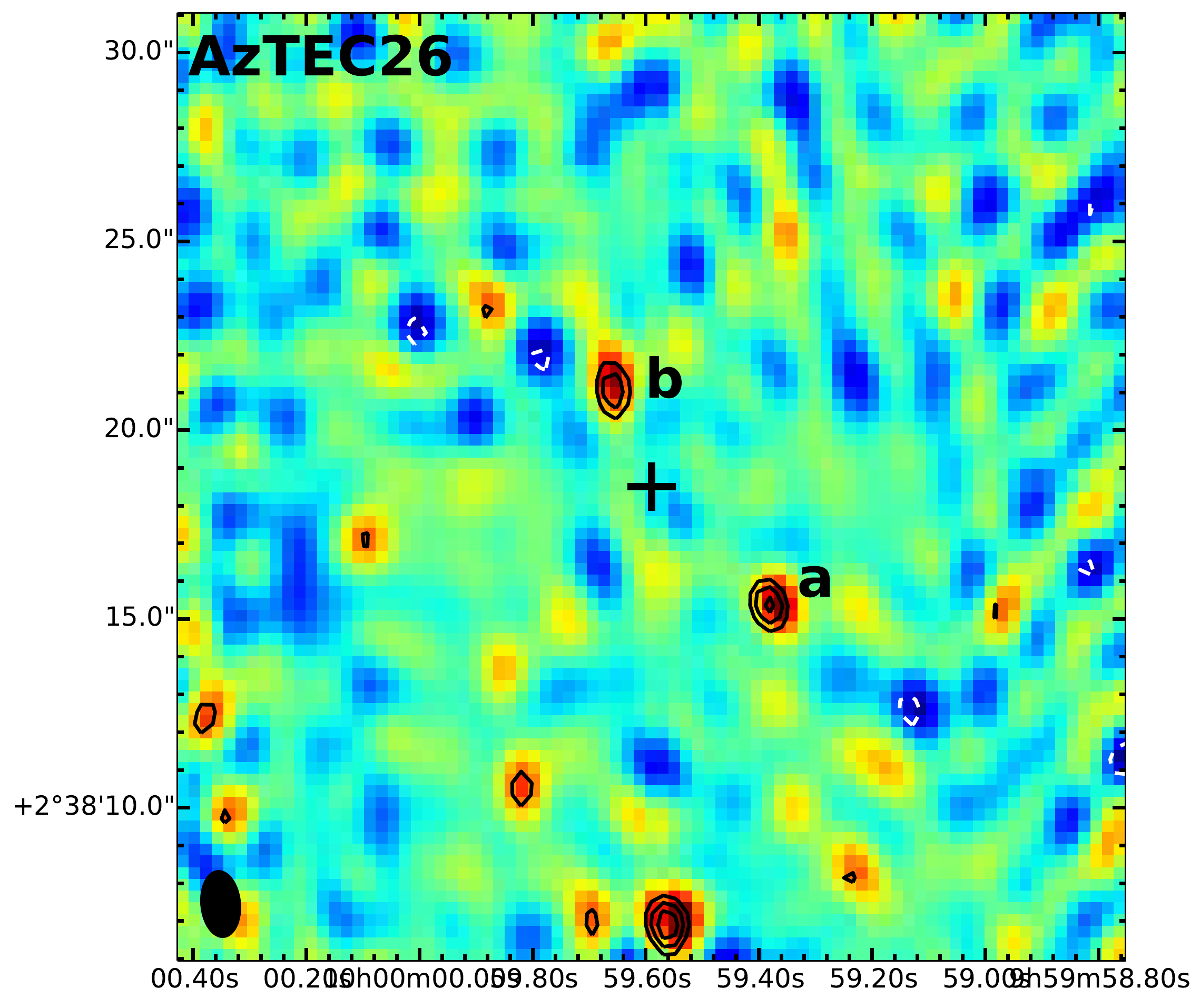}
\includegraphics[width=0.31\textwidth]{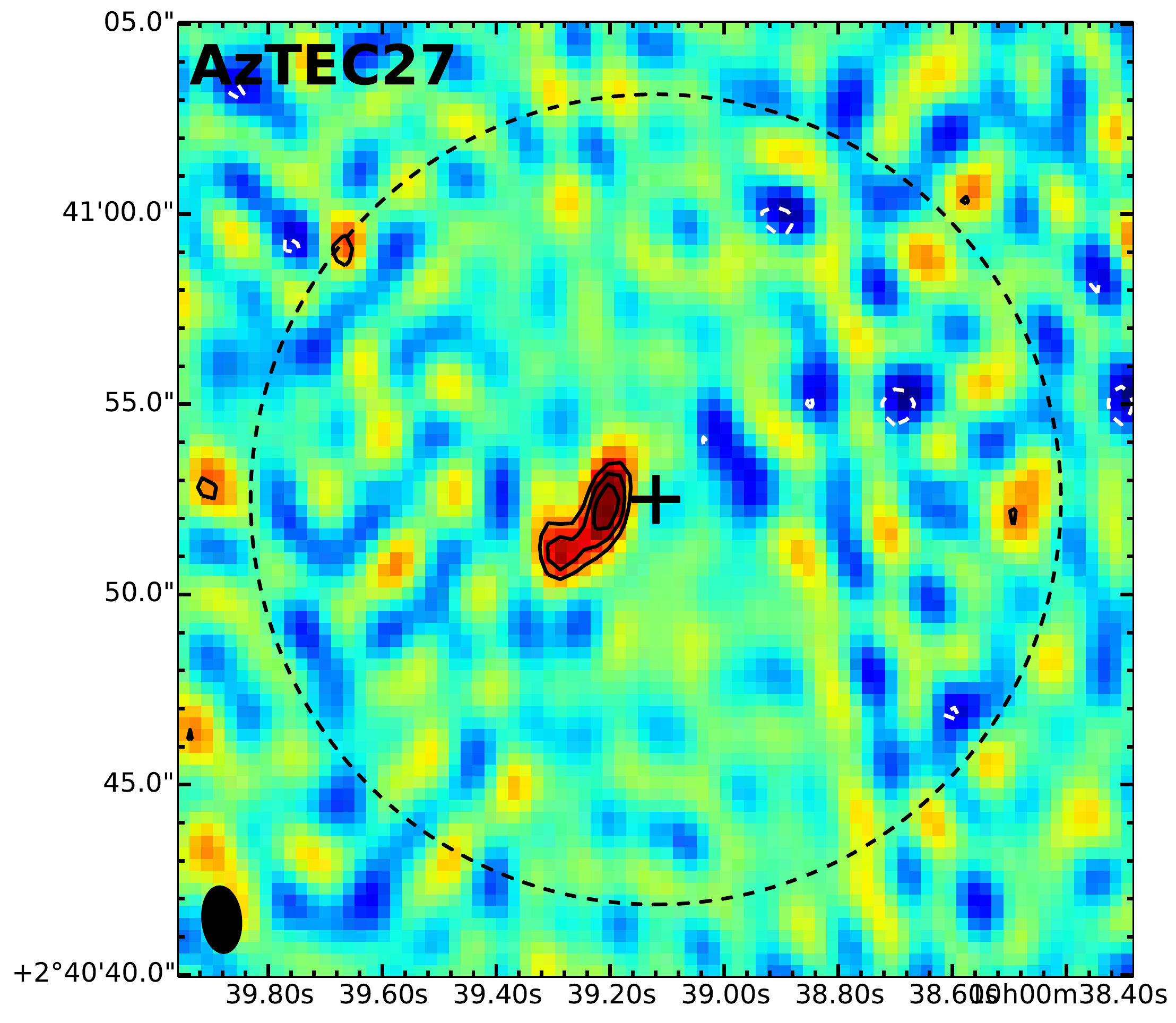}
\includegraphics[width=0.3195\textwidth]{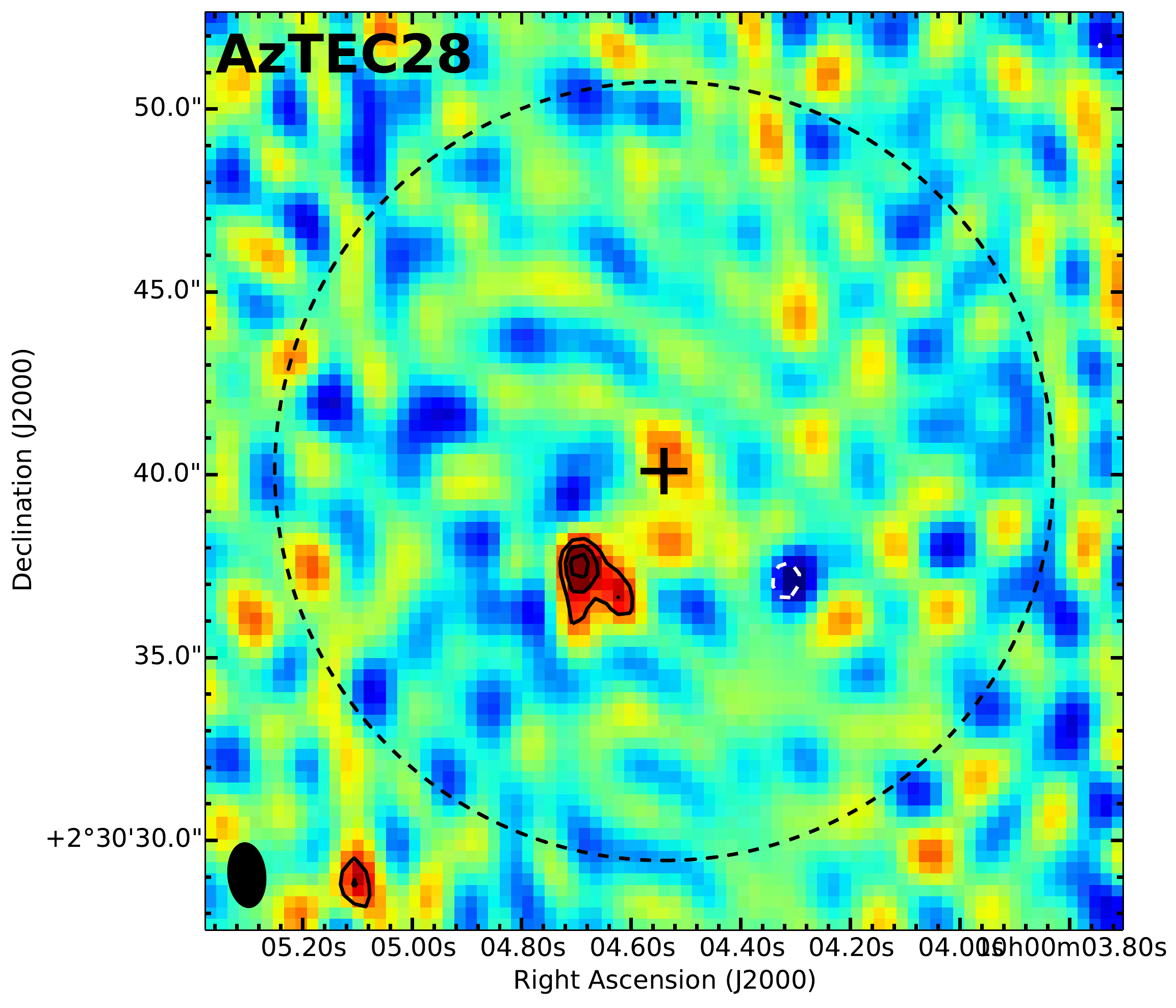}
\includegraphics[width=0.31\textwidth]{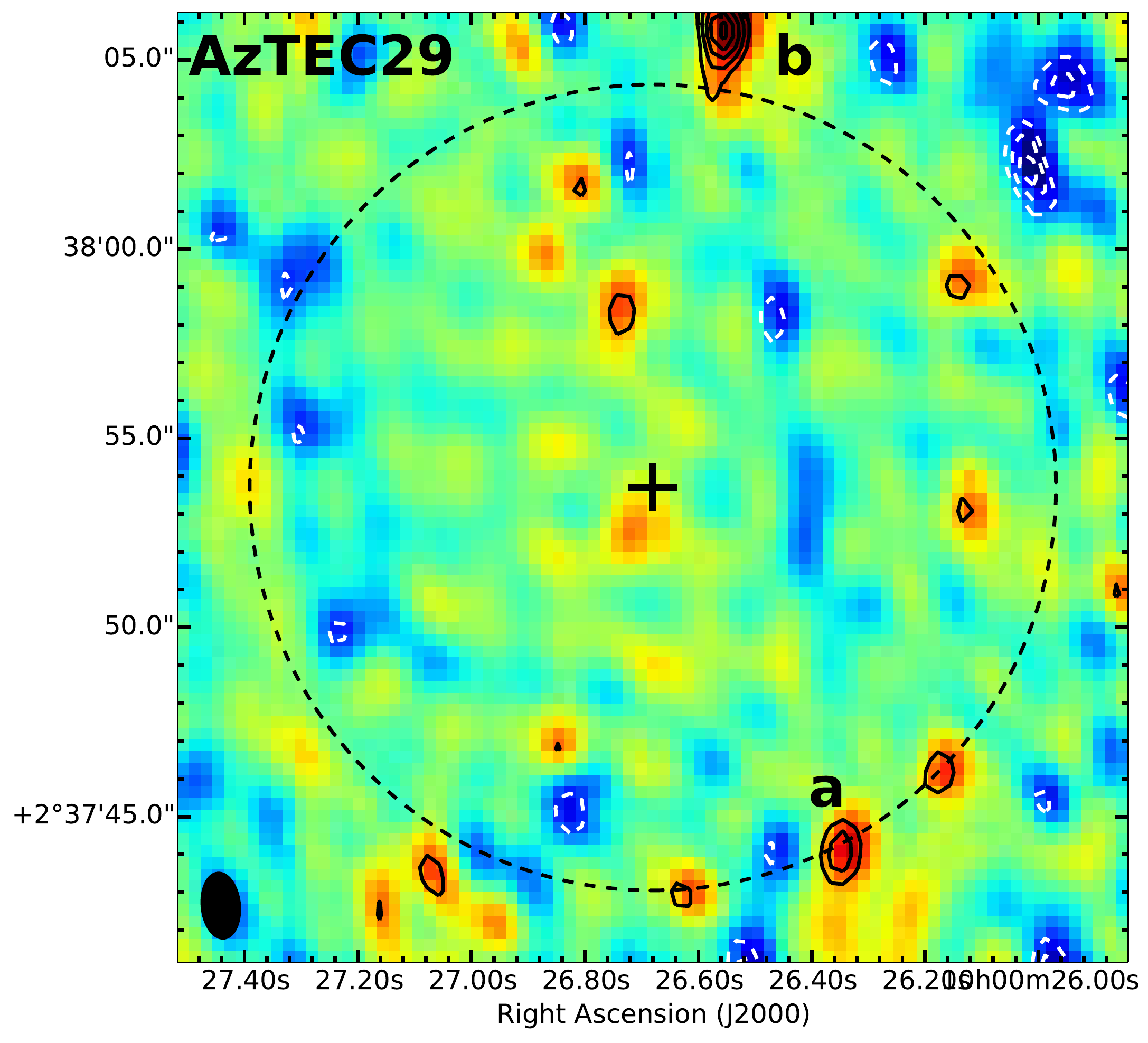}
\includegraphics[width=0.315\textwidth]{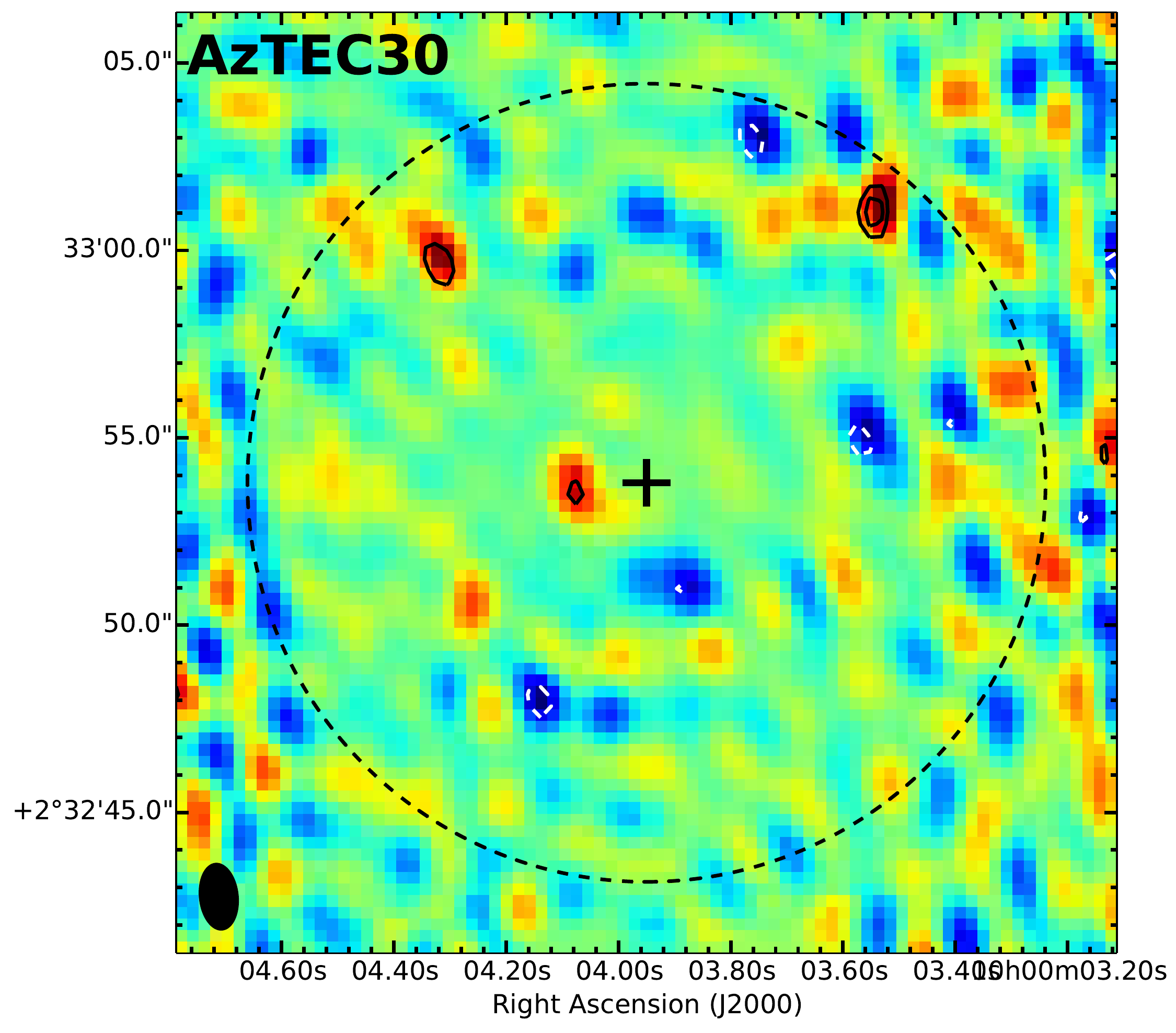}
\caption{The PdBI 1.3 mm continuum images towards AzTEC16--30, annotated with the source designations 
(AzTEC17a, 17b, etc.). The images are shown with linear scaling, and the overlaid solid contours start from $3\sigma$ and 
increase in steps of $1\sigma$. The white dashed contours show the corresponding negative features 
(starting from $-3\sigma$). Each image is centred on the phase centre position, 
i.e. the original AzTEC 1.1 mm centroid from Scott et al. (2008) marked by the plus 
sign. The dashed circle shows the primary beam FWHM of $21\farcs3$, while the filled 
ellipse in the bottom left corner shows the synthesised beam size. The images 
are not corrected for the primary beam attenuation.}
\label{figure:pdbi}
\end{center}
\end{figure*}

The PdBI 1.3 mm images towards AzTEC16--30 are shown in Fig.~\ref{figure:pdbi}. We note that the PdBI PB at the observed frequency, 
$21\farcs3$, closely resembles the size of the JCMT/AzTEC beam of about $18\arcsec$ (FWHM), and that our observation wavelength (1.3~mm) 
is very close to that of the original discovery observations (JCMT/AzTEC) by Scott et al. (2008; 1.1~mm), facilitating comparison between 
these two studies. 

To systematically search for sources in the final, cleaned maps, we followed Hodge et al. (2013) and Karim et al. (2013). 
Briefly, we used an Interactive Data Language ({\tt IDL})-based source-extraction routine that first finds pixel 
values above $2.5\sigma$, where $\sigma$ is the rms noise determined using non-overlapping rectangular apertures across the map. The size of each 
aperture was taken to be large compared to the synthesised beam so that each of them will yield a representative sampling of independent beams. Since some 
apertures contained physical signal (i.e. sources), the value of $\sigma$ was taken to be the median of all different rms values. The value of $\sigma$ 
derived this way is consistent with the {\tt GILDAS}-derived map rms noise given in Col.~(6) in Table~\ref{table:obs}: the first value was found to be 6\% higher 
on average than the second value (the median ratio between the two rms values is 1.05). The routine then models the source emission 
within a $3\arcsec \times 3\arcsec$ region using a Gaussian and the MCMC (Metropolis-Hastings Markov chain Monte Carlo) algorithm. 
Extended sources are fit for six parameters (peak surface brightness, peak position, minor axis, major-to-minor axis ratio, and position 
angle) while for point sources the size is fixed to that of the synthesised beam (leaving only three free parameters). 
To generate a robust catalogue of PdBI-detected sources we adopt the approach already used by Smol{\v c}i{\'c} et al. (2012b) and consider 
sources with ${\rm S/N}>4.5$ in the PdBI 1.3 mm maps as robust detections, while sources with $4<{\rm S/N}\leq4.5$ are considered real 
only if they have a multiwavelength counterpart within a search radius of $1\farcs5$ (within the COSMOS Ultra\-VISTA NIR, \textit{Spitzer} 
IRAC/MIPS, VLA 10 or 20~cm radio catalogues). When multiple PdBI source candidates are detected, we label them AzTEC17a, AzTEC17b, etc.

In total we find 22 sources, 11 of which are associated with multiwavelength counterparts (three 
additional sources have a nearby ACS $I$-band source but no ``counterparts'' at other wavelengths). 
We report their positions (J2000.0 equatorial coordinates and offset from the phase centre) and 
primary-beam corrected flux densities in Table~\ref{table:properties}. 
For the unresolved sources, we report the peak flux density, while for the (marginally) resolved sources, AzTEC21a, 27, and 28, 
we give the total flux density derived from the best-fit six-parameter model. The latter values were also 
independently derived by summing over all pixels within the $2.5\sigma$ contour of 1.3 mm emission, and 
were found to be consistent with the model values. The quoted flux density uncertainties are based on the 
rms noise values and the 20\% absolute calibration error (added in quadrature). We note that inspection of the residual maps of the model Gaussian fits showed 
that AzTEC21a could be well-fitted with a deconvolved FWHM size of $\theta_{\rm maj}\times \theta_{\rm min}=(2\farcs6 \pm 1\farcs2)\times(0\farcs3 \pm 0\farcs5)$. 
However, AzTEC27 and 28 are not as well modelled by a single Gaussian source model. 
For AzTEC27, only the major axis of the elliptical Gaussian could be determined ($\theta_{\rm maj}=3\farcs6$), while the size of AzTEC28 was determined to be 
$\theta_{\rm maj}\times \theta_{\rm min}=(1\farcs5 \pm 0\farcs8)\times(0\farcs6 \pm 0\farcs7)$. The large uncertainties in the sizes reflect that 
these sources are only marginally resolved and/or are not well-represented by a single Gaussian (see e.g. \cite{condon1997}). In all the three cases (AzTEC21a, 27, and 28), 
however, the peak flux density was found to be lower than the total flux density, supporting their marginally extended nature. 
The multiwavelength counterpart IDs of our PdBI sources are reported in Col.~(7) in Table~\ref{table:properties}, 
and the last column lists the projected angular offset from the PdBI source. 
Multiwavelength zoomed-in images towards our sources are provided in Fig.~\ref{figure:stamps}. 
Notes on individual sources are given in Appendix~\textbf{C}.

\subsection{Source catalogue validation}

\begin{figure}[!h]
\centering
\resizebox{\hsize}{!}{\includegraphics{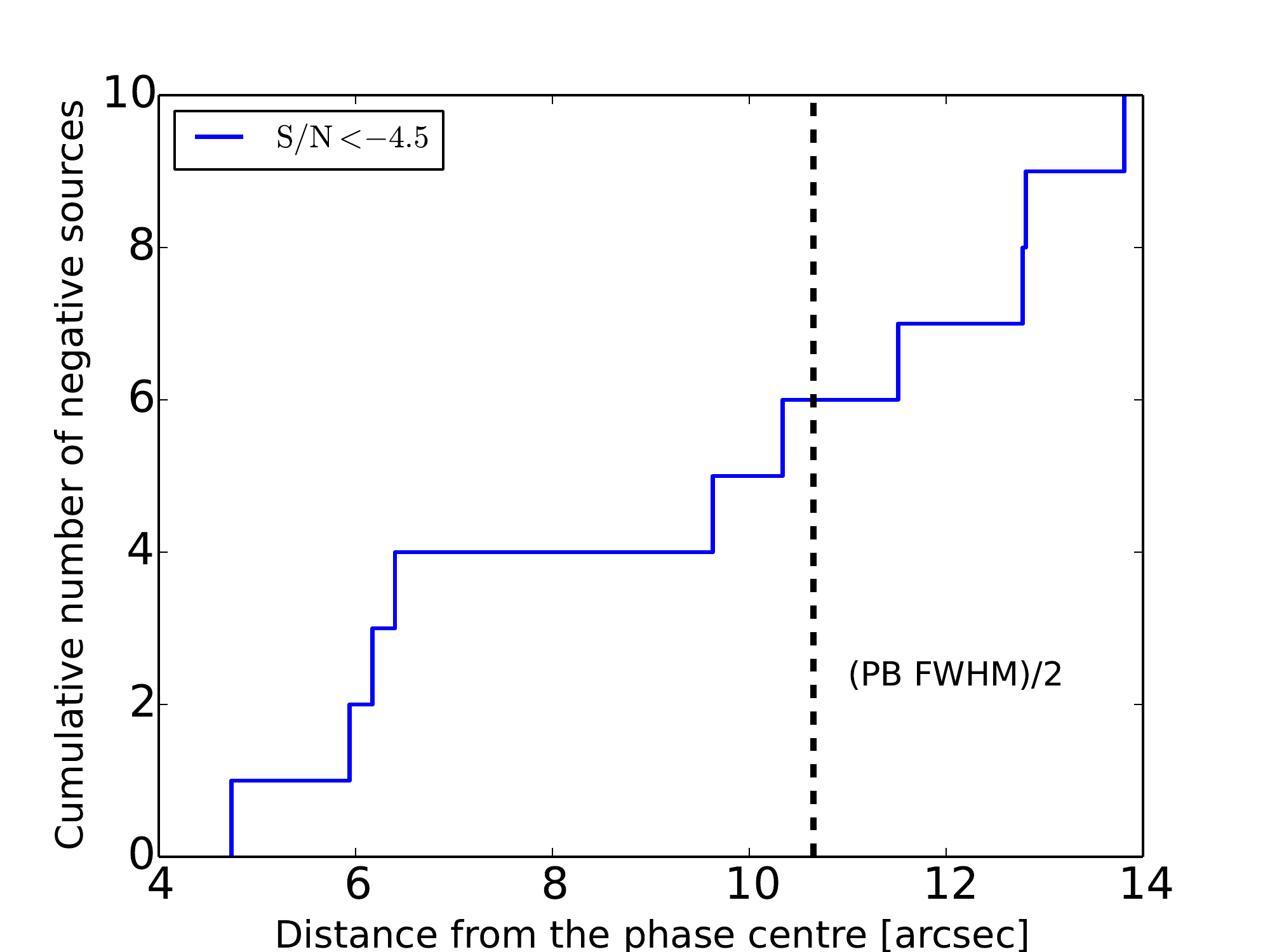}}
\caption{Cumulative number distribution of spurious sources identified in the negative parts of 
the PdBI maps (${\rm S/N }<-4.5$) as a function of angular distance [$\arcsec$] from the phase centre. 
The vertical dashed line marks the half-power radius of the PB.}
\label{figure:spurious}
\end{figure}

To test the robustness of our PdBI source identifications we quantified the number of expected spurious sources by searching for detections in the negative part 
of the PdBI maps in the same way as described above. We found no spurious sources associated with multiwavelength counterparts within a search radius of $1\farcs5$. 
This is consistent with the random match probability within this radius, based on the optical/IR/radio catalogue source densities, of only $\sim0.2\%$ (\cite{smolcic2012b}). 
It also implies that all our PdBI detections with multiwavelength counterparts are likely to be real. We also find that the number of spurious sources increases 
with increasing distance from the phase centre, as shown in Fig.~\ref{figure:spurious}. Out to a distance of $6\arcsec$ from the phase centre only one spurious source is 
expected. This suggests that potentially one of the three sources (AzTEC20, 26b, 28; Table~\ref{table:properties}) detected with ${\rm S/N}>4.5$, within $6\arcsec$ from 
the phase centre and with no multiwavelength counterparts may be spurious. At a distance of $9-13\arcsec$ from the phase centre a total of five spurious sources with 
${\rm S/N}>4.5$ is expected. This suggests that five sources (AzTEC22, 24abc, 30; see Table~\ref{table:properties}) we detect at ${\rm S/N}>4.5$ and within 
$9\farcs3-12\farcs8$ from the phase centre and with no multiwavelength counterparts may be spurious.

\begin{figure}[!h]
\centering
\resizebox{\hsize}{!}{\includegraphics{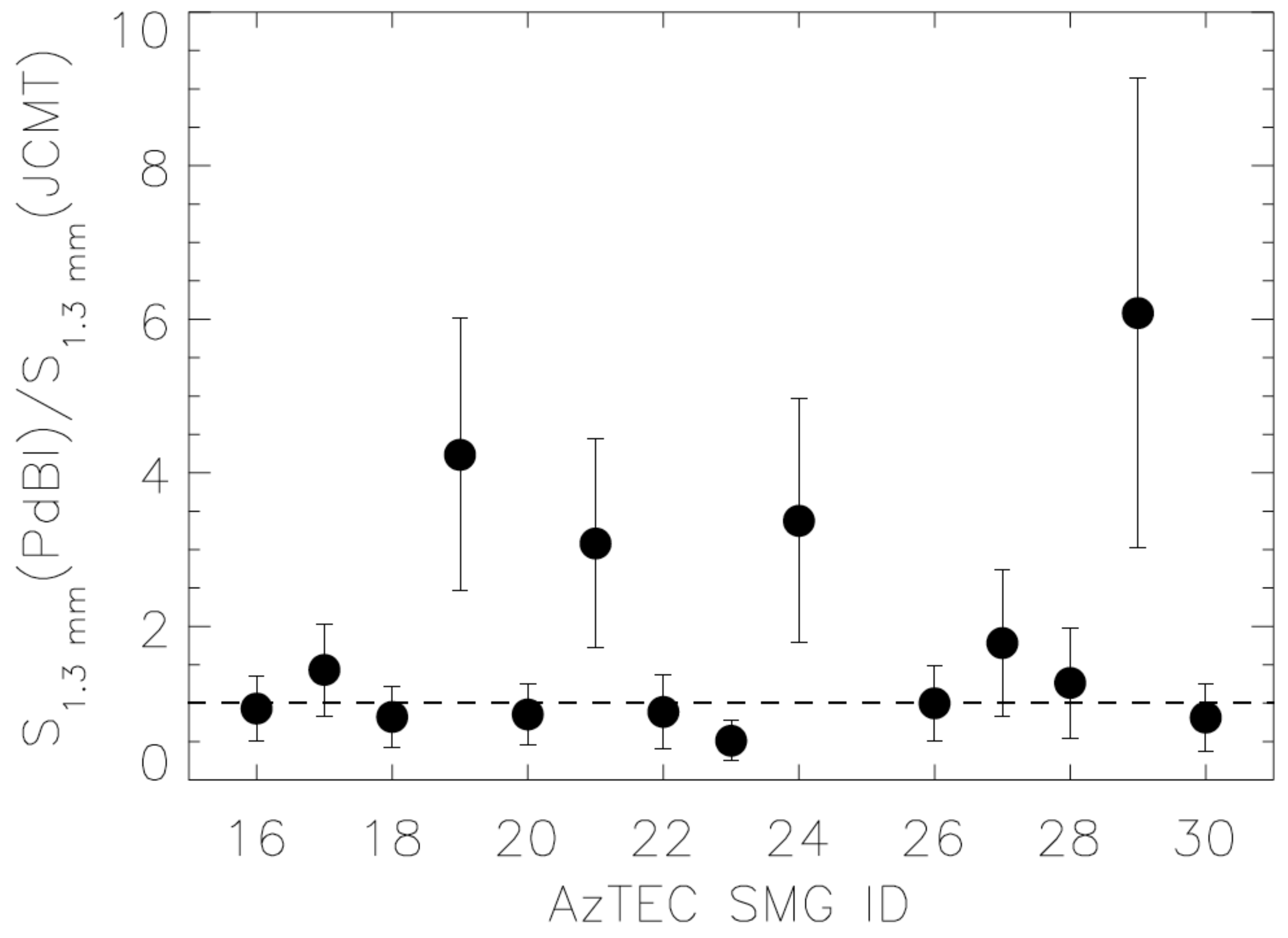}}
\caption{A plot showing the PdBI/JCMT 1.3 mm flux density ratio for the SMGs AzTEC16--30. 
The single-dish JCMT/AzTEC flux densities were scaled from the deboosted 1.1 mm 
flux densities (Table~\ref{table:sample}) by assuming that the dust emissivity index is $\beta=1.5$. 
The dashed line represents the line of equality. 
}
\label{figure:flux}
\end{figure}

To test the validity of our sources further, we compare for each PdBI source the PdBI 1.3~mm flux to the AzTEC/JCMT 1.1~mm flux in 
Fig.~\ref{figure:flux}. For this, the deboosted AzTEC 1.1 mm flux densities given in Table~\ref{table:sample} were scaled down using the common assumption 
that the dust emissivity index is $\beta=1.5$ (e.g. \cite{dunne2001}; \cite{barger2012} and references therein). 
In case multiple PdBI sources are extracted from the target field, the sum of their flux densities is plotted 
(AzTEC25 was not detected in the PdBI map). As can be seen, the two values are generally in reasonable agreement 
with each other. We note, however, that a one-to-one correspondence is not expected given observational limitations (such as flux deboosting methods and possible 
blending of multiple sources in the low-resolution single-dish data; see e.g. Fig.~2 in \cite{smolcic2012b} for a comparison of SMA and JCMT/AzTEC 890 $\mu$m flux 
densities for AzTEC1--15). The four sources showing higher PdBI flux densities than expected from single-dish measurements are AzTEC19, 21, 24, and 29. 
Given that AzTEC24 has no multiwavelength counterpart associated, \textit{and} that its PdBI flux is inconsistent with that from the JCMT/AzTEC survey may suggest that 
the three identified PdBI sources within the AzTEC24 field are spurious. On the other hand, emission features at the border of and/or just outside the PB FWHM 
(as in the cases of AzTEC19, 24, and 29) may not be contributing to the JCMT/AzTEC source flux densities detected at $18\arcsec$ resolution, and deboosted JCMT/AzTEC flux 
densities may also be invalid in this comparison for sources with widely separated components. 
In summary, we conclude that 4 out of the 22 PdBI identified sources may be spurious. 

\subsection{Multiplicity of single-dish detected, JCMT/AzTEC SMGs}

Submillimetre sources identified in single-dish studies can be composed of multiple components, and this multipli\-city can be revealed by
higher-resolution (interferometric) observations. These components can typically be associated with individual galaxies that might be physically related (and potentially
interacting), or might be just chance alignments of galaxies lying at different redshifts. As the multiplicity fraction of submm sources depends on the initial re\-solution 
of the single-dish observations and on the depth of the interferometric follow-up, it is not sensible to provide a simple definition for multiplicity. 
The present study benefits from the fact that the PB of our PdBI follow-up observations is well matched to the beam FWHM of the JCMT observations used for identifying 
the initial source sample ($21\farcs3$ versus $18\arcsec$; Sect.~3.1). This facilitates the analysis of source blending in the single-dish maps. 
Based on our PdBI source detections six out of the 15 AzTEC16--30 SMGs (or 5/15 if AzTEC24 is treated as spurious) separate into multiple components. 
This corresponds to $40\pm16\%$ ($33\pm15\%$), where the errors are Poisson errors. Within the sample of the brightest COSMOS JCMT/AzTEC SMGs (AzTEC1--15) only 
2/15 ($13\pm9\%$) separate into multiple components (\cite{younger2007}, 2009; \cite{smolcic2012b}). 
Combining these two samples then yields that 7--8 SMGs in our flux- and S/N-limited single-dish detected, COSMOS JCMT/AzTEC sample of 30 sources separate 
into multiple sources when observed at $\lesssim2\arcsec$ angular resolution. This corresponds to $\sim25\pm9\%$, and will be discussed further in Sect.~5.1.

\begin{table*}
\caption{Characteristics and optical-to-MIR and VLA 20 cm counterparts of the SMGs identified in the PdBI 1.3 mm maps.}
{\tiny
\begin{minipage}{2\columnwidth}
\centering
\renewcommand{\footnoterule}{}
\label{table:properties}
\begin{tabular}{c c c c c c c c}
\hline\hline 
Source & $\alpha_{2000.0}$ & $\delta_{2000.0}$ & $S_{\rm 1.3\, mm}$ & S/N$_{\rm 1.3\, mm}$ & Offset & Candidate & $r$ \\
       & [h:m:s] & [$\degr$:$\arcmin$:$\arcsec$] & [mJy] & & [$\arcsec$] & counterpart ID\tablefootmark{a}  & [$\arcsec$]\\ 
\hline
AzTEC16 & 09 59 50.069 & +02 44 24.50 & $2.07\pm0.62$ & 5.0 & 9.0 & 501 (ACS-$I$) & 0.35\\
AzTEC17a & 09 59 39.194 & +02 34 03.83 & $1.58\pm0.43$ & 6.2 & 4.5 & 613229 (\textit{GALEX}) & 1.29 \\
         &              &              &               &     &     & 1496 (ACS-$I$) & 1.36\\
         &              &              &               &     &     & 1475165 (COSMOS+UltraVISTA) & 1.38\\
         &              &              &               &     &     & 271694 (New UltraVISTA) & 1.47\\
         &              &              &               &     &     & 250117 (IRAC) & 0.30\\
         &              &              &               &     &     & J095939.19+023403.6 (VLA Deep) & 0.22\\
AzTEC17b & 09 59 38.904 & +02 34 04.69 & $1.53\pm0.46$ & 4.5 & 6.8 & 1475223 (COSMOS+UltraVISTA) & 1.25 \\
AzTEC18 & 09 59 42.607 & +02 35 36.96 & $1.78\pm0.54$ & 4.5 & 7.3 & 1044 (ACS-$I$) & 0.76\\
        &              &              &               &     &     & 1471053 (COSMOS+UltraVISTA) & 0.82\\
AzTEC19a & 10 00 28.735 & +02 32 03.84 & $3.98\pm0.91$ & 10.3 & 4.4 & 1593 (ACS-$I$) & 0.40 \\
         &              &              &               &     &     & 1455882 (COSMOS+UltraVISTA) & 0.20 \\
         &               &              &               &     &     & 262214 (New UltraVISTA) & 0.20 \\
         &               &              &               &     &     & 242501 (IRAC) & 0.63 \\
         &               &              &               &     &     & 2158 (MIPS 24 $\mu$m) & 0.15 \\
         &               &              &               &     &     & J100028.70+023203.7 (VLA Deep) & 0.53 \\
AzTEC19b & 10 00 29.256 & +02 32 09.82 & $5.21\pm1.30$ & 9.7 & 10.6 & 597821 (\textit{GALEX}) & 0.66 \\
         &              &              &               &     &     & 1486 (ACS-$I$) & 0.68 \\ 
         &               &              &               &    &     & 1455681 (COSMOS+UltraVISTA) & 0.63 \\
         &               &              &               &    &     & 262766 (New UltraVISTA) & 0.74\\ 
AzTEC20 & 10 00 20.251 & +02 41 21.66 & $1.85\pm0.49$ & 6.0 & 5.9 & \ldots & \ldots\\
AzTEC21a & 10 00 02.558 & +02 46 41.74 & $3.37\pm1.03$ & 5.8 & 4.3 & 711447 (\textit{GALEX}) & 0.35 \\
         &              &              &               &     &     & 1688587 (COSMOS+UltraVISTA) & 0.15 \\
         &              &              &               &     &     & 328878 (New UltraVISTA) & 1.00 \\
         &              &              &               &     &     & 297396 (IRAC) & 1.11\\
         &              &              &               &     &     & 7262 (MIPS 24 $\mu$m) & 1.46\\
AzTEC21b & 10 00 02.710 & +02 46 44.51 & $1.34\pm0.38$ & 4.2 & 0.7 & 711786 (\textit{GALEX}) & 0.53 \\
         &              &              &               &     &     & 1688585 (COSMOS+UltraVISTA) & 0.45\\
AzTEC21c & 10 00 02.856 & +02 46 40.80 & $1.27\pm0.40$ & 4.5 & 4.5 & 712026 (\textit{GALEX}) & 1.49 \\
         &              &              &               &     &     & 297223 (IRAC) & 1.46 \\
AzTEC22 & 09 59 50.681 & +02 28 19.06 & $1.82\pm0.59$ & 5.1 & 10.5 & \ldots & \ldots \\ 
AzTEC23 & 09 59 31.399 & +02 36 04.61 & $0.99\pm0.29$ & 4.9 & 4.0 & 1494354 (COSMOS+UltraVISTA) & 1.24 \\
AzTEC24a & 10 00 38.969 & +02 38 33.90 & $1.79\pm0.53$ & 4.9 & 10.6 & \ldots & \ldots\\
AzTEC24b & 10 00 39.410 & +02 38 46.97 & $1.72\pm0.53$ & 5.0 & 10.8 & \ldots & \ldots\\
AzTEC24c & 10 00 39.194 & +02 38 54.46 & $2.85\pm0.78$ & 5.1 & 12.8 & \ldots & \ldots\\
AzTEC25\tablefootmark{b} & \ldots & \ldots & \ldots & \ldots & \ldots & \ldots & \ldots\\
AzTEC26a & 09 59 59.386 & +02 38 15.36 & $0.98\pm0.28$ & 5.4 & 4.4 & 647670 (\textit{GALEX}) & 1.05 \\
         &              &              &               &     &     & 930 (ACS-$I$) & 1.07 \\
         &              &              &               &     &     & 1709726 (COSMOS+UltraVISTA) & 0.94 \\
         &              &              &               &     &     & 291786 (New UltraVISTA) & 0.83 \\
AzTEC26b & 09 59 59.657 & +02 38 21.08 & $0.90\pm0.26$ & 4.8 & 2.8 & \ldots & \ldots \\
AzTEC27 & 10 00 39.211 & +02 40 52.18 & $3.36\pm0.97$\tablefootmark{c} & 6.0 & 1.4 & 666 (ACS-$I$) & 1.15 \\
AzTEC28 & 10 00 04.680 & +02 30 37.30 & $2.38\pm0.77$ & 5.5 & 3.5 & \ldots & \ldots\\
AzTEC29a & 10 00 26.351 & +02 37 44.15 & $2.45\pm0.67$ & 4.7 & 10.8 & 736 (ACS-$I$) & 1.41\\
AzTEC29b & 10 00 26.561 & +02 38 05.14 & $9.01\pm2.39$ & 7.3 & 11.6 & 1685295 (COSMOS+UltraVISTA) & 0.76\\
AzTEC30 & 10 00 03.552 & +02 33 00.94 & $1.53\pm0.45$ & 4.6 & 9.3 & \ldots & \ldots\\
\hline 
\end{tabular} 
\tablefoot{The coordinates given in columns~(2) and (3) refer to the peak position determined by the three-parameter point-source model fit for 
all sources with peak surface brightness of ${\rm S/N}>4$. The flux densities given in column~(4) 
are primary-beam corrected, and the quoted errors include the 20\% calibration uncertainty. For the (marginally) resolved sources 
AzTEC21a, AzTEC27, and AzTEC28, the total flux density was derived from the best-fit six-parameter source model. 
The S/N ratio given in column~(5) refers to the extracted value 
in the non-primary-beam-corrected map. 
In column~(6), we give the PdBI source offset from the phase centre, i.e. the AzTEC 1.1~mm centroid. The last column gives the projected angular 
separation between the 1.3~mm peak position and the counterpart position.\tablefoottext{a}{The references for the different source catalogues are as follows: 
\textit{ACS} $I$-band (\cite{leauthaud2007}); \textit{GALEX} (M.~A.~Zamojski et al., in prep.); COSMOS photometry catalogue (\cite{capak2007}); UltraVISTA 
(\cite{capak2007}; \cite{mccracken2012}; \cite{ilbert2013}); \textit{Spitzer} IRAC/MIPS (S-COSMOS team); VLA Deep (\cite{schinnerer2010}).}\tablefoottext{b}{The 1.3~mm features in AzTEC25 did not fulfil our detection criteria.}\tablefoottext{c}{AzTEC27 is probably subject to gravitational lensing (see Appendices~C and D), and our lens modelling suggests a magnification factor of $\mu=2.04\pm0.16$. In this case, AzTEC27's intrinsic flux density at observing-frame 1.3 mm would be $1.65\pm0.49$~mJy.}}
\end{minipage} }
\end{table*}

\section{Redshift distribution of 1.1-mm selected SMGs in the COSMOS field}

\subsection{Redshifts for AzTEC1--15}

Among the SMA-detected SMGs AzTEC1--15, there are eight spectroscopic redshifts reported in the literature 
(AzTEC1, 2, 3, 5, 6, 8, 9, and 11; see \cite{smolcic2012b}; their Tables~1 and 4 and references therein). However, as described in Appendix~B, 
the spectroscopically determined redshifts for AzTEC5, 6, and 9 are uncertain because of the poor quality of the spectrum or contamination 
by foreground galaxies. Besides these cases, we discuss the updated redshifts among AzTEC1--15 in Appendix~B 
(the redshifts are listed in Table~\ref{table:redshifts}). 

\subsection{Redshifts for AzTEC16--30}

The optical/IR photometric redshifts of AzTEC16--30 were computed (when possible) by fitting 
optimised spectral templates to their spectral energy distributions (SEDs) using the {\tt HYPERZ} code 
(\cite{bolzonella2000})\footnote{{\tt http://webast.ast.obs-mip.fr/hyperz/}}. 
The redshift was treated as a free parameter and determined using a $\chi^2$ mi\-nimisation method, i.e. the most likely redshift 
was determined statistically (see also \cite{smolcic2012a},b). We used the Calzetti et al. (2000) extinction law, 
and the $V$-band extinction, $A_{\rm V}$, was varied from 0 to 5 mag. The allowed redshift range was $z \in [0,\,7]$. 
The spectral templates used were generated with the {\tt GRASIL} radiative-transfer code (\cite{silva1998}; \cite{iglesis2007})
and optimised for SMGs using the method described by Micha{\l}owski et al. (2010). 
When comparing results for their tested spectral templates, Smol{\v c}i{\'c} et al. (2012a) found that 
the best agreement (i.e. the tightest $\chi^2$ distribution) between the spectroscopic and photometric redshifts 
(sample of eight COSMOS SMGs) is obtained when employing the templates derived by Micha{\l}owski et al. (2010). 
Smol{\v c}i{\'c} et al. (2012b; see their Fig.~7) repeated 
the analysis using a larger source sample, and their similar results lend further support to the reliability of 
the Micha{\l}owski et al. (2010) spectral-template library. We therefore decided to perform our photo-$z$ analysis 
using this library of templates. The optical/IR SEDs for the identified SMG counterparts are shown in Fig.~\ref{figure:sed}. 
The template-fitting method of finding the best photo-$z$ solution is based on the minimisation of the reduced chi-square ($\chi_{\rm red}^2$) value, which 
is the chi-square divided by the number of degrees of freedom (dof) [see Eq.~(1) in \cite{bolzonella2000}]. 
The {\tt HYPERZ} program yields the probability associated with the minimum $\chi_{\rm red}^2$ for each redshift step, 
$P(z)=\exp(-\chi_{\rm red}^2/2)$. The absolute (total) chi-square ($\chi_{\rm tot}^2$) distribution for each source as a function of redshift is presented 
in a panel next to the corresponding SED plot in Fig.~\ref{figure:sed}. We computed the formal lower and 
upper $99\%$ confidence limits for the best-fit photo-$z$ value. Formally, these were calculated from 
the $\chi^2$ probability distribution $P(\chi^2 \leq \Delta_{\chi^2 }\vert \nu)=0.99$ (\cite{avni1976}), where 
$\Delta_{\chi^2}$ is the increment in $\chi^2$ required to cover the parameter space region 
with a $z$ confidence of $99\%$, and $\nu$ is the number of dof. The confidence interval (CI) equals the set of 
all $z$ values that satisfy the condition $\chi^2(z)-\chi_{\rm min}^2\leq \Delta_{\chi^2 }$.

For those sources with no optical/IR counterparts, the above method could not be used to derive their redshift. 
The sources that are not spurious are likely to lie at high redshift or/and be heavily obscured by dust. 
Since the radio and submm continuum have very different $K$-corrections, 
the radio/submm flux-density ratio strongly depends on the source redshift. 
As proposed by Carilli \& Yun (1999, 2000), the spectral index between 860~$\mu$m (350~GHz) and 20~cm (1.4~GHz), 
$\alpha_{1.4}^{350}$, can be used to estimate the redshift. The 860 $\mu$m flux densities were estimated from 
the 1.3~mm values by assuming that the dust emissivity index is $\beta=1.5$. 
On the basis of this, we used the mean radio-to-submm spectral index, $\langle \alpha_{1.4}^{350} \rangle$, 
predicted from 17 low-redshift star-forming galaxies by Carilli \& Yun (2000), 
to set constraints on the source redshift. For 1.4~GHz non-detections, 
we used the $3\sigma$ upper limit to $S_{\rm 1.4\, GHz}$ (typically $\lesssim0.05$~mJy) to 
derive a lower limit to $\alpha_{1.4}^{350}$, hence a lower limit to $z$. 
The uncertainty in the radio/submm redshift was determined from those associated with the flux densities. 
For AzTEC27, which is subject to gravitational lensing, the differential lensing effects were assumed to be negligible 
(i.e. the boost in flux density was assumed to be independent of wavelength), hence the radio/submm flux 
density ratio was assumed to be independent of the magnification factor. We note that a 1.4~GHz non-detection could simply 
be related to a spurious PdBI 1.3 mm source. Therefore, some of the derived lower limits to $z$ should be taken with caution. 

The redshifts of AzTEC16-30 are discussed in more detail in Appendix~C. In summary, for one source (AzTEC17a) we have a spectroscopic redshift, 
for nine sources we have derived photometric redshifts, and for 12 sources we have submm-radio flux ratio based redshift estimates.

\subsection{Redshift distribution of AzTEC1--30}

All the derived and adopted redshifts for AzTEC1--30 are listed in Table~\ref{table:redshifts}. n total, for six (five among AzTEC1-15, 
one among AzTEC16-30) sources in the sample we have a spectroscopic redshift, for 17 a photo-$z$ (eight among AzTEC1-15, nine among AzTEC16-30), while 
for 15 (three for AzTEC1-15, 12 for AzTEC16-30) we have a redshift estimated from the submm-radio flux-density ratio. By combining the up-dated
redshifts of AzTEC1--15 with the present results, we derived the redshift distribution for the combined sample of 
AzTEC1--30. The constructed redshift distributions are shown in the top panel of Fig.~\ref{figure:redshift}.
The bottom panel of Fig.~\ref{figure:redshift} shows the probability density distribution [$P \propto \exp(-\chi^2/2)$; a kernel density estimate], 
of our total SMG sample constructed using the same redshift data as in the histograms in the top panel. 
The probability distribution functions (PDFs) were summed for \textit{i)} sources with $z_{\rm spec}$ values, 
where the PDF was assumed to be a delta function [$\delta=\delta(z_{\rm spec})$], \textit{ii)} sources with $z_{\rm phot}$ values 
(PDFs derived using {\tt HYPERZ}), and \textit{iii)} sources that had only lower limits to $z$, in which case the PDF was assumed 
to be a flat function from the lower $z$ limit up to $z=6.5$. Before calculating the overall PDF, the individual 
PDFs were normalised so that their integral becomes unity. From this distribution we derived a median redshift of $\tilde{z}=3.20$ and 68\% CI of 
$z=1.39-5.26$. 

We further calculated the statistical parameters independently using the {\tt R} program package called Nondetects 
And Data Analysis for environmental data (NADA; \cite{helsel2005}), which is an implementation of the statistical methods provided 
by the Astronomy Survival Analysis (ASURV; \cite{feigelson1985}) package. This method robustly takes lower redshift limits into account 
(e.g. \cite{yun2012}). It was assumed that the censored data follow the same
distribution as the actual values, and we used the Kaplan-Meier (K-M) method to construct a model of 
the input data. After applying the K-M survival estimator, we found that the mean, median, standard deviation (std), 
and 95\% CI of the redshifts of AzTEC1--15 are
$\langle z \rangle=3.16\pm0.37$, $\tilde{z}=3.05\pm 0.44$, ${\rm std}=1.48$, and ${\rm CI}=2.44-3.89$, respectively. 
For AzTEC16--30, the corresponding values are $\langle z \rangle =3.02\pm0.20$, 
$\tilde{z}=3.20\pm0.25$, ${\rm std}=0.92$, and ${\rm CI}=2.63-3.40$. 
The combined sample (open grey histrogram in Fig.~\ref{figure:redshift}) has the values 
$\langle z \rangle=3.19\pm0.22$, $\tilde{z}=3.17\pm0.27$, ${\rm std}= 1.35$, and ${\rm CI}=2.76-3.62$. 
We note that the median redshift of 3.20 derived from the redshift PDF shown in the bottom panel of Fig.~\ref{figure:redshift} is 
in excellent agreement with the value we derived using the survival analysis.
The median redshift, $\tilde{z}=3.17\pm0.27$, corresponds to a luminosity distance of 
$d_{\rm L}~=~27.6^{+2.8}_{-2.7}$ Gpc. Finally, we performed a two-sided Kolmogorov-Smirnov (K-S) test between 
the $z$ distributions of AzTEC1--15 and AzTEC16--30. By excluding the lower limits from 
the samples\footnote{In the survival analysis it was assumed that the lower limits 
and exact values have a common distribution.}, the maximum difference between 
the cumulative distributions, i.e. the $D$ statistic, was found to be $D=0.2615$, 
while the associated probability that the two distributions are drawn from 
the same parent distribution is $p=73.4\%$. 
Moreover, the Welch's two-sample $t$-test under the null hypothesis that the two means are equal yields 
a $p$-value of about 0.338 (when excluding the lower limits), meaning that there is no 
evidence for a difference in the mean redshifts of AzTEC1--15 and AzTEC16--30. 
However, as shown in Fig.~\ref{figure:redshift}, the highest redshift SMGs ($z\gtrsim 4.3$) 
in our sample are found among AzTEC1--15. The redshift distributions derived in other SMG surveys, 
and how they compare to the present results, will be discussed in Sect.~5.3.

\begin{figure*}
\begin{center}
\includegraphics[width=0.45\textwidth]{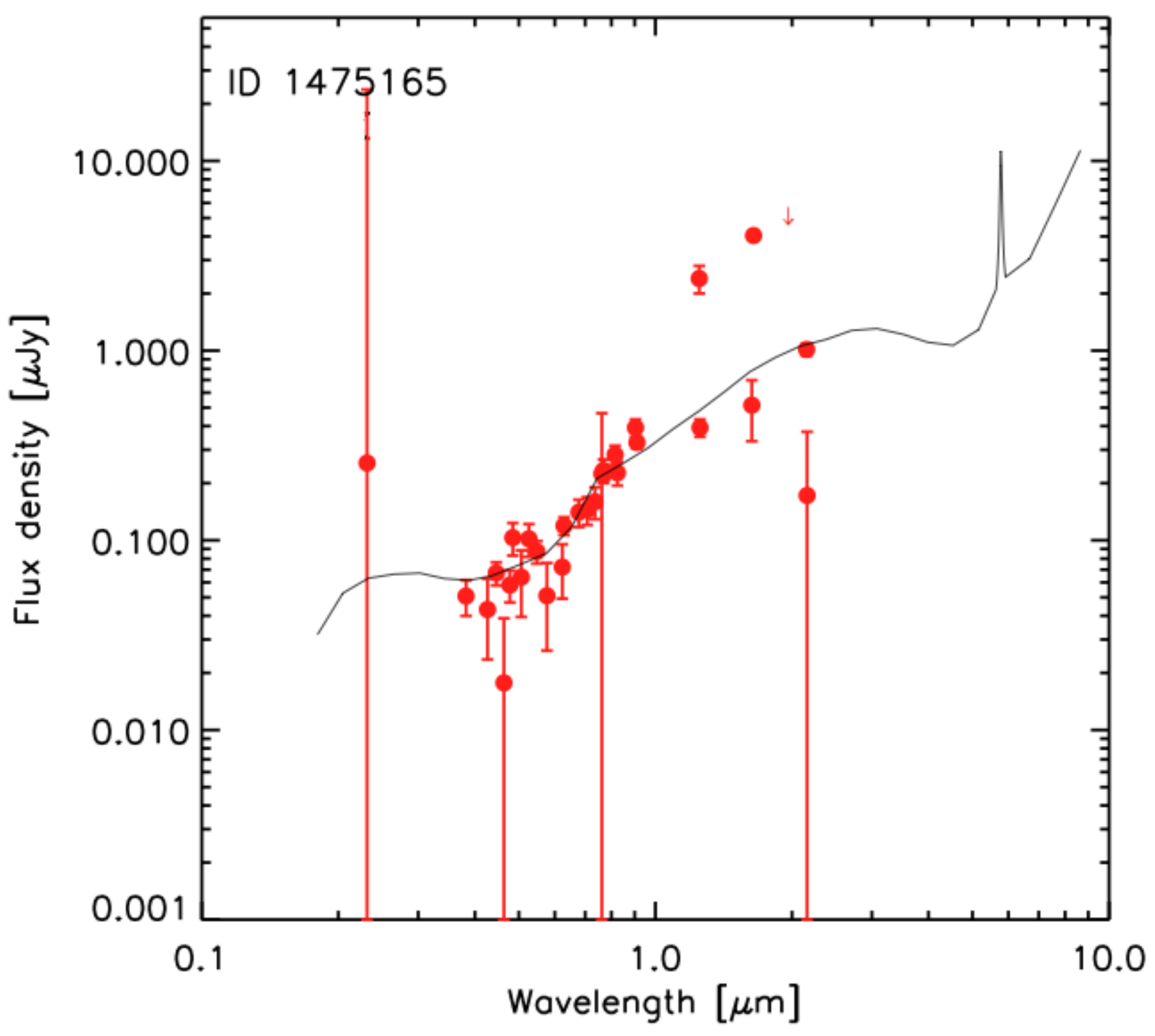}
\includegraphics[width=0.45\textwidth]{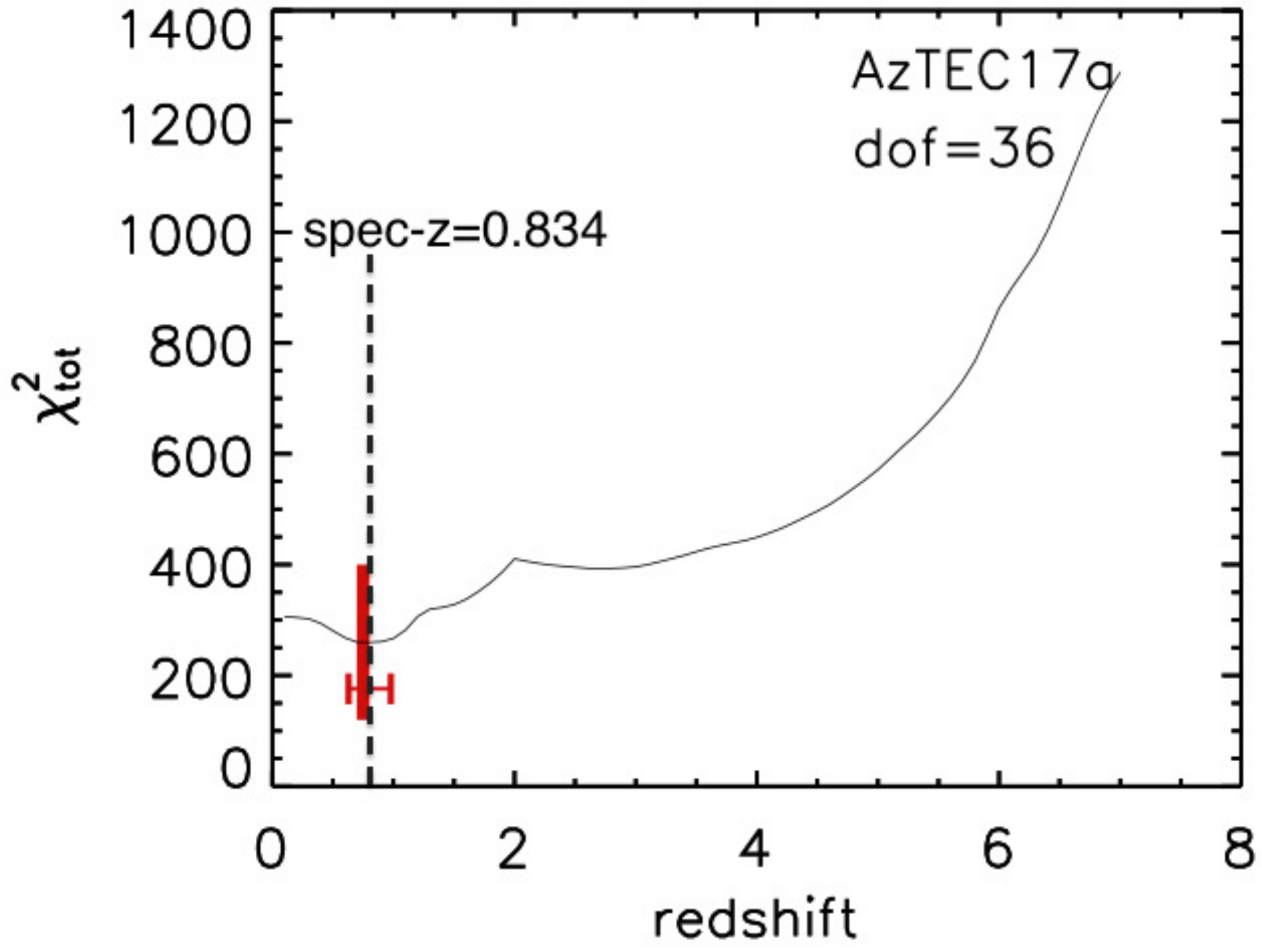}
\includegraphics[width=0.45\textwidth]{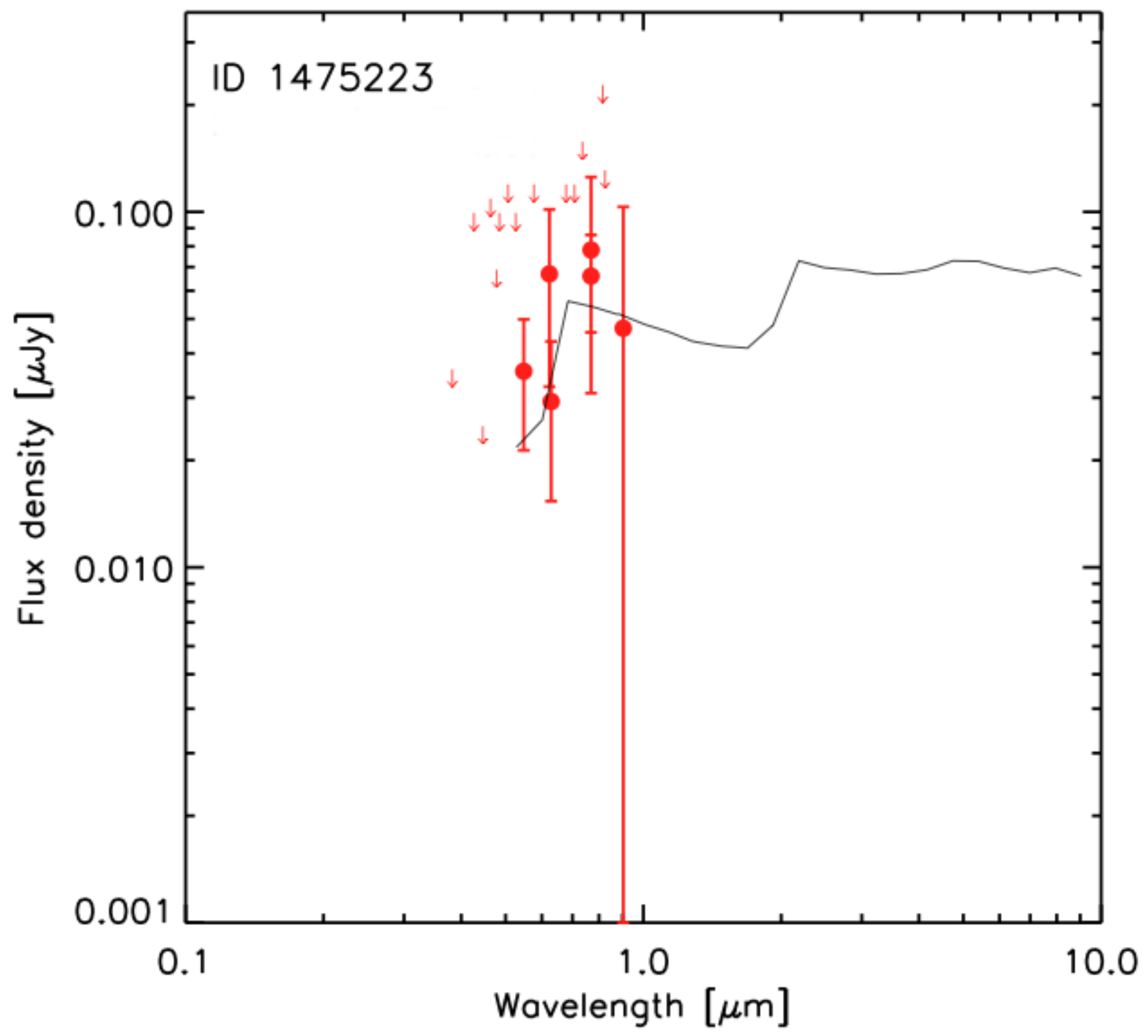}
\includegraphics[width=0.45\textwidth]{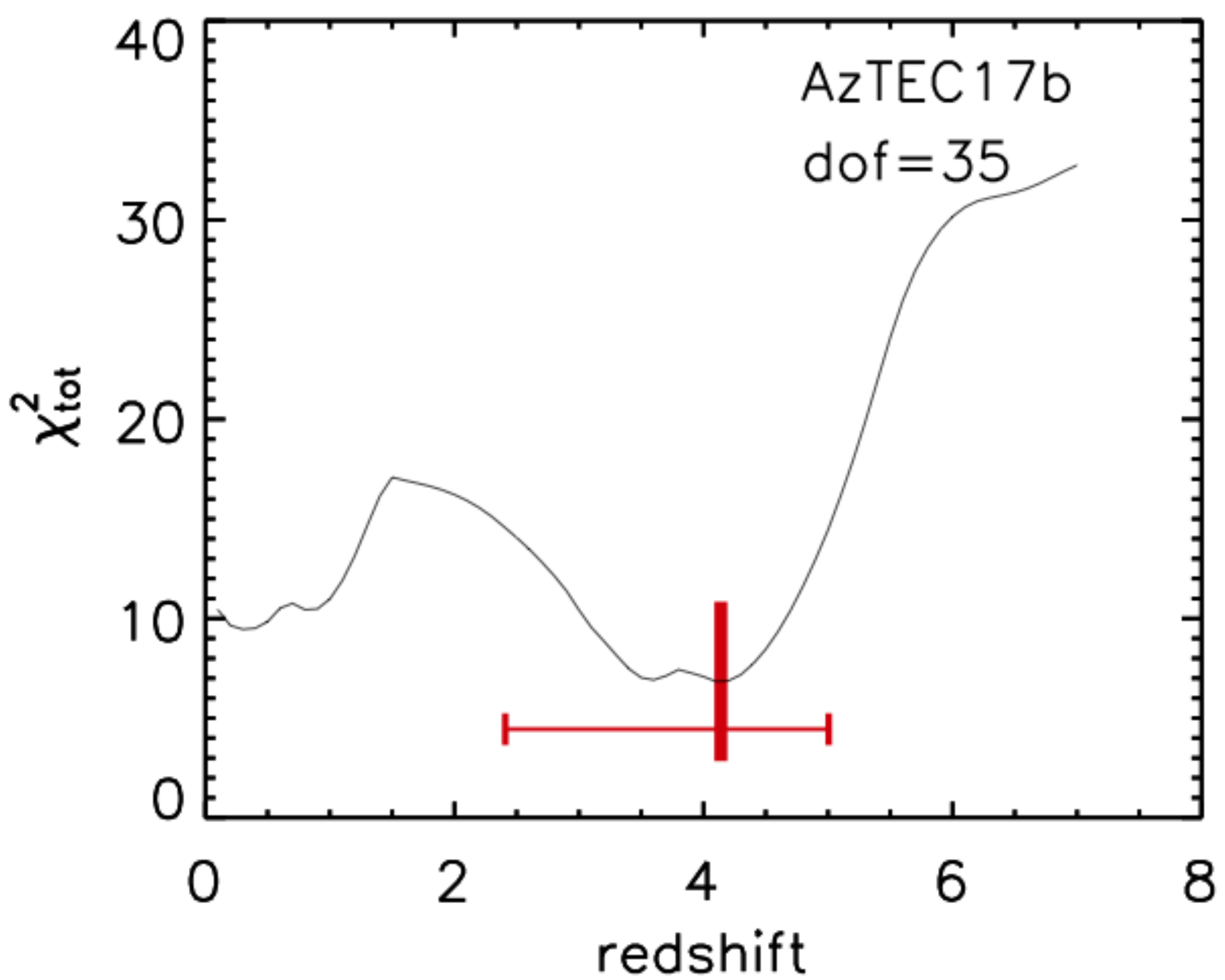}
\includegraphics[width=0.45\textwidth]{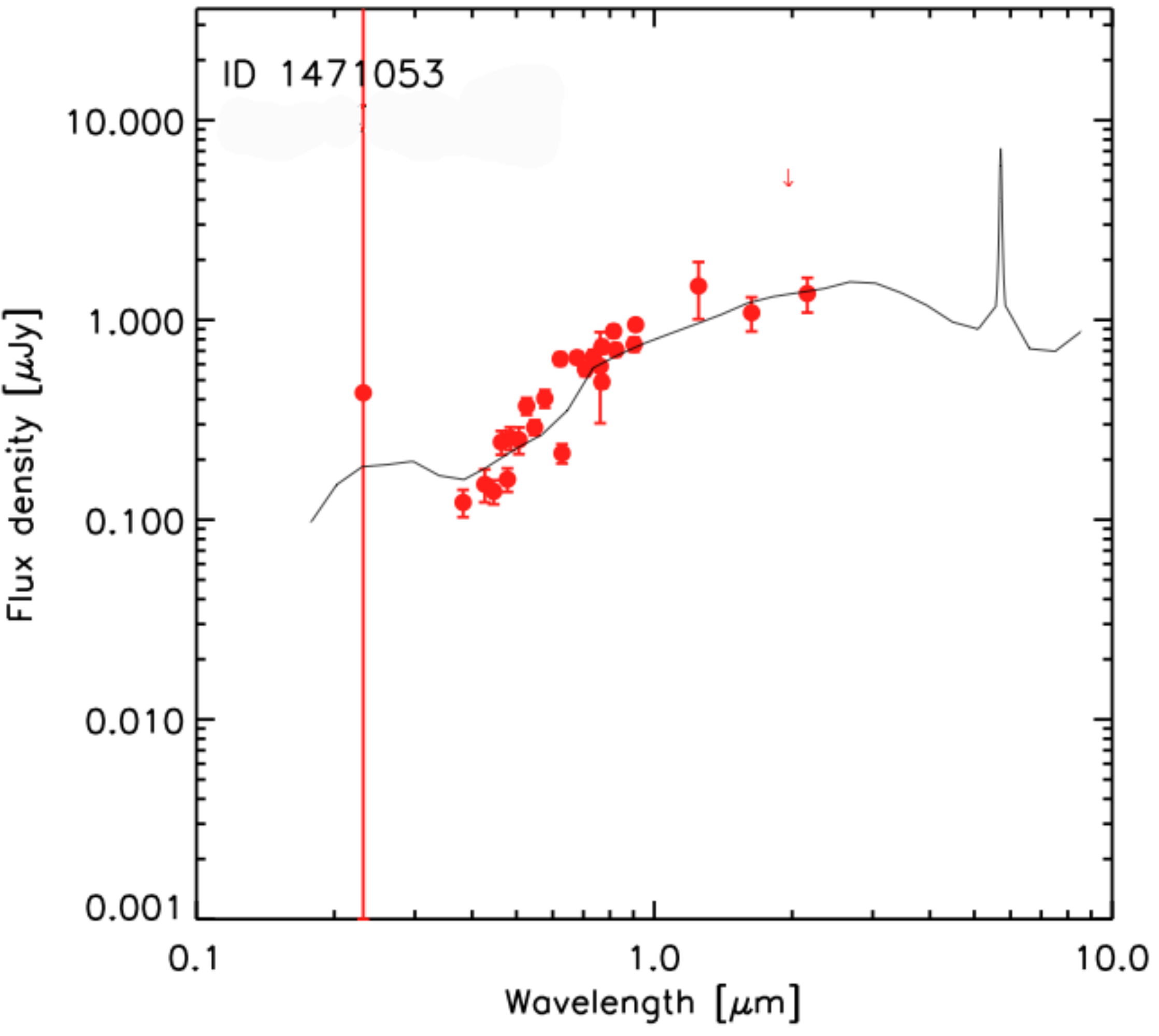}
\includegraphics[width=0.45\textwidth]{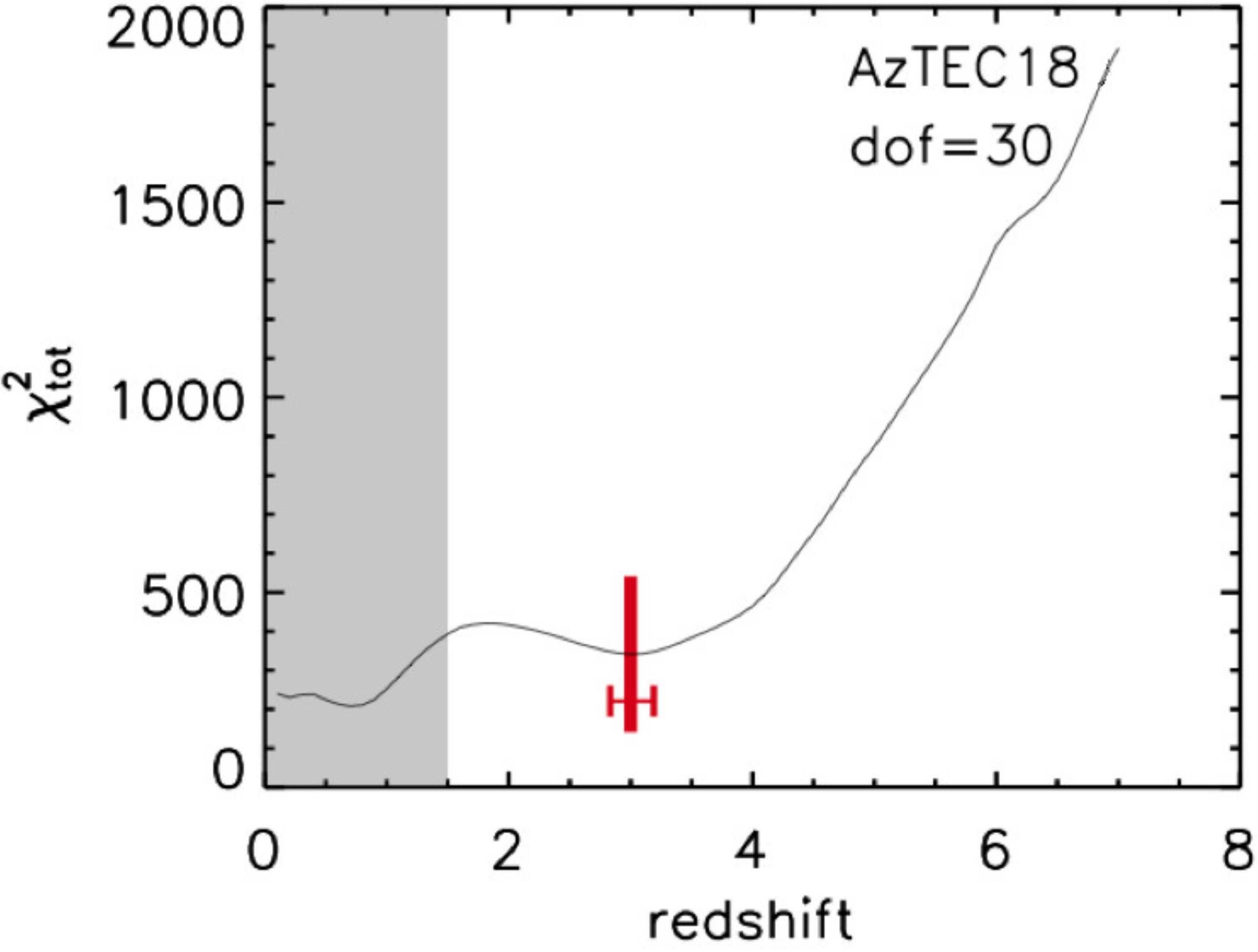}
\caption{Spectral energy distributions of the optical-to-IR counterparts of our SMGs. 
The red filled circles with error bars represent the photometric data points, while upper limits are denoted by 
downward pointing arrows [some of the data points with large error bars, such as the \textit{GALEX} NUV 
($\lambda_{\rm eff}\simeq0.2316$ $\mu$m), are also upper limits]. The solid lines represent the best-fit {\tt HYPERZ} 
model to these data from the spectral model library of Micha{\l}owski et al. (2010). 
The panels on the right side show the corresponding total $\chi^2$ distributions as a function of redshift. 
The number of degrees of freedom (dof) in the $\chi^2$ minimisation is indicated in the top right corner of each panel. 
The grey shaded area in the AzTEC18, 19a, 21a, 21b, 23, and 26a panels indicates the $z$ range ignored in the determination of the 
best-fit photo-$z$ solution (two almost equally probable $z_{\rm phot}$ solutions, where the higher $z_{\rm phot}$ is supported 
by the radio non-detection; see Appendix~\textbf{C}). The thick red vertical line marks the best-fit $z_{\rm phot}$ value, 
and the thin red horizontal line shows the 99\% CI. The vertical dashed line in the AzTEC17a panel marks the spectroscopic redshift 
$z_{\rm spec}=0.834$ (Appendix~\textbf{C}).}
\label{figure:sed}
\end{center}
\end{figure*}

\addtocounter{figure}{-1}
\begin{figure*}
\begin{center}
\includegraphics[width=0.4\textwidth]{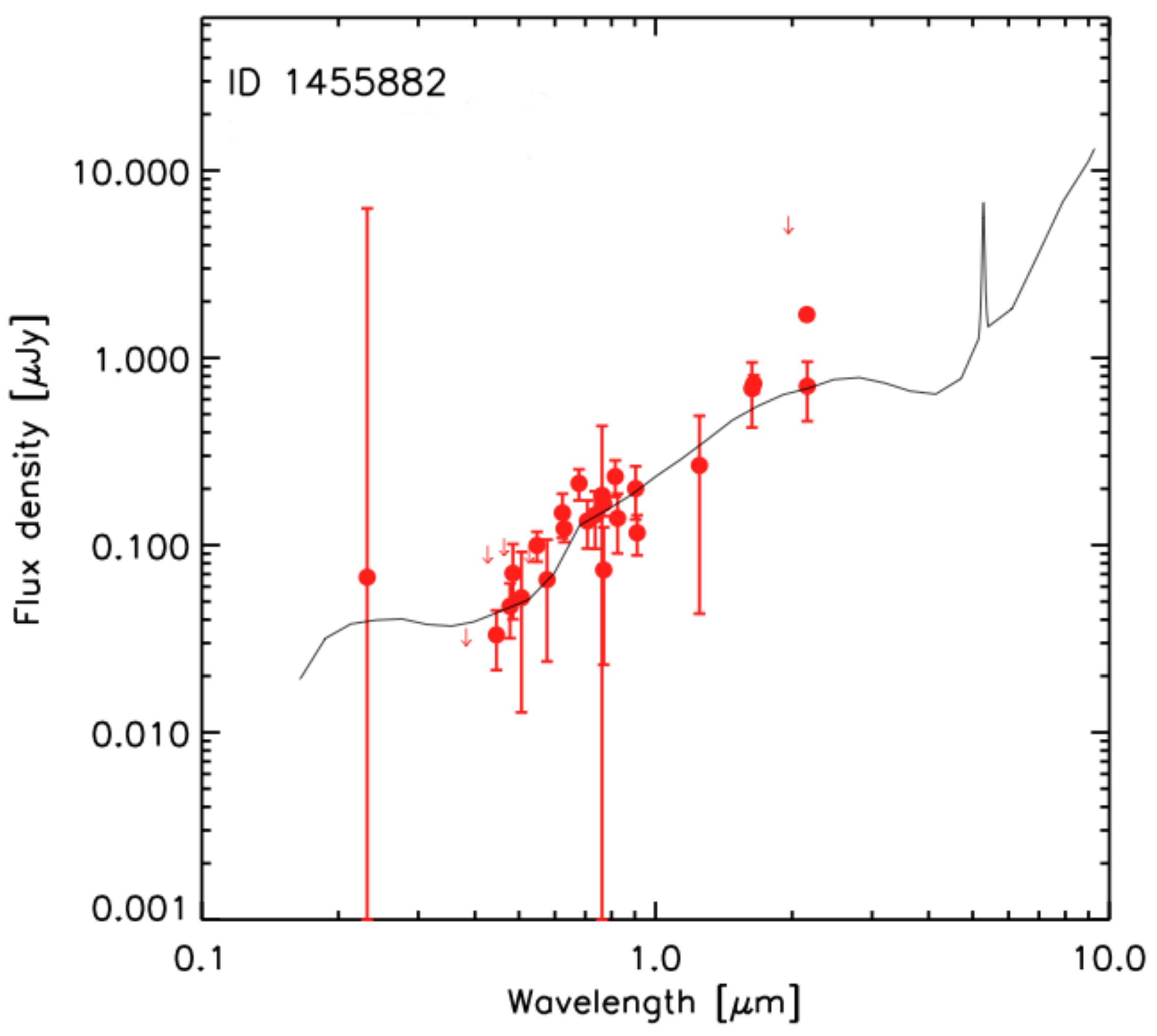}
\includegraphics[width=0.4\textwidth]{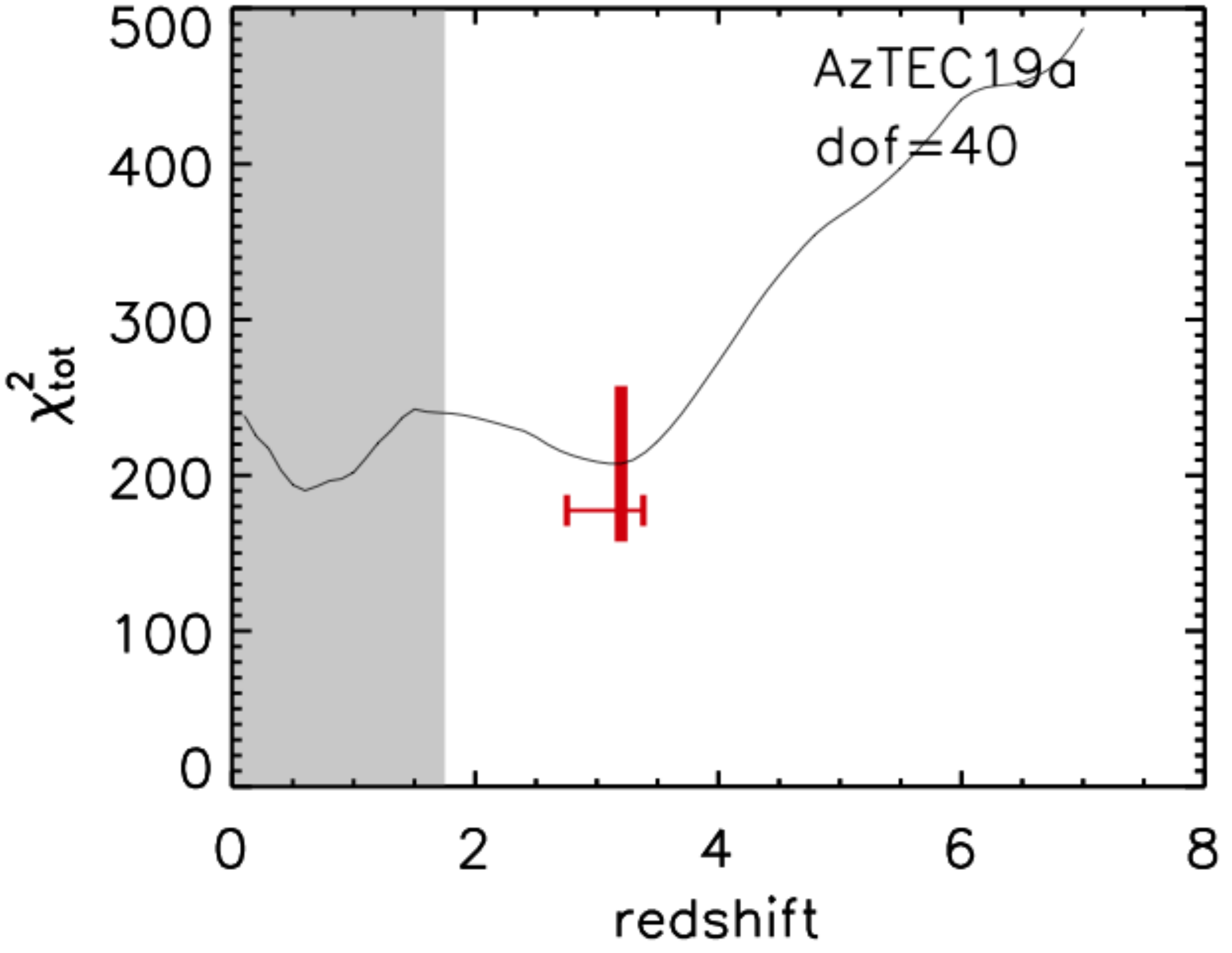}
\includegraphics[width=0.4\textwidth]{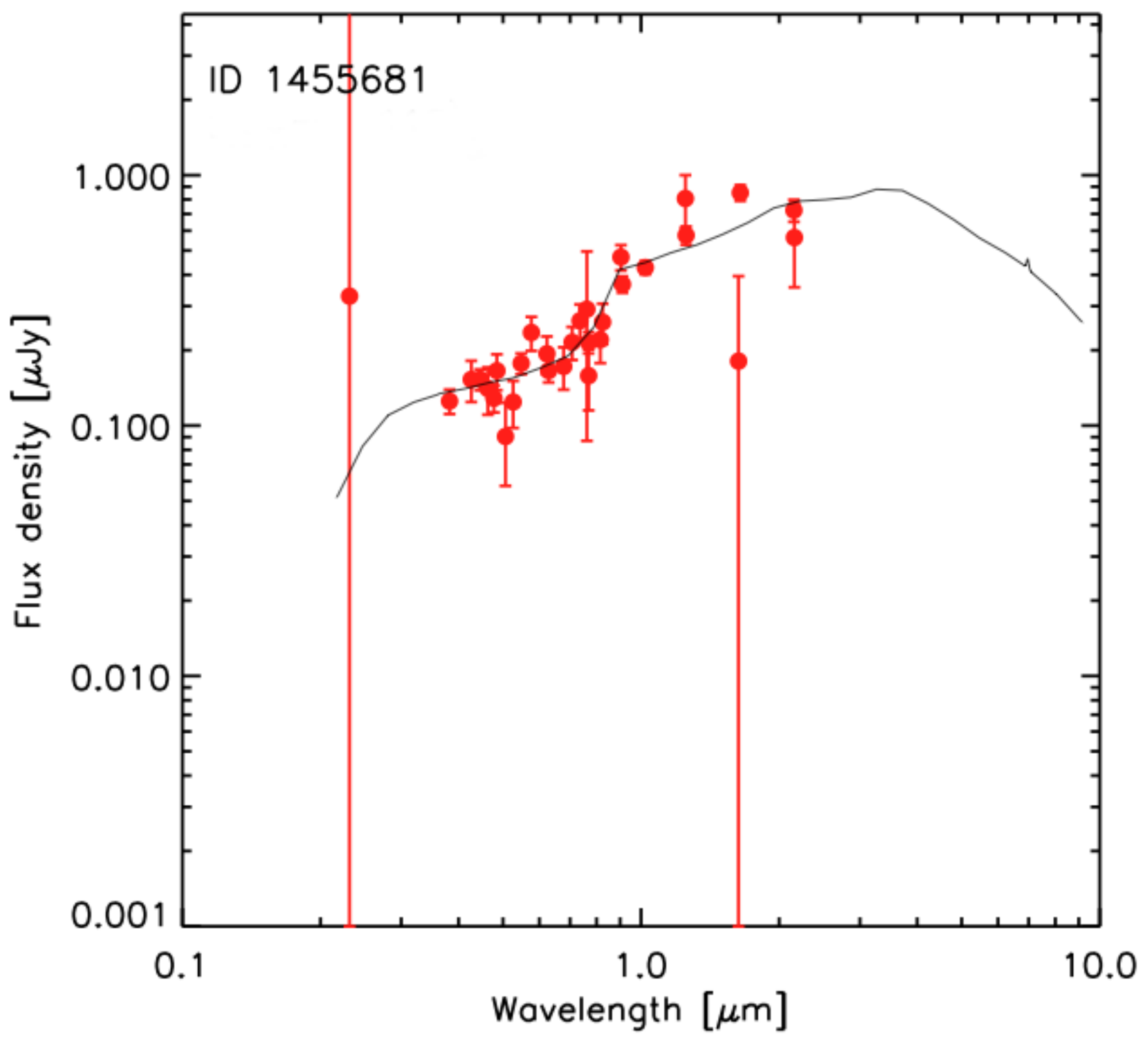}
\includegraphics[width=0.4\textwidth]{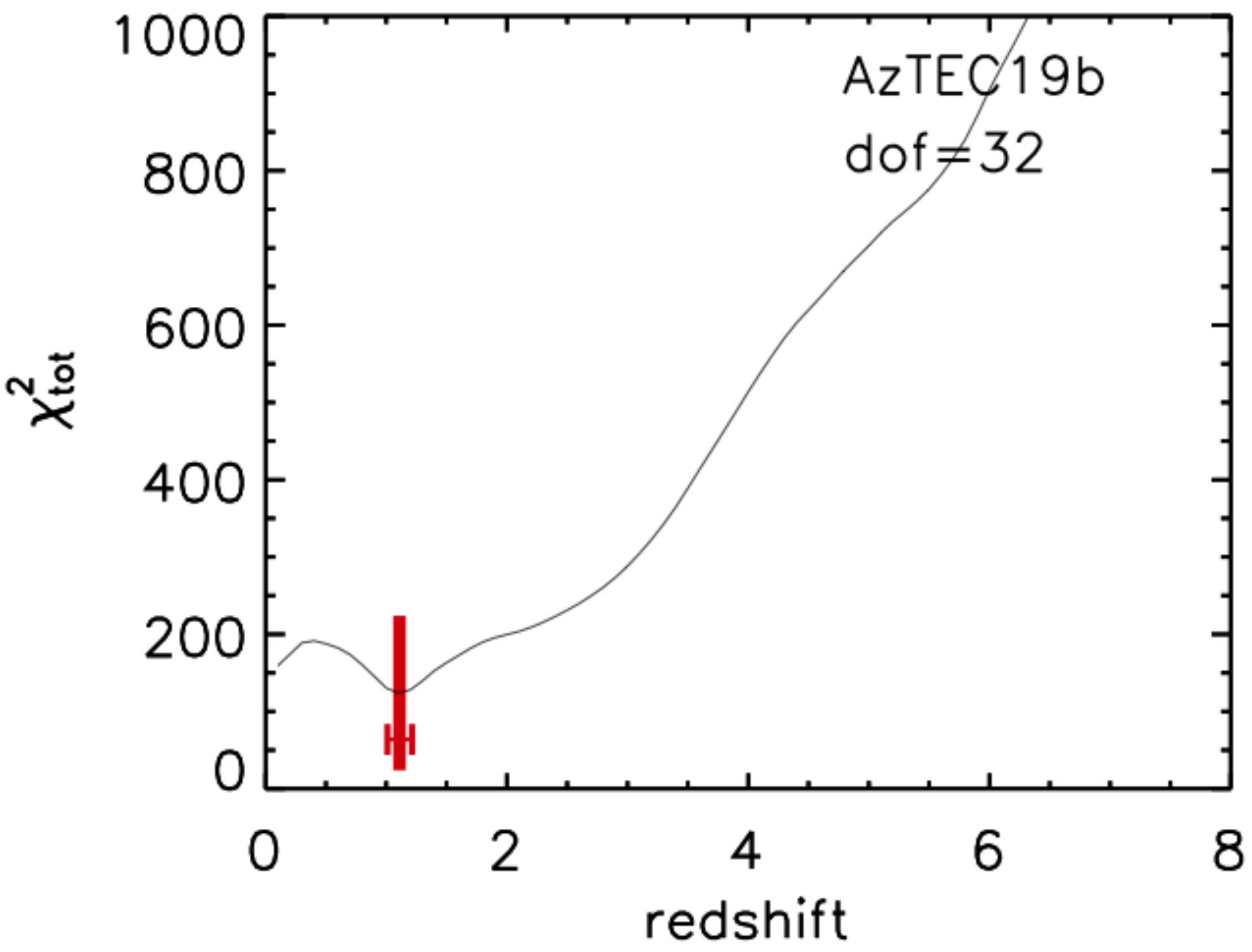}
\includegraphics[width=0.4\textwidth]{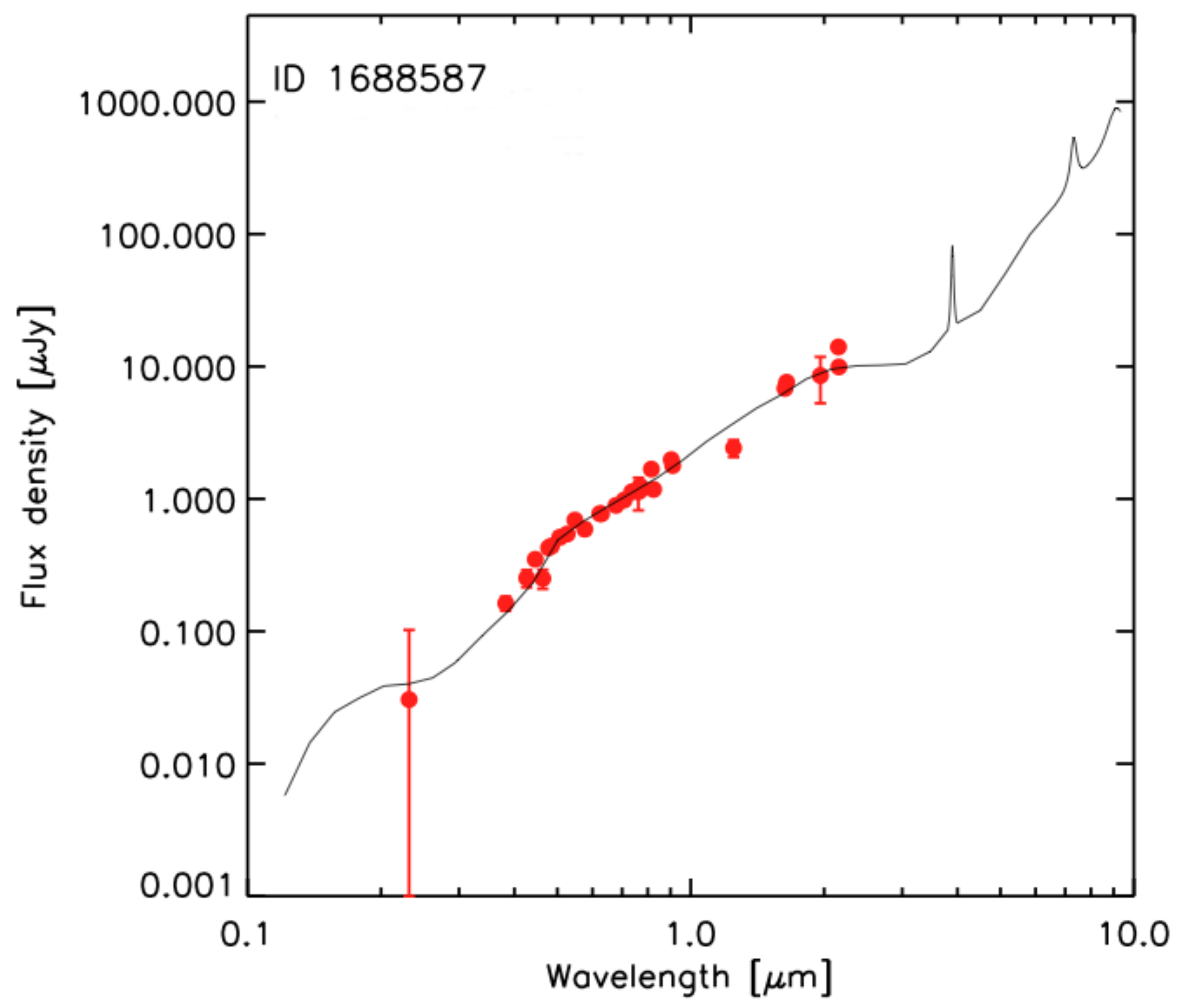}
\includegraphics[width=0.4\textwidth]{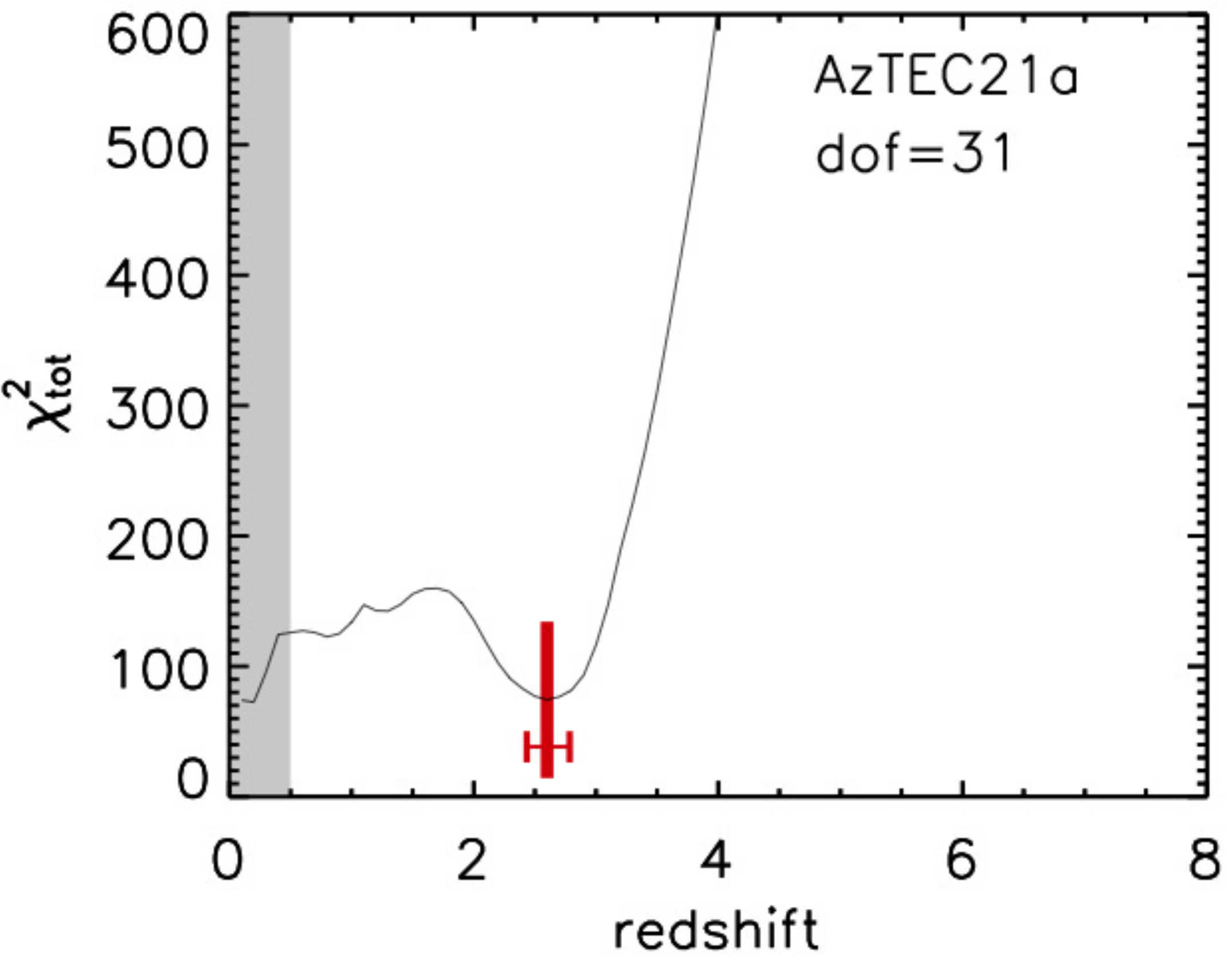}
\caption{continued.}
\label{figure:sed}
\end{center}
\end{figure*}

\addtocounter{figure}{-1}
\begin{figure*}
\begin{center}
\includegraphics[width=0.4\textwidth]{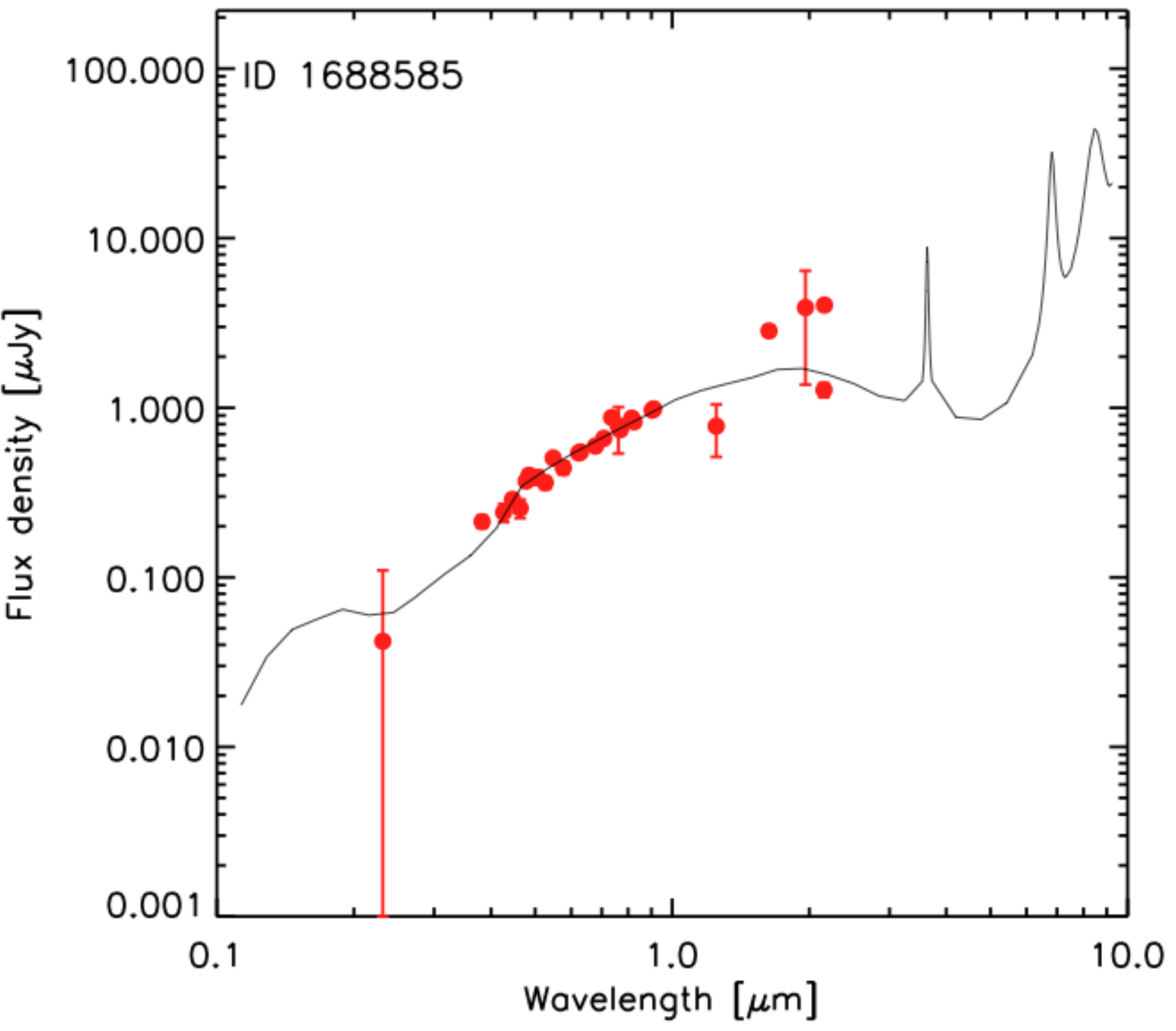}
\includegraphics[width=0.4\textwidth]{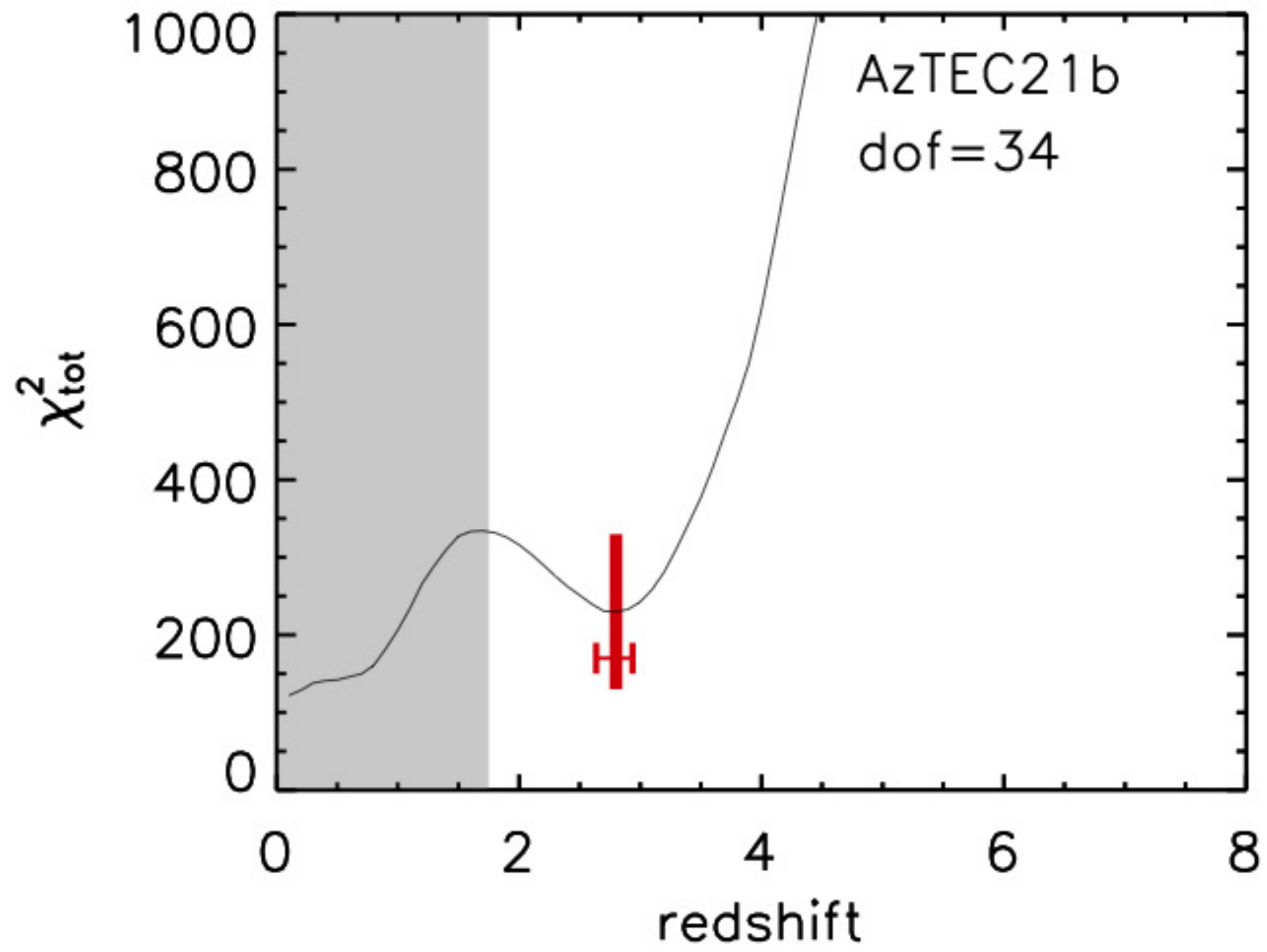}
\includegraphics[width=0.4\textwidth]{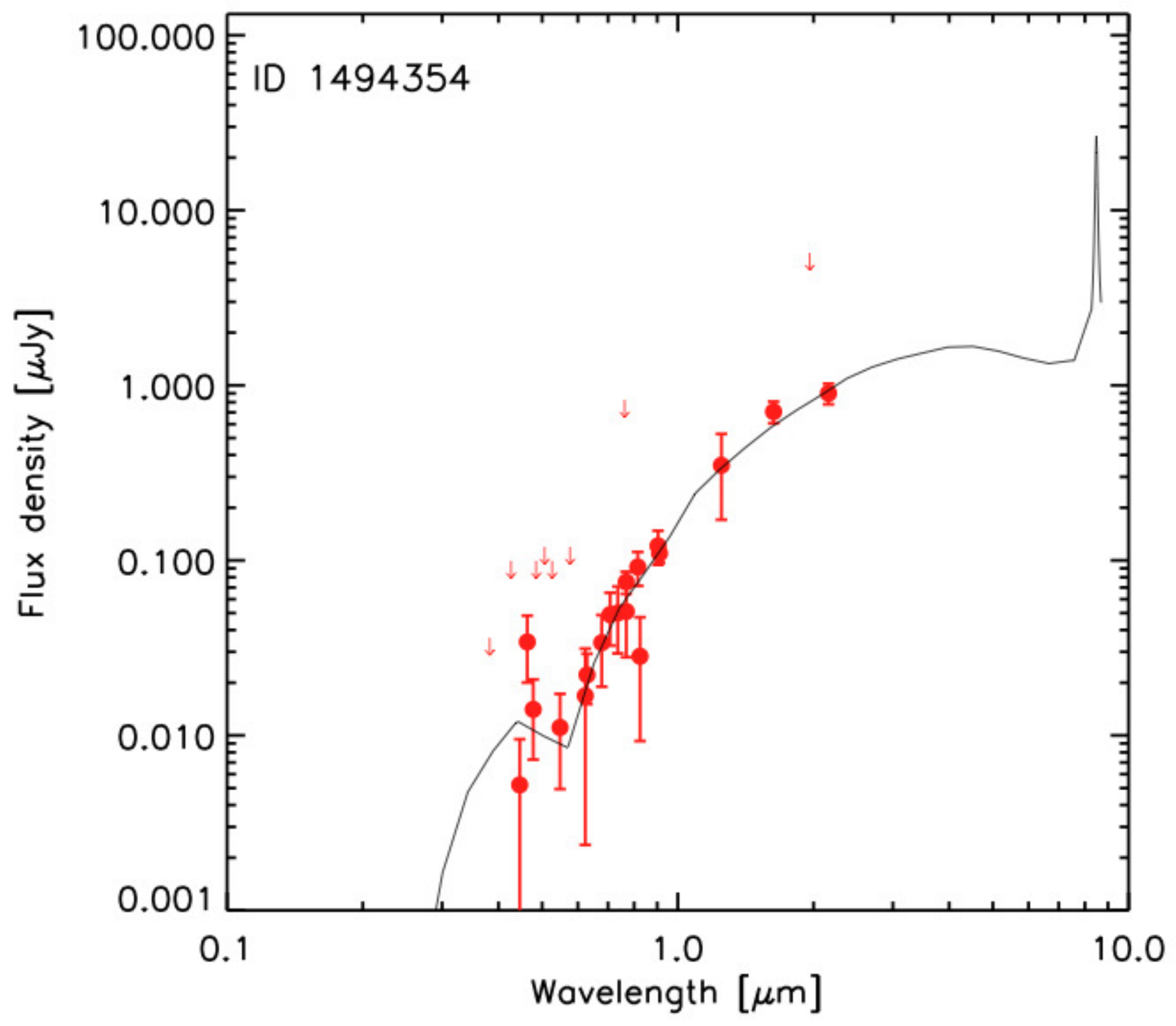}
\includegraphics[width=0.4\textwidth]{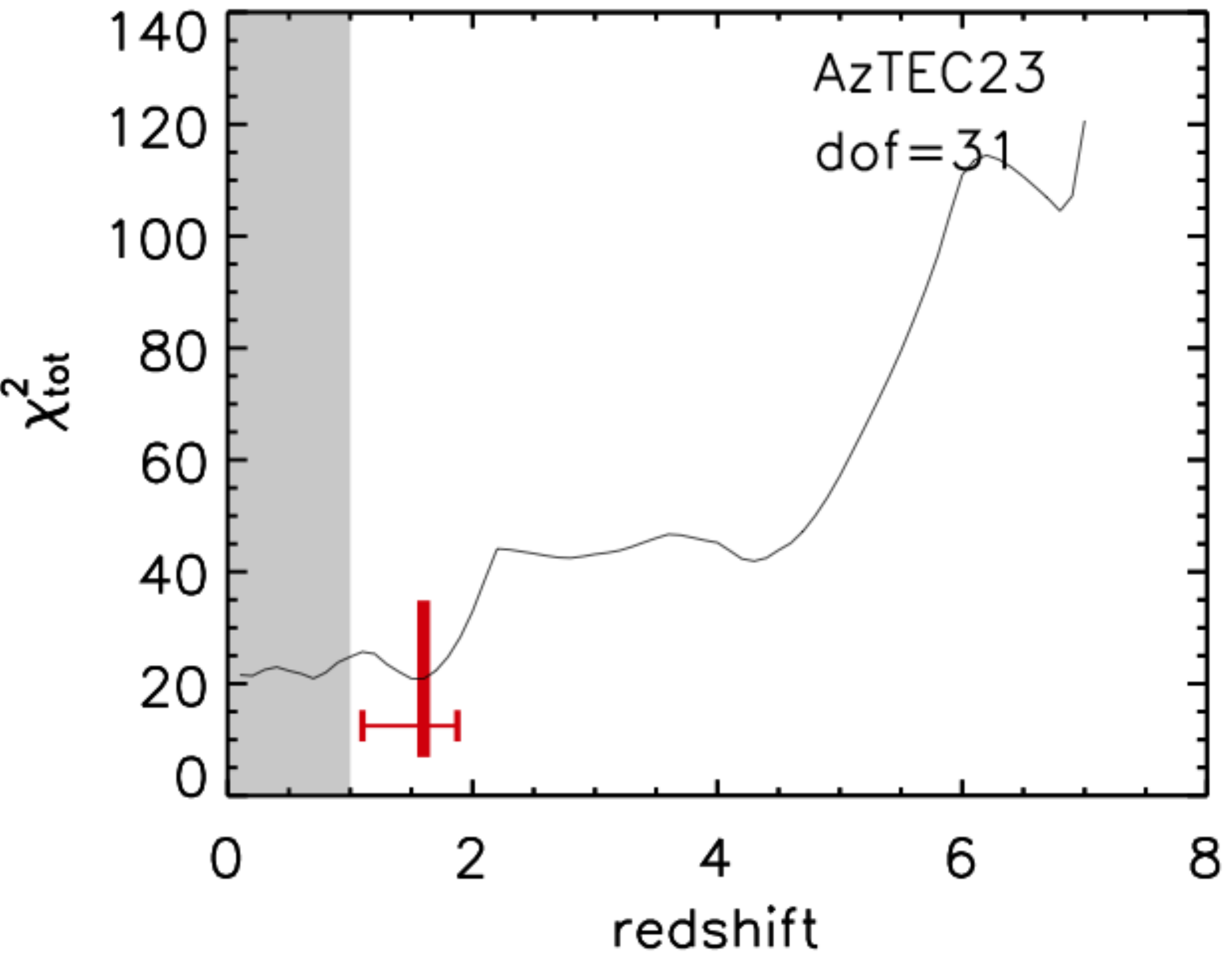}
\includegraphics[width=0.4\textwidth]{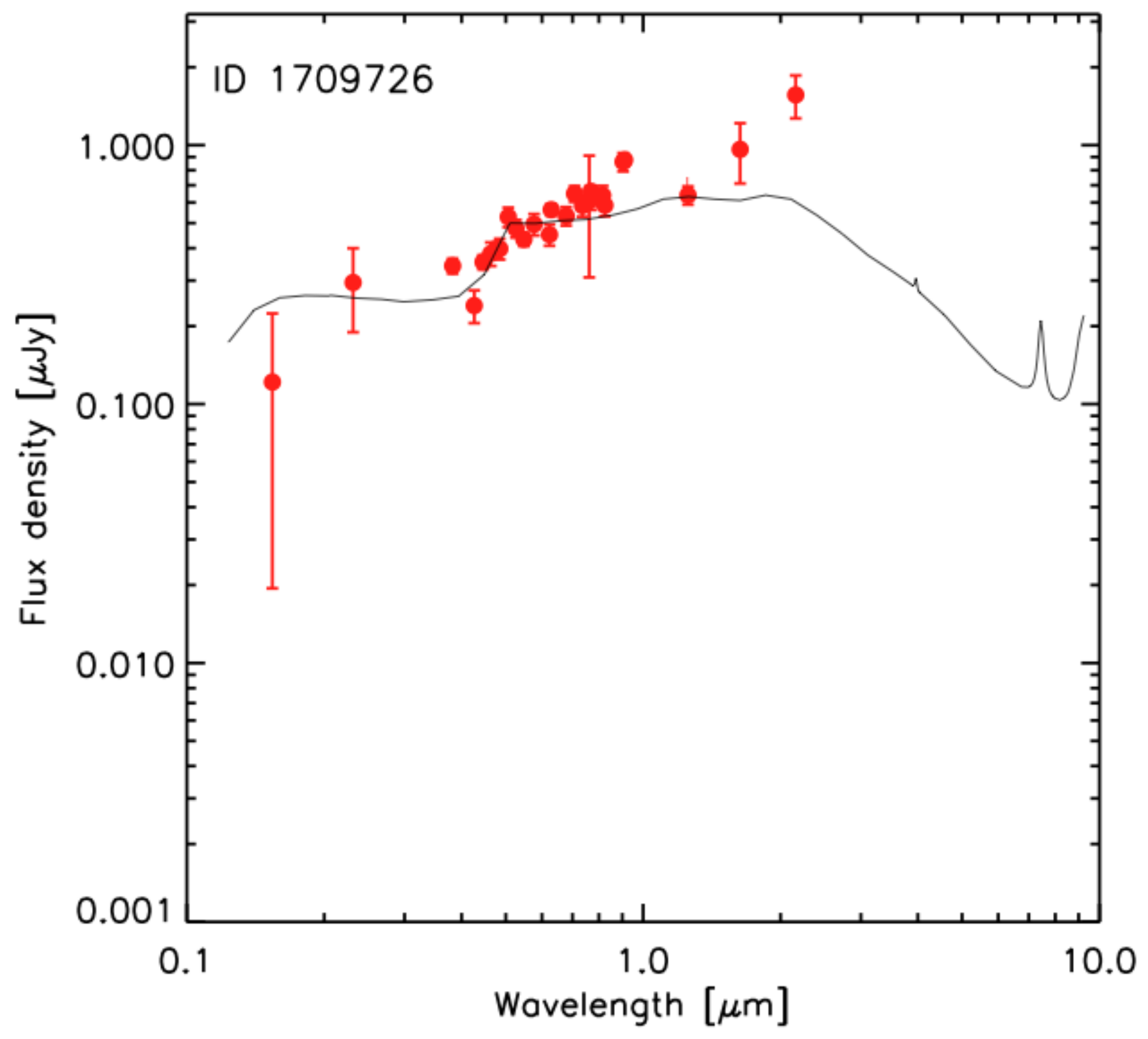}
\includegraphics[width=0.4\textwidth]{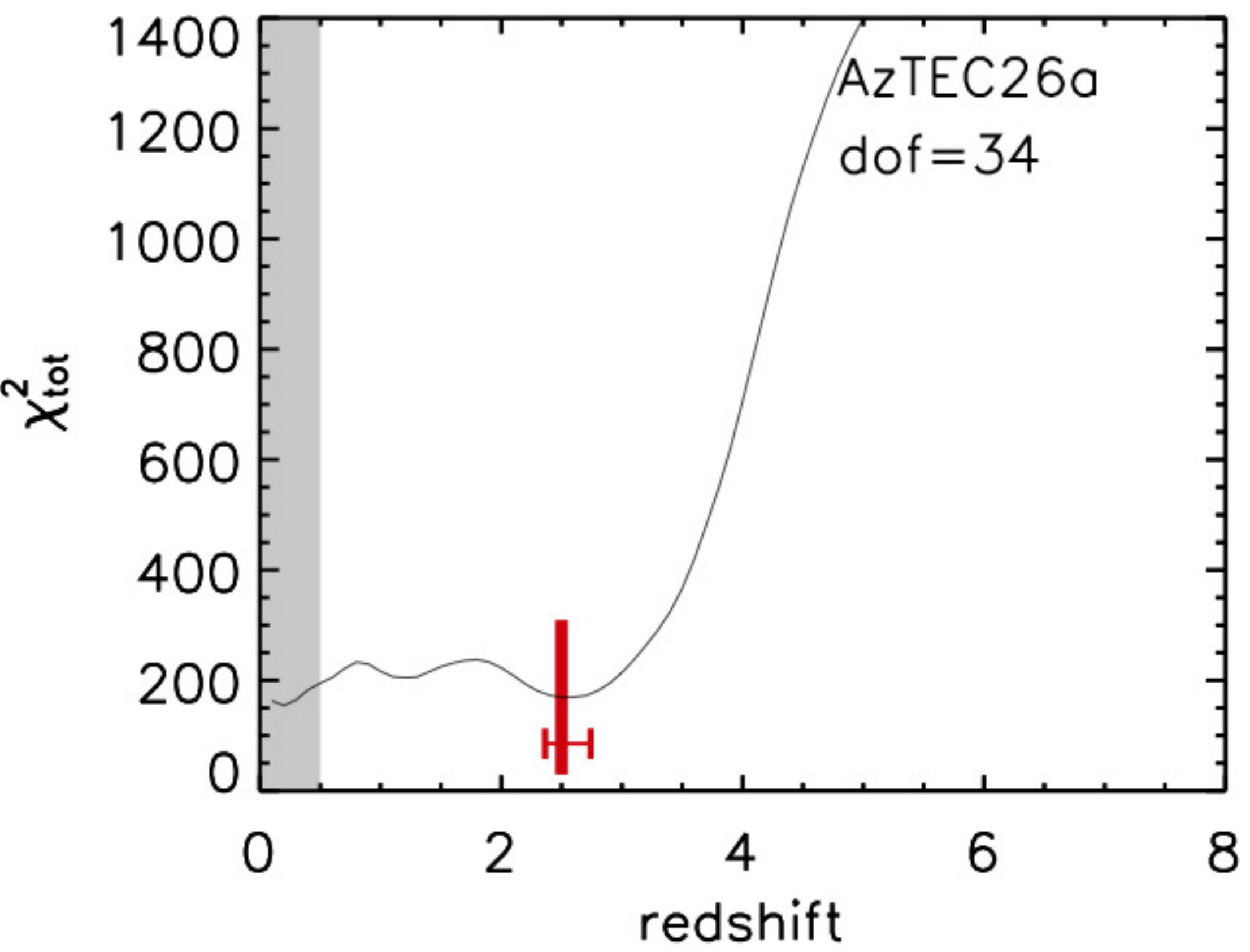}
\caption{continued.}
\label{figure:sed}
\end{center}
\end{figure*}

\addtocounter{figure}{-1}
\begin{figure*}
\begin{center}
\includegraphics[width=0.4\textwidth]{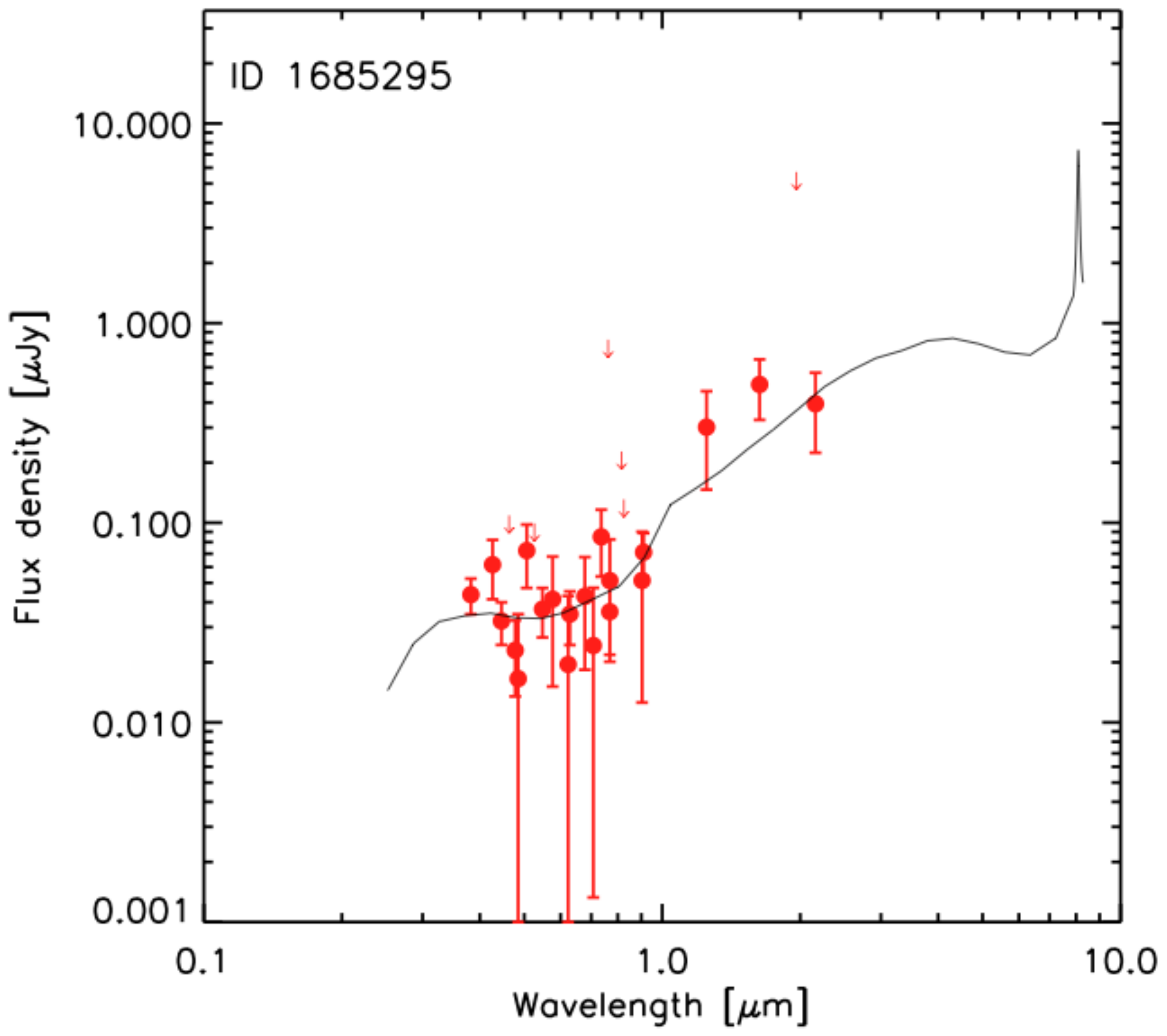}
\includegraphics[width=0.4\textwidth]{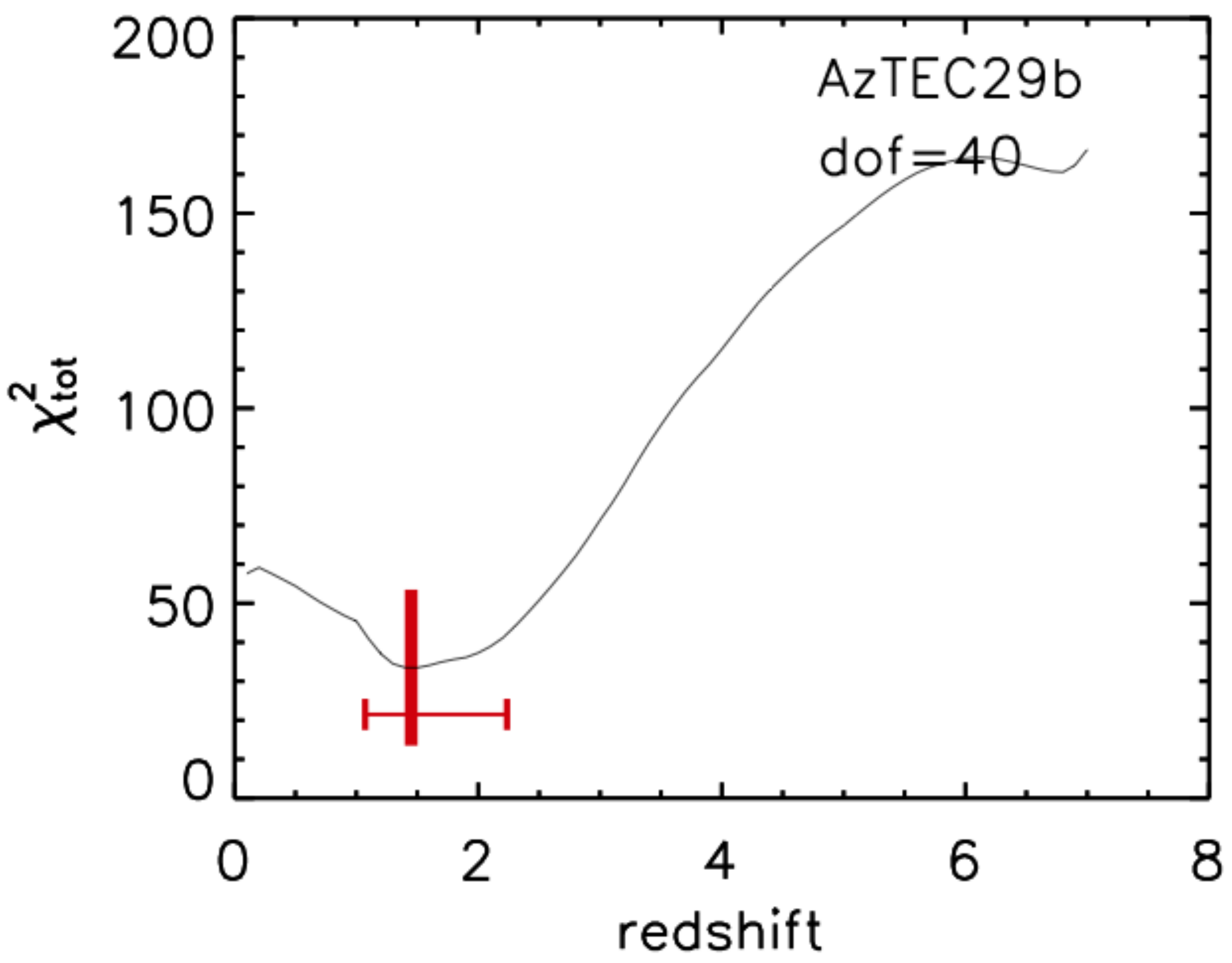}
\caption{continued.}
\label{figure:sed}
\end{center}
\end{figure*}

\begin{table}
\renewcommand{\footnoterule}{}
\caption{Best available redshifts for the 30 brightest JCMT/AzTEC-detected SMGs.}
{\scriptsize
\begin{minipage}{1\columnwidth}
\centering
\label{table:redshifts}
\begin{tabular}{c c c c}
\hline\hline 
Source & Redshift & Comment and\\
       & & reference \\
\hline
AzTEC1 & $4.3415$ & spec-$z$; Yun et al., in prep. \\
AzTEC2\tablefootmark{a} & 1.125 & spec-$z$; Balokovi\'c et al., in prep.\\
                        & ($4.28\pm0.82$) & radio/submm-$z$; this work\\
                        & ($3.60_{-0.18}^{+0.13}$) & radio/submm-$z$;\\
                        &                       & \cite{koprowski2014}    \\
AzTEC3 & 5.298 & spec-$z$; \cite{riechers2010}; \\
                        &             & \cite{capak2011}\\
AzTEC4 & $4.93_{-1.11}^{+0.43}$ & photo-$z$; \cite{smolcic2012b} \\
AzTEC5 & $3.05_{-0.28}^{+0.33}$ & photo-$z$; \cite{smolcic2012b} \\
       & ($1.85\pm0.23$) & radio/submm-$z$; this work\\
AzTEC6\tablefootmark{a} & $>3.52$ & radio/submm-$z$; this work\\
                        & ($3.86_{-0.92}^{+4.91}$) & radio/submm-$z$; \\
                        &                      & \cite{koprowski2014} \\
AzTEC7 & $2.30\pm0.10$ & photo-$z$; \cite{smolcic2012b} \\
AzTEC8 & 3.179 & spec-$z$; Riechers et al., in prep. \\
AzTEC9\tablefootmark{a} & $ 1.07_{-0.10}^{+0.11}$ & photo-$z$; \cite{smolcic2012b} \\
                        & (1.357) & spec-$z$; Salvato et al., in prep.  \\
                        & ($2.82\pm0.76$) & radio/submm-$z$; this work\\
                        & ($4.85_{-0.15}^{+0.50}$) & photo-$z$; \\
                        &                        & \cite{koprowski2014}    \\
AzTEC10\tablefootmark{a} & $2.79_{-1.29}^{+1.86}$ & photo-$z$; \cite{smolcic2012b} \\
                         &  ($5.00_{-0.50}^{+2.00}$) & photo-$z$;  \\
                         &                        & \cite{koprowski2014} \\
AzTEC11\tablefootmark{b} & 1.599 & spec-$z$; Salvato et al., in prep. \\
AzTEC12 & $2.54_{-0.33}^{+0.13}$ & photo-$z$; \cite{smolcic2012b} \\
AzTEC13 & $>4.07$ & radio/submm-$z$; this work\\
        & ($4.70_{-1.04}^{+1.25}$) & radio/submm-$z$; \\
        &                      & \cite{koprowski2014}    \\
AzTEC14-E\tablefootmark{c} & $>2.95$ & radio/submm-$z$; this work\\
                           & ($3.38_{-0.54}^{+1.00}$) & radio/submm-$z$;\\
                           &                        & \cite{koprowski2014}    \\
AzTEC14-W\tablefootmark{c} & $1.30_{-0.36}^{+0.12}$ & photo-$z$; \cite{smolcic2012b} \\
AzTEC15 & $3.17_{-0.37}^{+0.29}$ & photo-$z$; \cite{smolcic2012b} \\
\hline
AzTEC16 & $>2.42$ & radio/submm-$z$; this work\\
\textbf{AzTEC17a} & $0.834$ & spec-$z$; Salvato et al., in prep. \\
         & ($0.75_{-0.12}^{+0.23}$) & photo-$z$; this work\\
         & ($2.29\pm0.42$) & radio/submm-$z$; this work\\
AzTEC17b & $4.14_{-1.73}^{+0.87}$ & photo-$z$; this work\\
         & ($>2.49$) & radio/submm-$z$; this work\\
AzTEC18 & $3.00_{-0.17}^{+0.19}$ & photo-$z$; this work\\
         & ($>2.20$) & radio/submm-$z$; this work\\
\textbf{AzTEC19a} & $3.20_{-0.45}^{+0.18}$ & photo-$z$; this work\\
         & ($4.22\pm0.91$) & radio/submm-$z$; this work\\
\textbf{AzTEC19b} & $1.11\pm0.10$ & photo-$z$; this work\\
         & ($>6.57$) & radio/submm-$z$; this work\\
\textbf{AzTEC20} & $>2.35$ & radio/submm-$z$; this work\\
\textbf{AzTEC21a} & $2.60_{-0.17}^{+0.18}$ & photo-$z$; this work\\
         & ($>3.45$) & radio/submm-$z$; this work\\
AzTEC21b & $2.80_{-0.16}^{+0.14}$ & photo-$z$; this work\\
         & ($>2.47$) & radio/submm-$z$; this work\\
AzTEC21c & $>1.93$ & radio/submm-$z$; this work\\
AzTEC22 & $>3.00$ & radio/submm-$z$; this work\\
AzTEC23 & $1.60_{-0.50}^{+0.28}$ & photo-$z$; this work\\
        & ($>2.06$) & radio/submm-$z$; this work\\
AzTEC24a & $>2.35$ & radio/submm-$z$; this work\\
AzTEC24b & $>2.28$ & radio/submm-$z$; this work\\
AzTEC24c & $>3.17$ & radio/submm-$z$; this work\\
AzTEC26a & $2.50_{-0.14}^{+0.24}$ & photo-$z$; this work\\
         & ($>1.87$) & radio/submm-$z$; this work\\
AzTEC26b & $>1.79$ & radio/submm-$z$; this work\\
\textbf{AzTEC27} & $>4.17$ & radio/submm-$z$; this work \\
\textbf{AzTEC28} & $>3.11$ & radio/submm-$z$; this work \\
AzTEC29a & $>2.96$ & radio/submm-$z$; this work \\
\textbf{AzTEC29b} & $1.45_{-0.38}^{+0.79}$ & photo-$z$; this work\\
         & ($>7.25$) & radio/submm-$z$; this work \\
AzTEC30 & $>2.51$ & radio/submm-$z$; this work \\
\hline 
\end{tabular} 
\tablefoot{When multiple values are given for the redshift, the one not enclosed in parentheses 
has been adopted in the present study. Among AzTEC16-30 the source names highlighted in bold-face indicate 
detections with ${\rm S/N_{\rm 1.3 \, mm}}\geq5.5$.\tablefoottext{a}{See text for details about the difference between our 
redshifts and those from Koprowski et al. (2014).}\tablefoottext{b}{AzTEC11 was resolved into two 890~$\mu$m sources 
(N and S) by Younger et al. (2009). Here, for the redshift analysis, we treat it as a single source because the two 
components are probably physically related (\cite{koprowski2014}).}\tablefoottext{c}{AzTEC14 was resolved 
into two 890~$\mu$m sources (E and W) by Younger et al. (2009). The eastern component appears to lie at 
a higher redshift than the western one (\cite{smolcic2012b}).}}
\end{minipage} 
}
\end{table}

\section{Discussion}

\begin{figure}[!h]
\centering
\resizebox{\hsize}{!}{\includegraphics{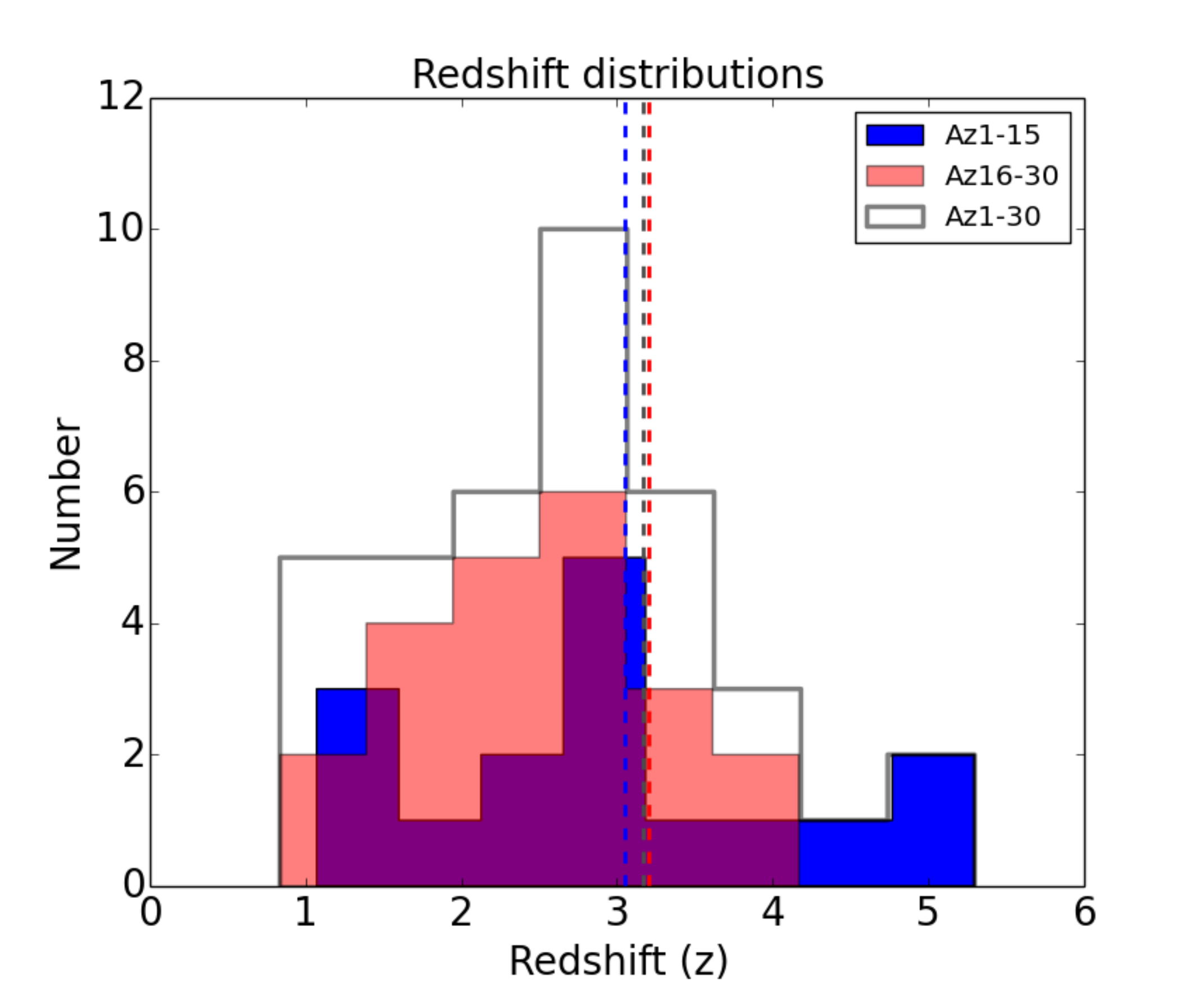}}
\resizebox{0.9\hsize}{!}{\includegraphics{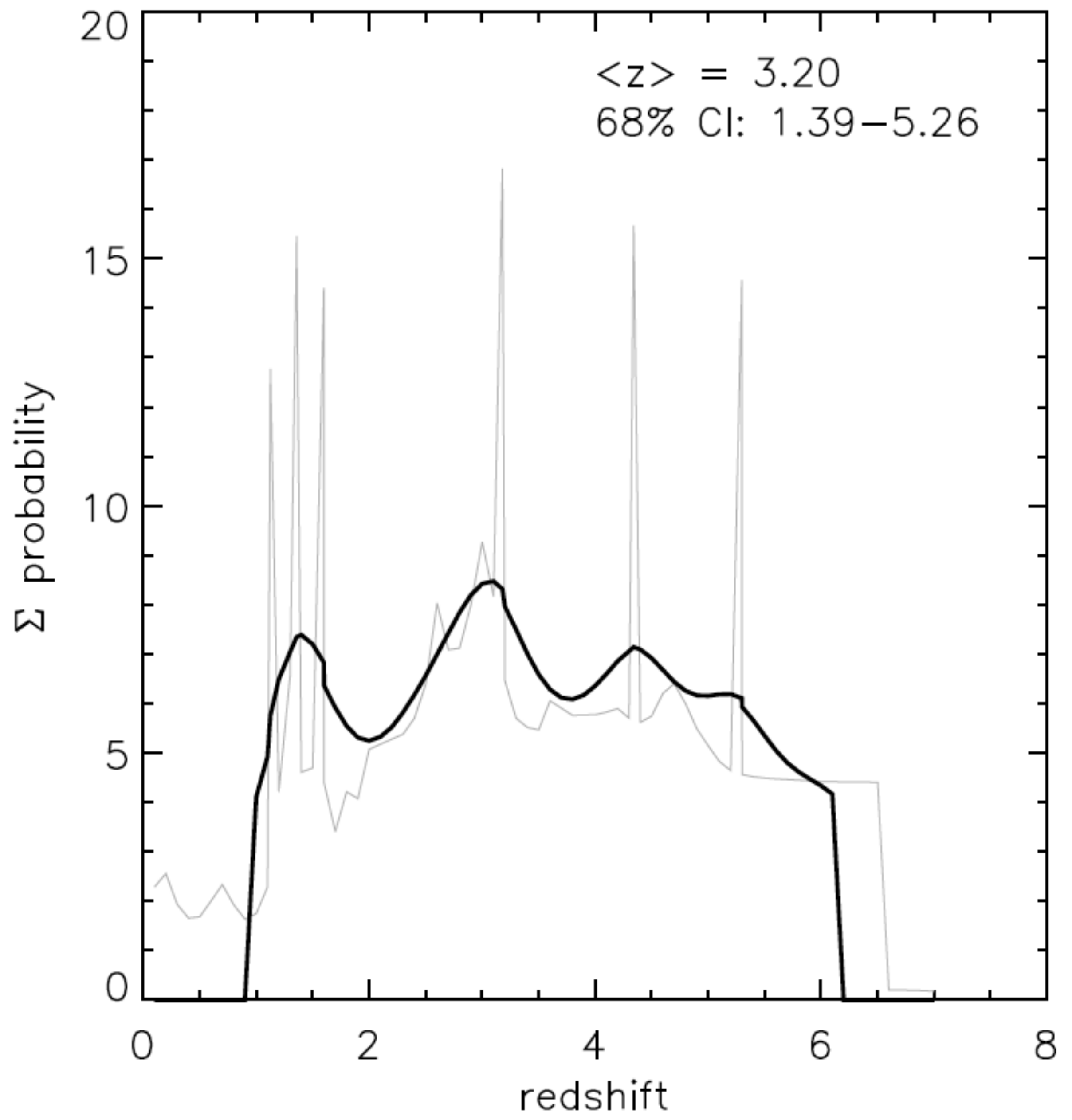}}
\caption{\textbf{\textit{Top}:} The distribution of redshifts of the studied SMGs. The blue filled histogram shows 
the redshift distribution of the SMGs AzTEC1--15, while the red filled histogram shows that 
of AzTEC16--30. The open grey histogram indicates the redshift distribution of the combined sample (AzTEC1--30). 
The vertical dashed lines mark the corresponding median redshifts (blue: $\tilde{z}=3.05$; red: $\tilde{z}=3.20$; grey: $\tilde{z}=3.17$). 
The lower redshift limits (see Table~\ref{table:redshifts}) were placed in the bins corresponding to those values, but the indicated median 
redshifts were properly derived through survival analysis. \textbf{\textit{Bottom}:} The probability density distribution of the redshifts of 
AzTEC1--30. The light grey curve shows the unsmoothed distribution, and the black curve represents the Gaussian-smoothed kernel density 
estimate (see text for details). The median redshift and the 68\% CI are indicated.}
\label{figure:redshift}
\end{figure}

\subsection{PdBI 1.3 mm imaging results and source multiplicity}

Even though our SMGs were detected at $4-4.5\sigma$ significance in the JCMT/AzTEC 1.1 mm 
survey (\cite{scott2008}), not all of them are (clearly) detected in the present higher-resolution 
PdBI 1.3 mm imaging. It is possible that some of the weak/non-detected AzTEC sources are actually composed 
of multiple objects that are too faint to be detected at the current detection limit 
[see Large APEX BOlometer CAmera (LABOCA) compared to Atacama Large Millimetre/submillimetre Array 
(ALMA); \cite{hodge2013}; and SCUBA compared to SMA; \cite{chen2014}]. As can be seen in Fig.~\ref{figure:flux}, 
only one source (AzTEC23) appears to have $S_{\rm 1.3\, mm}^{\rm PdBI}/S_{\rm 1.3\, mm}^{\rm JCMT}<1$, so we are not generally 
missing extended emission in our SMGs. Another reason for some of our PdBI non-detections could be that some of the original JCMT/AzTEC detections are spurious. 
Out of the 50 JCMT/AzTEC SMGs rerported by Scott et al. (2008), 48 (96\%) lie within the region mapped with AzTEC on the 
10~m Atacama Submillimetre Telescope Experiment (ASTE) by Aretxaga et al. (2011, 2012). However, only 16 JCMT/AzTEC-detected sources 
are common to the ASTE/AzTEC 1.1 mm catalogue (Table~1 in \cite{aretxaga2011}). For example, the ASTE/AzTEC 1.1 mm image extracted 
from the position of AzTEC30 shows emission only in the eastern part of the target region. 
Although the difference is, at least partly, caused by the difference in angular resolutions of ASTE 
($34\arcsec$; \cite{aretxaga2011}) and JCMT ($18\arcsec$), it is possible that the JCMT/AzTEC source is spurious. 
The expected false-detection rate in the JCMT/AzTEC survey at S/N$_{\rm 1.1\, mm}\geq4.0$ is $\lesssim2$ sources 
(\cite{scott2008}; Fig.~7 therein). Because all the 15 brightest sources AzTEC1--15 are interferometrically confirmed 
(\cite{younger2007}, 2009), we could expect to find about two spurious sources among AzTEC16--30.

When observed with single-dish telescopes, multicomponent source systems can be blended and give an 
impression of a single source. This can be the case even if the sources are not physically 
related to each other, i.e. they can lie at signi\-ficantly different redshifts (e.g. \cite{cowley2015}). 
Among our target fields, AzTEC17, 19, 21, 24, 26, and 29 appear to show two to three source components. Of the 
15 observed AzTEC single-dish detected SMGs, this would mean that $40\pm16\%$ are multiple systems (or, as explained in Sect.~3.3, 
$33\pm15\%$ if AzTEC24 is not included).\footnote{In most cases, namely AzTEC17, 19, 21, 24, and 26, most of the components exhibit 
comparable 1.3 mm flux densities, i.e. the single-dish measured flux density from these sources appears to include contributions from equally 
bright components.} Among AzTEC1--15, only two sources (or $13\pm9\%$), AzTEC11 and AzTEC14, 
were resolved into two distinct components by Younger et al. (2007, 2009) in their $2\arcsec$ resolution
SMA 890~$\mu$m imaging. We note that the northern and southern components of AzTEC11
could just belong to an extended object (\cite{koprowski2014}). The 890 $\mu$m flux density ratio for the 
two components of AzTEC11 is $44\pm23\%$, and even higher, $77\pm35\%$, for AzTEC14 (\cite{younger2009}; Table~1 therein).
The low observed multiplicity fraction among AzTEC1--15 could be the result of a sensitivity too low to reveal 
the real multiplicity (see \cite{wang2011}). However, as our angular resolution is only slightly better 
(and the observing wavelengths are quite similar, i.e. 1.3~mm compared to $\sim0.9$~mm), 
our observations provide a hint that the multiplicity fraction could be somewhat higher among 
the fainter SMGs AzTEC16--30 (i.e. $\sim30-40\%$ compared to 13\%). Considering the combined sample AzTEC1--30, 
the multiplicity fraction, based on the currently available data, appears to be $\sim25\pm9\%$.

Besides the present work, it has been found that interferometric observations have 
the potential to resolve SMGs into separate components (e.g. \cite{daddi2009a}; \cite{wang2011}; 
\cite{smolcic2012b}; \cite{barger2012}; \cite{karim2013}; \cite{hodge2013}). 
For example, Smol{\v c}i{\'c} et al. (2012b) used PdBI 1.3 mm observations 
at $\sim1\farcs5$ resolution of 28 SMGs in the COSMOS field in conjunction 
with earlier interferometrically identified COSMOS SMGs, and concluded that $\gtrsim15\%$, 
and possibly up to $\sim40\%$ of single-dish detected SMGs (at $18\arcsec$ with AzTEC and at $27\farcs6$ with LABOCA), 
consist of multiple sources. Consistent with this, Hodge et al. (2013) found that 24 out of 
their sample of 69 SMGs ($\sim35\%$) detected with LABOCA at 870~$\mu$m ($19\arcsec$ resolution) are split into 
multiples when observed with ALMA at the same wavelength [the ALMA-identified SMGs from the LABOCA Extended
\textit{Chandra} Deep Field South (ECDFS) Submillimetre Survey (LESS), i.e. the ALESS SMGs; see also \cite{karim2013}; \cite{simpson2014}]. 
We also examined the multiplicity fraction among those LESS SMGs (\cite{weiss2009}) that have LABOCA 870~$\mu$m flux densities 
corresponding to our AzTEC 1.1~mm flux density range, i.e. 3.3~mJy $\leq S_{\rm 1.1\, mm} \leq9.3$~mJy. Assuming that $\beta=1.5$, this flux density range 
is 7.7 mJy $\leq S_{\rm 870\, \mu m} \leq21.8$~mJy. Based on their main and supplementary source samples, altogether 20 SMGs from Hodge et al. (2013) have 
LABOCA 870 $\mu$m flux densities in this range. Among these SMGs, five were found to have multiple (2--3) components (ALESS SMGs), 
resulting in a multiplicity fraction of $25\pm11\%$, which is very similar to our value. Dividing these LESS SMGs into two subsamples corresponding to 
the flux densities of AzTEC1--15 and 16--30 (i.e. six SMGs with 9.8 mJy $\leq S_{\rm 870\, \mu m} \leq21.8$~mJy, and nine SMGs with 
7.7 mJy $\leq S_{\rm 870\, \mu m} \leq9.1$~mJy), we derived the corresponding multiplicity fractions to be $50\pm29\%$ and $44\pm22\%$. 
These two values are similar within the counting uncertainties and hence, unlike what we found among AzTEC1--30, the fainter LESS SMGs do not appear to 
exhibit a higher multiplicity fraction than the brighter SMGs.

As recently discussed by Koprowski et al. (2014), there is some controversy about how common SMG multiplicity 
actually is. The multiplicity statistics reported in the literature so far seem to suggest that the fraction of single-dish detected SMGs 
being composed of more than one SMG can be quite high (values ranging from $\sim15\%$ to $\sim40\%$). 
The multiplicity fraction also depends on the angular resolution of both the single-dish observations of the initial SMG detection and 
the follow-up observations (the higher the former is the lower the multiplicity fraction is expected to be for a given follow-up resolution).  
However, whether it is a common feature (several tens of percent) is an important knowledge when studying the number counts of SMGs, 
and future high-resolution observations of large, well-defined samples of SMGs are required to better understand the multiplicity 
fraction of submm-emitting galaxies. The completed ALMA array is well suited for this purpose.

\subsection{Counterpart associations of the SMGs AzTEC16--30}

Some of the detected (candidate) SMGs appear to have no counterparts at optical-to-IR wavelengths. 
These include AzTEC20, 22, 24a--c, 26b, 28, and 30 [moreover, the $I$-band sources lying $0\farcs35$ from 
AzTEC16, $1\farcs15$ from AzTEC27, and $1\farcs41$ from AzTEC29a might be unrelated to the (candidate) SMGs because no sources 
at other wavelengths are identified there]. In particular, the case of AzTEC28, a clearly detected PdBI 1.3~mm source (${\rm S/N}=5.5$) 
without multiwavelength counterparts, shows that SMGs can be so heavily obscured by dust and/or at high redshift that only 
FIR-to-mm continuum emission can be detected (this is probably true also for AzTEC27).
Given the deep multiwavelength data available for the COSMOS field, for example the 1st Ultra\-VISTA data release (DR1) going down to $K_{\rm s}<24$~mag 
(\cite{mccracken2012}), the fraction of sources that lack shorter-wavelength (and radio) counterparts seems high 
($8/22\sim36\%$, or $11/22=50\%$ if the three additional SMGs having only a nearby ACS $I$-band source are counted). 
More sensitive observations could reveal the presence of faint emission at optical-to-MIR 
wavelengths, such as Ultra\-VISTA DR2 reaching $K_{\rm s}\sim25$ mag (\cite{mccracken2013}), and new IRAC 3.6 and 
4.5~$\mu$m observations (reaching $\sim25.5$~mag) from the \textit{Spitzer} Large Area Survey with Hyper-Suprime-Cam 
(SPLASH) (PI: P.~Capak; \cite{steinhardt2014}). 
Younger et al. (2009) found that AzTEC13 and AzTEC14-E and 14-W are not coincident 
with any optical, \textit{Spitzer}, or VLA sources. Therefore, among AzTEC1--15, altogether comprising 17 SMGs, 
the fraction of SMGs that lack multiwavelength counterparts is 3/17 or 18\%, which is lower than for the fainter SMGs AzTEC16--30. 
For comparison, 45\% of the ALESS SMGs were found to lack MIR/radio counterparts (\cite{hodge2013}; 
see also \cite{simpson2014}; cf.~\cite{biggs2011}). This fraction is the same ($\sim45\%$) if we consider the 20 LESS SMGs 
that have similar flux densities as our AzTEC SMGs (see Sect.~5.1); the total number of ALMA-detected components in these SMGs is 
22 (main and supplementary samples in \cite{hodge2013}), and 12 of them were found to have robust MIR/radio counterparts.

The lack of multiwavelength counterparts means that we are not able to place strong constraints on the source's 
photometric redshift. In particular, the absence of an optical counterpart suggests that the source is highly obscured 
by interstellar dust, which conforms to the fact that SMGs represent very dusty galaxies. More generally, 
the non-detections at optical, NIR, and cm wavelengths suggest that the source lies at a high redshift ($z>3-4$), 
so that the flux density at a wavelength other than (sub)mm dims because of a large luminosity distance 
(i.e. the radiation suffers from the positive $K$-correction). In contrast, the (sub)mm flux density stays 
almost the same over the redshift range $z \sim 1-8$ because of the negative $K$-correction of 
the thermal dust emission (\cite{blain1993}). 

As shown in Table~\ref{table:properties}, for AzTEC17a, 17b, and 23, the projected angular offset between 
the optical-to-NIR candidate counterpart (from the COSMOS/UltraVISTA catalogue) and 
the 1.3 mm emission peak is quite large, $1\farcs24-1\farcs38$. Moreover, for AzTEC18, 26a, and 29b 
the above offset is also relatively large ($0\farcs82$, $0\farcs94$, and $0\farcs76$, respectively).  
Such an offset could be the effect of complex source morphology, 
expected in the case of galaxy mergers (\cite{daddi2009a}), and/or be the result 
of strong differential dust obscuration (e.g. \cite{carilli2010}).
With respect to our sample of 22 detections, $\sim14-27\%$ (three to six sources) exhibit a considerable ($\gtrsim 0\farcs8$) projected 
separation between the PdBI and UltraVISTA emission peaks. Even though the positional error of our PdBI sources is much lower 
($\sim0\farcs2$; Sect.~2.2), the above mentioned angular separations are still within the beam FWHM of $\sim1\farcs8$. 
Moreover, towards AzTEC17a, showing the largest offset between the PdBI peak emission and the UltraVISTA source in our sample ($1\farcs38$),  
the \textit{Spitzer}/IRAC counterpart lies only $0\farcs3$ from the PdBI peak, strongly indicating that the SMG is emitting at observing-frame 
wavelengths $\geq3.6$ $\mu$m. The photo-$z$ value of $0.75_{-0.12}^{+0.23}$ we derived for AzTEC17a is in good agreement with the spec-$z$ 
of 0.834, further strengthening our counterpart identification (Appendix~C).

\subsection{Redshift distribution of the JCMT/AzTEC-detected SMGs in the COSMOS field, and comparison with other surveys}

\begin{figure}[!h]
\centering
\resizebox{\hsize}{!}{\includegraphics{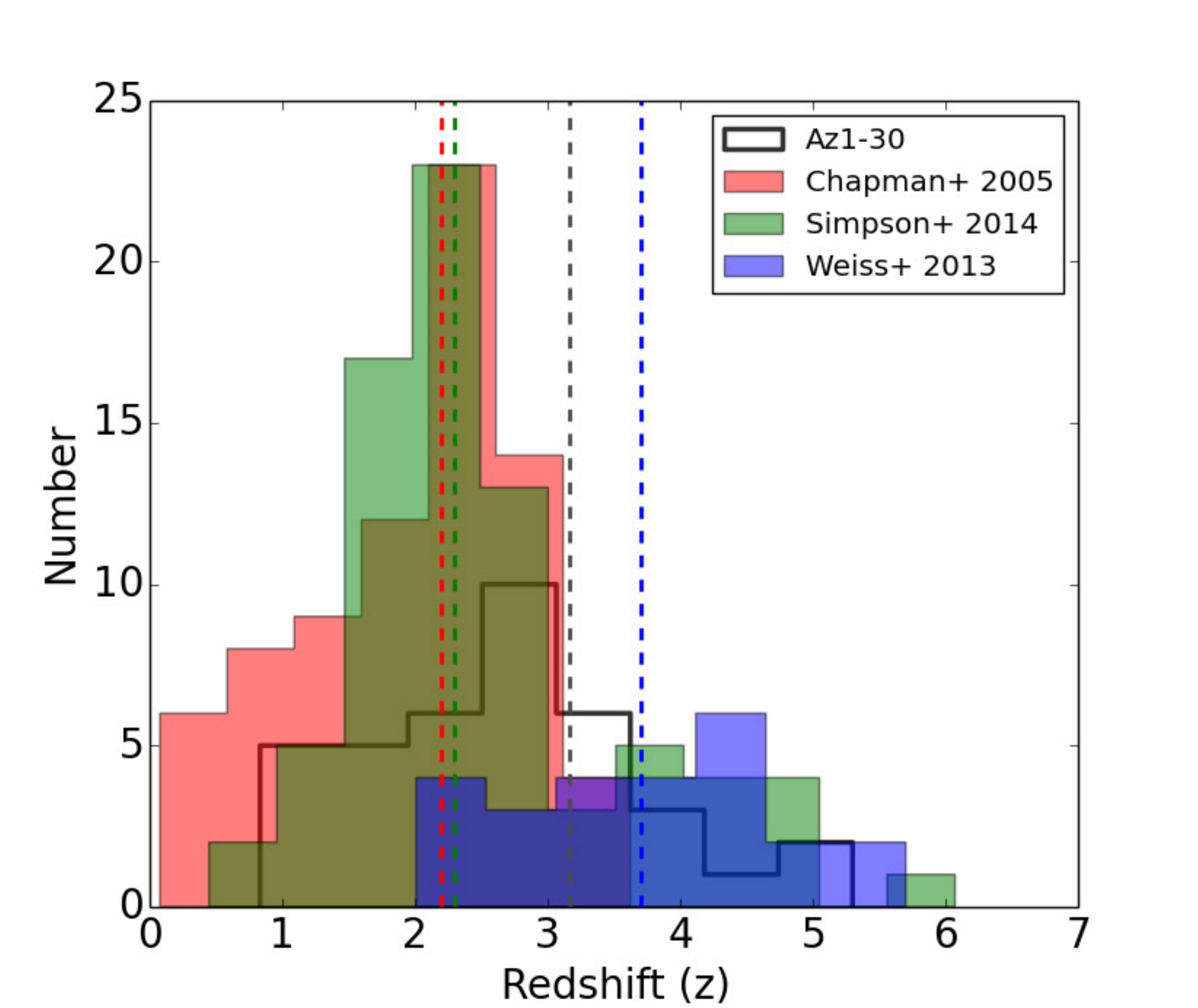}}
\caption{The SMG redshift distributions discussed in the text. Besides our result for AzTEC1--30, the plot shows the $z$ distributions for the radio-identified SCUBA SMGs 
from Chapman et al. (2005), ALESS SMGs from Simpson et al. (2014), and SPT SMGs from Wei{\ss} et al. (2013). The vertical dashed lines show the corresponding median 
redshifts [$\tilde{z}=3.17$ for AzTEC1--30, $\tilde{z}=2.2$ for the Chapman et al. (2005) SMGs, $\tilde{z}=2.3$ for the ALESS SMGs, and $\tilde{z}=3.7$ for the SPT SMGs].}
\label{figure:redshifts}
\end{figure}

The median redshift of the SMGs studied here (AzTEC16-30) is found to be $\tilde{z}=3.20\pm0.25$, 
while that for the SMGs AzTEC1--15 is $\tilde{z}=3.05\pm0.44$. The combined sample of these JCMT/AzTEC-detected SMGs, 
i.e. AzTEC1--30, has a median redshift of $\tilde{z}=3.17\pm 0.27$, which corresponds to an age of the universe of 
$2.06^{+0.21}_{-0.18}$ Gyr or about $15^{+2}_{-1}\%$ of its current age. 
A two-sample K-S test of the null hypothesis that the two subsamples, AzTEC1--15 and AzTEC16--30, 
are drawn from the same underlying parent distribution yielded a $p$-value of 0.7342. 
This suggests that the two subsamples are probably sampled from a common distribution. 
The $t$-test also suggests that the mean redshifts of the subsamples are similar to each other. 
In Fig.~\ref{figure:redshifts}, we show the redshift distribution of AzTEC1--30 and, for comparison, 
those derived in other SMG surveys discussed below.

We note that our combined SMG sample contains a source (AzTEC17a) at a redshift of $\simeq0.8$. 
This redshift is quite similar to that of the $\sim25$ Mpc long filamentary COSMOS large-scale structure (the COSMOS Wall) 
at $z\simeq0.73$ (\cite{guzzo2007}). However, cross-correlation with the redshift survey catalogue consisting of 1\,023 
galaxies belonging to the COSMOS Wall did not yield any cross matches within a $1\farcs5$ search radius 
(A.~Iovino, priv. comm.). Although AzTEC17a appears to be a member of a $z\sim0.8$ galaxy overdensity (V.~Smol{\v c}i{\'c} et al., in prep.), 
our redshift survey is not subject to strong cosmic variance arising from the COSMOS large-scale structure, 
and the different results compared to other cosmological survey fields have their origin in other effects (e.g. observing wavelength used, 
inclusion of radio-faint SMGs, etc.).

As demonstrated in the present study, radio-dim SMGs are probably lying at high redshifts (cf.~\cite{chapman2005}). 
For example, the $z\simeq 4.3$ SMG AzTEC1 is associated (near the SMA position) with only a weak 20~cm radio source 
($S_{\rm 20\, cm}=48\pm14$~$\mu$Jy), and the $z\simeq5.3$ SMG AzTEC3 has no 20~cm counterpart (\cite{younger2007}). 
These results are based on the VLA 1.4~GHz imaging down to a mean $1\sigma$ rms depth of $\sim10.5$~$\mu$Jy~beam$^{-1}$ (\cite{schinnerer2007}).
However, both AzTEC1 and AzTEC3 are associated with VLA 10~cm emission where the corresponding maps have 
a $1\sigma$ noise of 4.5~$\mu$Jy~beam$^{-1}$, which, for a typical radio spectral index of $-0.7$, corresponds to the equivalent 20~cm rms 
noise of about $\sim1.4$ times lower than the aforementioned 1.4 GHz sensitivity level (\cite{smolcic2015}). For comparison, the $1\sigma$ rms 
noise at 1.4~GHz in the phase centre of the seven fields analysed by Chapman et al. (2005) was 4--15~$\mu$Jy~beam$^{-1}$, 
while the rms sensitivity in the 1.4~GHz imaging of the ALESS SMGs was 6~$\mu$Jy~beam$^{-1}$ (\cite{thomson2014}).

Some earlier studies of SMGs have suggested that more luminous SMGs lie, on average, at higher redshifts 
compared to less luminous SMGs (e.g. \cite{ivison2002}; \cite{pope2006}; \cite{younger2007}; \cite{biggs2011}; 
\cite{smolcic2012b}). Our redshift analysis suggests that the brighter SMGs (AzTEC1--15) have a similar mean 
redshift ($\langle z \rangle=3.16\pm0.37$) within the errors to the less bright sources in our sample (AzTEC16--30; 
$\langle z \rangle=3.02\pm 0.20$). Furthermore, as noted above, the median redshifts of the two subsamples are similar 
to each other within the uncertainties. Wall et al. (2008) suggested that there might be two SMG subpopulations, divided by their luminosity 
(dividing line being at $L_{\rm 850\, \mu m}=3 \times10^{12}$~L$_{\sun}$): these would evolve 
in different ways, and the corresponding luminosity functions have different shapes. However, in agreement with our
result, Wardlow et al. (2011) found no significant correlation between the redshift and 870~$\mu$m submm 
flux density for their extensive sample of LESS SMGs, although it should be noted that many ($\sim35\%-50\%$) of the 
LESS SMGs have been resolved into multiple sources with ALMA (\cite{karim2013}; \cite{hodge2013}). 
Moreover, as discussed by Hodge et al. (2013; Sect.~5.5 therein), some of the LESS SMGs suffered from missed/misidentified 
multiwavelength counterparts, which means that they had incorrect photometric redshifts. Overall, $\sim45\%$ of the ALESS SMGs 
were missed by the sophisticated counterpart association utilising multiwavelength information by Biggs et al. (2011), and of the reported counterparts 
$\sim1/3$ were found to be incorrect (\cite{hodge2013}). However, the ALESS SMGs also do not exhibit any significant 
trend between the redshift and $S_{\rm 870\, \mu m}$ (\cite{simpson2014}; their Fig.~7). The recent semi-analytic modelling of 850~$\mu$m 
SMG surveys by Cowley et al. (2015) predicted that the bright SMG population ($S_{\rm 850\, \mu m}>5$~mJy) has 
a lower median redshift than the faint SMG population (1~mJy $<S_{\rm 850\, \mu m}<5$~mJy). We note that all our 1.1~mm 
SMGs AzTEC1--30 would belong to the bright SMG population of Cowley et al. (2015), i.e. all our SMGs have $S_{\rm 1.1\, mm}>2$~mJy 
when scaling the $S_{\rm 850\, \mu m}>5$~mJy threshold by assuming that $\beta=1.5$. From the 50 mock surveys of 0.5~deg$^2$ 
in angular size by Cowley et al. (2015), where SMGs were gene\-rated out to $z=8.5$, the median redshift for 
the bright SMGs was derived to be 2.05, while that for the faint SMGs was found to be 2.77. 
The authors also compared their model predictions with the ALESS SMG survey, and found that the model successfully reproduces 
the median redshift of the ALESS photo-$z$ distribution (see below). The opposite redshift trend predicted by Cowley et al. (2015) 
compared to some earlier observational results can, as suggested by the authors, be tested with future interferometric SMG surveys. 
They also pointed out that field-to-field variance can play a role when comparing theoretical model predictions with observational survey results.

In the following, we investigate the origin of differences in mean/median redshift for differently selected SMG samples. 
Wei{\ss} et al. (2013) carried out a blind redshift survey with ALMA towards 26 strongly lensed SMGs originally 
detected with the 10~m South Pole Telescope (SPT) at 1.4~mm. Their sample consisted of sources with high 1.4~mm flux densities of 
$>20$~mJy, and the average redshift of the sample was found to be $\langle z\rangle=3.5$, significantly higher 
than what is found for radio-identified SMGs, but quite similar to that of AzTEC1--30 selected at 1.1~mm
($\langle z \rangle=3.19\pm0.22$). Simpson et al. (2014) presented the first photo-$z$ distribution for the ALESS SMGs 
derived using {\tt HYPERZ} SED fitting with the spectral templates of Bruzual \& Charlot (2003) rather than those optimised for 
SMGs by Micha{\l}owski et al. (2010) we have used (see \cite{smolcic2012a},b for the comparison of these model libraries in the photo-$z$ analysis). 
For their sample of 77 SMGs with broadband photometry, Simpson et al. (2014) found the median redshift to be 
$\tilde{z}=2.3\pm0.1$ ($2.5\pm0.2$ when the 19 sources with poorer photometry were 
included in the analysis). This is very similar to the median spectroscopic redshift of $\tilde{z}=2.2$ derived by Chapman et al. (2005) 
for a sample of 73 radio-identified (VLA 1.4~GHz) SCUBA SMGs compiled from seven separate fields, 
but lower than the median redshift $\tilde{z}=3.17\pm0.27$ we derived for AzTEC1--30. 
To perform a more quantitative comparison with the ALESS SMG redshift distribution, we selected only 
those LESS SMGs (\cite{weiss2009}; \cite{hodge2013}) with LABOCA 870 $\mu$m flux densities corresponding to our 
AzTEC 1.1 mm flux density range (see Sect.~5.1). Altogether 17 ALESS SMGs from Simpson et al. (2014) satisfy this criterion 
(7.7 mJy $\leq S_{\rm 870\, \mu m} \leq21.8$~mJy). For this 870 $\mu$m flux limited sample we derived the following photometric redshift statistics: 
$\langle z \rangle=2.99\pm0.31$, $\tilde{z}=2.85\pm0.39$, ${\rm std}=1.27$, and ${\rm CI=2.39-3.59}$ (95\%). 
As can be seen the median value is higher than that for the original sample of 77 ALESS SMGs (\cite{simpson2014}), but lower than the median 
redshift for AzTEC1-30. We also carried out a K-S test between our sample (excluding the lower limits 
in which case $\langle z \rangle=2.61\pm0.26$) and the ALESS SMGs with comparable flux densities, and found that 
they could have a common underlying parent distribution ($D=0.2379$, $p=0.6379$). 
Furthermore, the $t$-test suggests that these two samples have a comparable 
average redshift ($p=0.3573$ under the null hypothesis that the two $\langle z \rangle$ values are equal). 
The facts that the LESS and JCMT/AzTEC SMGs were selected at different wavelengths (870 $\mu$m compared to 1.1~mm) 
from different fields, and that the ALESS sample is larger than ours make a direct comparison between the two questionable. 
Given that the lower redshift limits to some of our SMGs raise the total sample average to $\langle z \rangle=3.19\pm0.22$ 
could be an indication that the 1.1~mm wavelength selects somewhat higher-redshift SMGs. Moreover, the ALESS sample was drawn from the Extended
Chandra Deep Field South that has a size of $30\arcmin \times 30\arcmin$, or 0.25~$\sq\degr$, while our SMGs were selected from the 0.15 
$\sq\degr$ COSMOS JCMT/AzTEC field. The intrinsic target field properties, or cosmic variance might therefore also play a role 
(cf.~\cite{cowley2015}).

Swinbank et al. (2014) analysed the ALESS SMGs detected in at least two \textit{Herschel}/SPIRE bands. 
They found that the SMGs exhibiting the peak dust emission at 
$\lambda=250$, 350, and 500~$\mu$m have median redshifts of $\tilde{z}=2.3\pm0.2$, $2.5\pm0.3$, and $3.5\pm0.5$, 
respectively (the SPIRE non-detected sources were found to have a median photo-$z$ of $3.3\pm0.5$). 
Although the SED peak position is not always well constrained and the subsamples clearly overlap with each other 
(Fig.~6 in \cite{swinbank2014}), this suggests that there is a positive correlation between the source redshift and the 
SED peak wavelength. 
Within the errors the 500 $\mu$m peakers have a median redshift comparable to that of our SMGs. 
The highest redshift SMG known to date, HFLS3 at $z=6.34$, is also a 500~$\mu$m peaker and was originally found 
from the HerMES survey as having a very high 500~$\mu$m flux density of 
$S_{\rm 500\, \mu m}=1.46 \times S_{\rm 350\, \mu m}=47.3\pm2.8$~mJy (\cite{riechers2013}). 
Similarly, Dowell et al. (2014) selected dusty star-forming galaxies from the HerMES survey on the basis of their 
\textit{Herschel}/SPIRE colours, and found most of the 500 $\mu$m peakers (including HFLS3) to lie at very high redshifts ($z>4$).
The relatively high redshifts among the AzTEC1--30 COSMOS SMGs could be a selection effect in the sense that they were 
originally discovered at $\lambda=1.1$ mm, although cosmic variance can also play a role because the COSMOS field is known 
to contain a relatively large number of very high-$z$ SMGs. 
Zavala et al. (2014) carried out simulations of the SMGs' redshift distributions, 
and they studied how different selection effects affect the derived distributions. 
Their simulated data showed the increase in the median redshift as a function of 
wavelength (changing from $\tilde{z}=2.06\pm0.10$ at 450 $\mu$m to $\tilde{z}=2.91\pm0.12$ at 2 mm). 
However, they demonstrated that the differences reported in the literature can be explained by 
the observing wavelength (related to the SED temperature) used and, to a lesser degree, by the map noise 
level and angular resolution, and that some of the redshift distributions suggested to be different from 
each other can actually be drawn from the same parent distribution.

As discussed above, the derived SMG redshift distribution can be highly affected by the wavelength selection 
and source sample under study. A well established example is the radio preselection that biases the sample towards
lower-redshift ($z < 3$) SMGs (\cite{chapman2005}). However, spectroscopic observations have shown that the $z>4$ SMGs 
are more common than originally thought (see references in Sect.~1). 
A mix of different methods to derive the redshifts, such as spectroscopic and photometric method can also lead to a biased distribution of 
redshift values. For galaxies in the COSMOS field, however, it has been shown that the photo-$z$ values agree well with those derived 
through spectroscopic observations (\cite{ilbert2013}). Considering only the most secure spec-$z$ values at $K_{\rm s} < 24$ 
(a sample of 12\,482 galaxies), Ilbert et al. (2013) found that the photo-$z$ accuracy is $\sigma_{\Delta z/(1+z)}=9.6\times10^{-3}$ 
and only 2.1\% are catastrophic failures with $\vert z_{\rm phot}- z_{\rm spec}\vert / (1+z_{\rm spec})>0.15$. 
The different methods of deriving the photo-$z$ values (e.g. varied assumptions and spectral templates) can also lead 
to differing results, but our photo-$z$ values derived from {\tt HYPERZ} using the SMG SED templates from Micha{\l}owski et al. (2010) 
are expected to be reliable (\cite{smolcic2012a},b); among our new SMG sample, this is supported by the case of 
AzTEC17a ($z_{\rm spec}=0.834$ compared to $z_{\rm phot}=0.75_{-0.12}^{+0.23}$). In some cases the best photo-$z$ solution is uncertain because 
the corresponding $\chi_{\rm tot}^2$ distribution is complex having a broad minimum or multiple dips of comparably low $\chi_{\rm tot}^2$ value. 
Moreover, our SMG redshift distribution is partly based on lower limits only that were derived using the Carilli-Yun radio-submm redshift 
indicator (\cite{carilli1999}, 2000). This method is subject to a degeneracy between $T_{\rm dust}$ and $z$, and can suffer from 
large uncertainties. Another caveat in determining the photo-$z$ values is the possible contamination by AGNs. 
The reason for this is that methods based on stellar libraries might suffer from short-wavelength 
(UV to MIR) AGN emission (see \cite{smolcic2012b} for further discussion). 
However, as mentioned earlier our sources do not exhibit any strong X-ray signatures and are therefore unlikely to 
contain bright AGNs. 

To summarise, our new interferometric observations have enabled us to pinpoint the multiwavelength counterparts of our SMGs, and therefore 
to derive the photo-$z$ values for these SMGs. For this type of analysis, interferometry provides an important improvement because 
the usage of single-dish (sub)mm data of $\sim10-30\arcsec$ resolution can result in a wrong counterpart identification, 
and therefore also wrong redshift of the SMG. For five SMGs among AzTEC1--15, we have a secure spectroscopic redshift available, 
but only one spec-$z$ among AzTEC16--30. In the ideal case, all the SMG redshifts would be based on spectroscopic data. 
This way one could carry out a completely fair comparison between our two subsamples of AzTEC1--15 and AzTEC16--30. 

\section{Summary and conclusions}

We have used the IRAM/PdBI to carry out an interferometric 1.3~mm continuum follow-up study of a sample of 
15 SMGs originally detected in the COSMOS field with the JCMT/AzTEC bolometer at 1.1~mm ($S_{\rm 1.1\, mm}\simeq 3.3-3.9$~mJy)
by Scott et al. (2008). The good angular resolution of about $1\farcs8$ allowed us to pinpoint the positions of the actual 
SMGs giving rise to the single-dish detected mm emission. We combined these new observations with results from the literature to study 
the ensemble properties of the 30 most significant (${\rm S/N}=4.0-8.3$) SMGs detected by JCMT/AzTEC in the COSMOS field.
Our main results and conclusions are summarised as follows:

\begin{enumerate}
\item The total number of sources detected in this survey is 22, where the sample does consist of S/N$_{\rm 1.3\, mm}>4.5$ detections 
(whether or not having a counterpart) and sources detected with $4<{\rm S/N}_{\rm 1.3\, mm}\leq4.5$ that have multiwavelength counterparts.
AzTEC19 is found to be the most significant 1.3~mm emitter (in the observed frame) with S/N$_{\rm 1.3\, mm}=10.3$.
\item Visual inspection of the 1.3~mm images reveals that AzTEC19, 21, 27, and 28 have elongated/clumpy morphologies, 
a possible manifestation of galaxy merging. AzTEC27 appears to be a gravitationally lensed SMG, where two intervening galaxies 
are warping and magnifying the radiation (see Fig.~\ref{figure:lens}).
\item AzTEC17, 19, 21, 24, 26, and 29 appear to split into two to three sources of 1.3 mm emission. Considering our new SMG sample (15 single-dish detected 
sources), this would mean that the multiplicity fraction is $40\pm16\%$ ($33\pm15\%$ if AzTEC24 is spurious). In all the other cases except AzTEC29, 
the source components have comparable observing-frame 1.3 mm flux densities. 
Among AzTEC1--15 there are two SMGs that are known to be multiple systems. Combining these statistics we conclude that the multiplicity fraction among AzTEC1--30 is 
$\sim25\pm9\%$. Deep, high-resolution (sub)mm surveys of large SMG samples are required to unambiguously determine how common multiplicity is among SMGs.
\item Besides the spectroscopic redshift of AzTEC17a, the redshifts of AzTEC16--30 were derived using either 
optical/IR photometric data or the Carilli-Yun redshift indicator (\cite{carilli1999}, 2000). 
In many cases only lower limits could be estimated, and the median redshift was found to be 
$\tilde{z}=3.20\pm0.25$. We identified some high-redshift candidates; in particular, AzTEC17b has a photo-$z$ of 
$4.14_{-1.73}^{+0.87}$, and a lower limit to $z_{\rm radio/submm}$ of AzTEC27 was derived to be as high as $>4.17$. For the 
15 brightest JCMT/AzTEC 1.1~mm detected SMGs, namely AzTEC1--15, the median redshift is $\tilde{z}=3.05\pm 0.44$ 
(partly based on secure spectroscopic redshifts). For the combined sample of AzTEC1--30, the median redshift was 
found to be $\tilde{z}=3.17\pm 0.27$. This is higher than what is usually reported for SMGs, 
but in agreement with mm-selected SMG samples. 
\item The redshift analysis does not support the earlier observational result that brighter SMGs (our sources 
AzTEC1--15) would lie at higher redshifts than the fainter SMGs (AzTEC16--30). Instead, besides the median 
redshifts, the mean redshifts of AzTEC1--15 and AzTEC16--30 are similar to each other within the errors 
($\langle z \rangle=3.16\pm0.37$ and $\langle z \rangle=3.02\pm0.20$, respectively). The $t$-test also supports the 
similarity between the mean redshift values. Finally, the K-S test suggests that the two 
subgroups are probably drawn from a common parent population, but we note that the highest redshift ($z\gtrsim 4.3$) SMGs 
are found among the strongest millimetre emitters. The absence of any significant trend between the source redshift and 
millimetre flux density is in agreement with that found for the ALESS SMGs at 870 $\mu$m wavelength (\cite{simpson2014}).
\end{enumerate}

Some of the great challenges in detailed observational studies of SMGs is to reliably identify their 
multiwavelength counterparts. While interferometric (sub)millimetre imaging is a prerequisite for secure counterpart 
identifications, faint SMGs, flux-boosted in single-dish observations, might not be detected in shallow interferometric maps. 
Moreover, as the present study demonstrates, the sample might suffer from spurious sources that may or may not have nearby sources detected 
at other wavelengths. The knowledge of secure counterparts is required to obtain accurate estimates of the sources' photometric redshifts. 
Spectral line observations of atoms (such as the $\lambda_{\rm rest}=158$~$\mu$m forbidden 
C$^+$ fine-structure line) or molecules (particularly rotational lines of $^{12}$CO) are needed to obtain the most reliable 
redshifts [cf.~ALMA observations towards SPT SMGs by Wei{\ss} et al. (2013)], and hence to push forward our understanding of 
high-redshift, submillimetre-luminous starburst galaxies, their role in galaxy formation and evolution, and cosmic star formation history.

\begin{acknowledgements}

We thank the referee for providing helpful and constructive comments.
We would also like to thank C.~M.~Casey and S.~Toft for their comments and suggestions.
This research was funded by the European Union's Seventh Framework programme 
under grant agreement 337595 (ERC Starting Grant, 'CoS-Mass'). 
AK acknowledges support by the Collaborative Research Council 956,
sub-project A1, funded by the Deutsche Forschungsgemeinschaft (DFG). 
We would like to thank the IRAM staff for carrying out the PdBI observations presented in this paper. 
This work is partly based on data products from observations made with ESO Telescopes at the La Silla Paranal 
Observatory under ESO programme ID {\tt 179.A-2005} and on data 
products produced by TERAPIX and the Cambridge Astronomy Survey Unit on behalf of the 
UltraVISTA consortium. This research has made use of NASA's Astrophysics Data System, and the NASA/IPAC Infrared Science Archive, which is operated by the JPL, 
California Institute of Technology, under contract with the NASA. 
This study also made use of {\tt APLpy}, an open-source plotting package for {\tt Python} hosted at 
{\tt http://aplpy.github.com}, and {\tt TOPCAT}, an interactive graphical tool for analysis and manipulation of tabular data 
available at {\tt http://www.star.bristol.ac.uk/$\sim$mbt/topcat/} (\cite{taylor2005}). 
We greatfully acknowledge the contributions of the entire COSMOS collaboration consisting of more than 100 scientists. 
More information on the COSMOS survey is available at {\tt http://www.astro.caltech.edu/$\sim$cosmos}. 

\end{acknowledgements}

\appendix

\section{Multiwavelength images}

A selection of zoomed-in multiwavelength views towards AzTEC16--30 is 
shown in Fig.~\ref{figure:stamps}. In the first (top left) panel of each source 
we show the PdBI 1.3~mm image overlaid with the same
contour levels as in Fig.~\ref{figure:pdbi}. The PdBI images are also annotated with the source designations.
The positive 1.3~mm contours are overlaid on the other wavelength images to guide the eye.

\begin{figure*}
\begin{center}
\includegraphics[width=\textwidth]{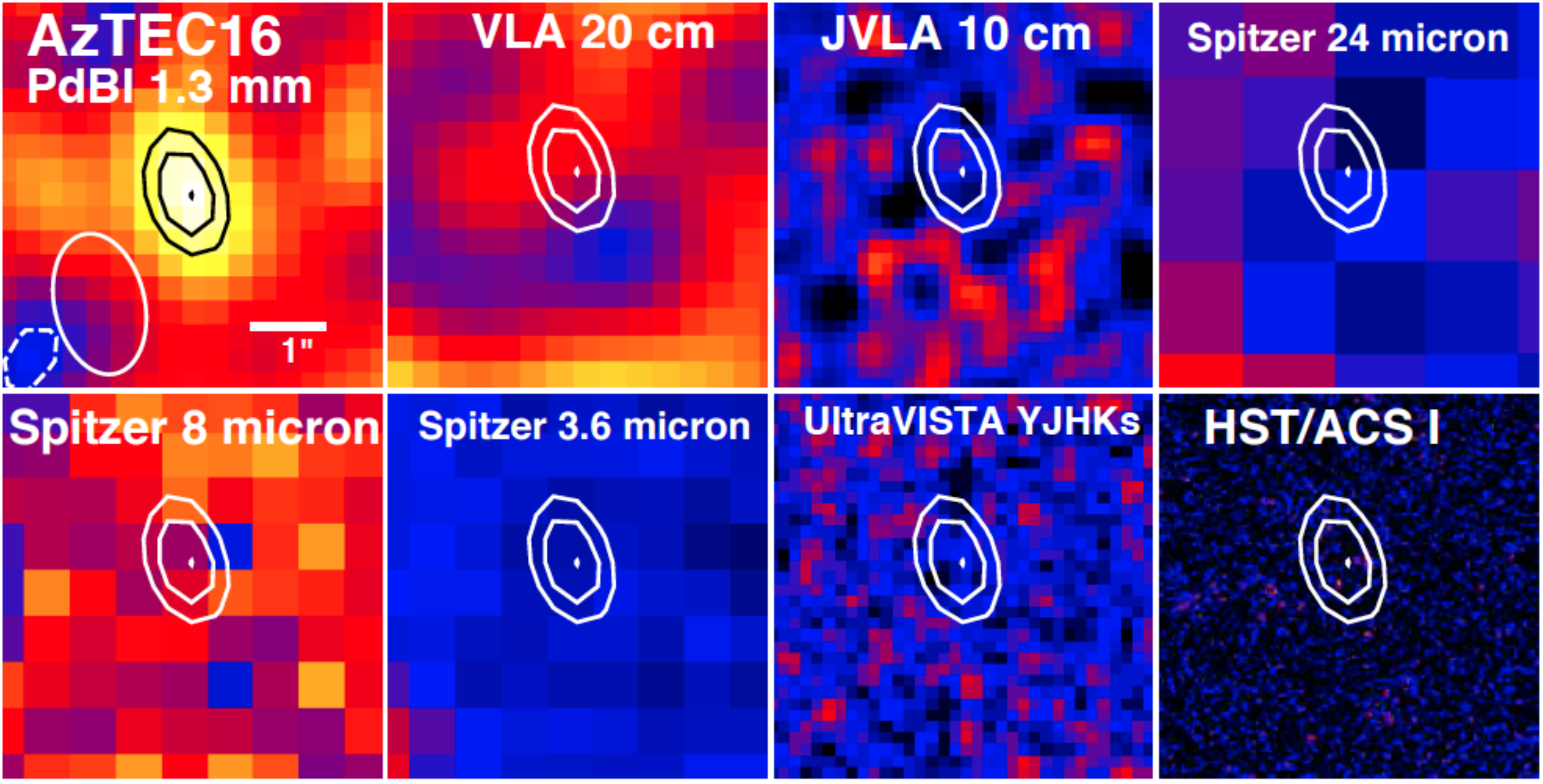}

\vspace{5mm}

\includegraphics[width=\textwidth]{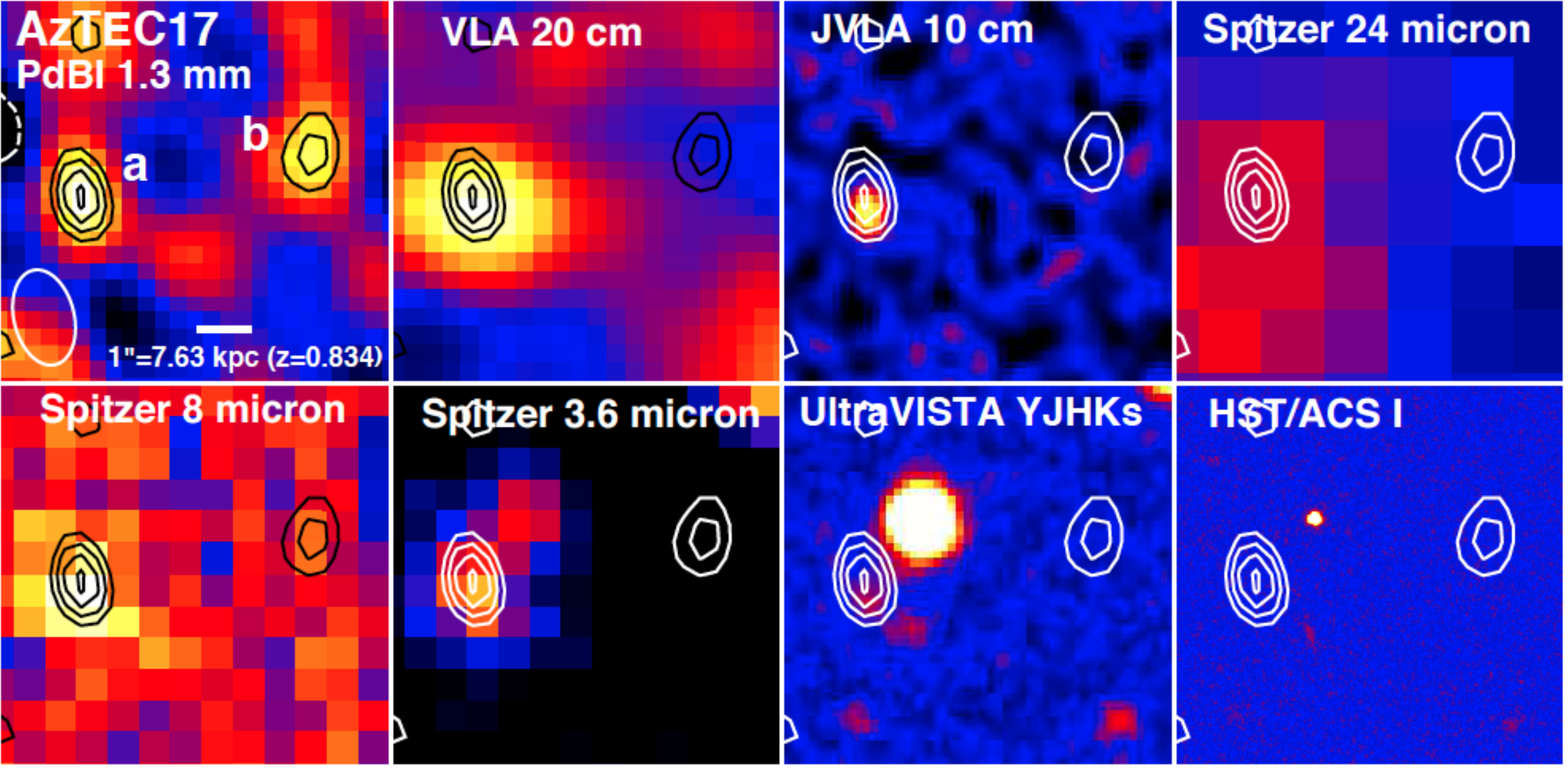}
\caption{Multiwavelength views towards AzTEC16--30. 
The panels from top left to bottom right for each source are as follows: PdBI 1.3 mm, VLA 20 cm, 
VLA 10 cm, \textit{Spitzer} 24 $\mu$m, \textit{Spitzer} 8 $\mu$m, 
\textit{Spitzer} 3.6 $\mu$m, UltraVISTA $YJHK_{\rm s}$ colour composite, and \textit{HST}/ACS $I$-band. 
The overlaid 1.3 mm contours are as in Fig.~\ref{figure:pdbi}, and positive 1.3 mm contours are shown in all panels. The
synthesised beam of the PdBI data is shown in the bottom left corner in the first panel for each source. 
A scale bar indicating the $1\arcsec$ projected length is shown in the PdBI panel, 
and the corresponding proper length [kpc] at the indicated redshift is also denoted (except when only a lower limit to $z$ could be derived).
The catalogue positions of the \textit{Herschel}/SPIRE 250 $\mu$m sources are 
marked with plus signs in the PdBI images towards AzTEC19, 20, and 24. The diamond symbol in the PdBI image towards AzTEC24 
indicates the position of the ASTE/AzTEC 1.1 mm source AzTEC/C48 from Aretxaga et al. (2011).}
\label{figure:stamps}
\end{center}
\end{figure*}

\addtocounter{figure}{-1}
\begin{figure*}
\begin{center}




\includegraphics[width=\textwidth]{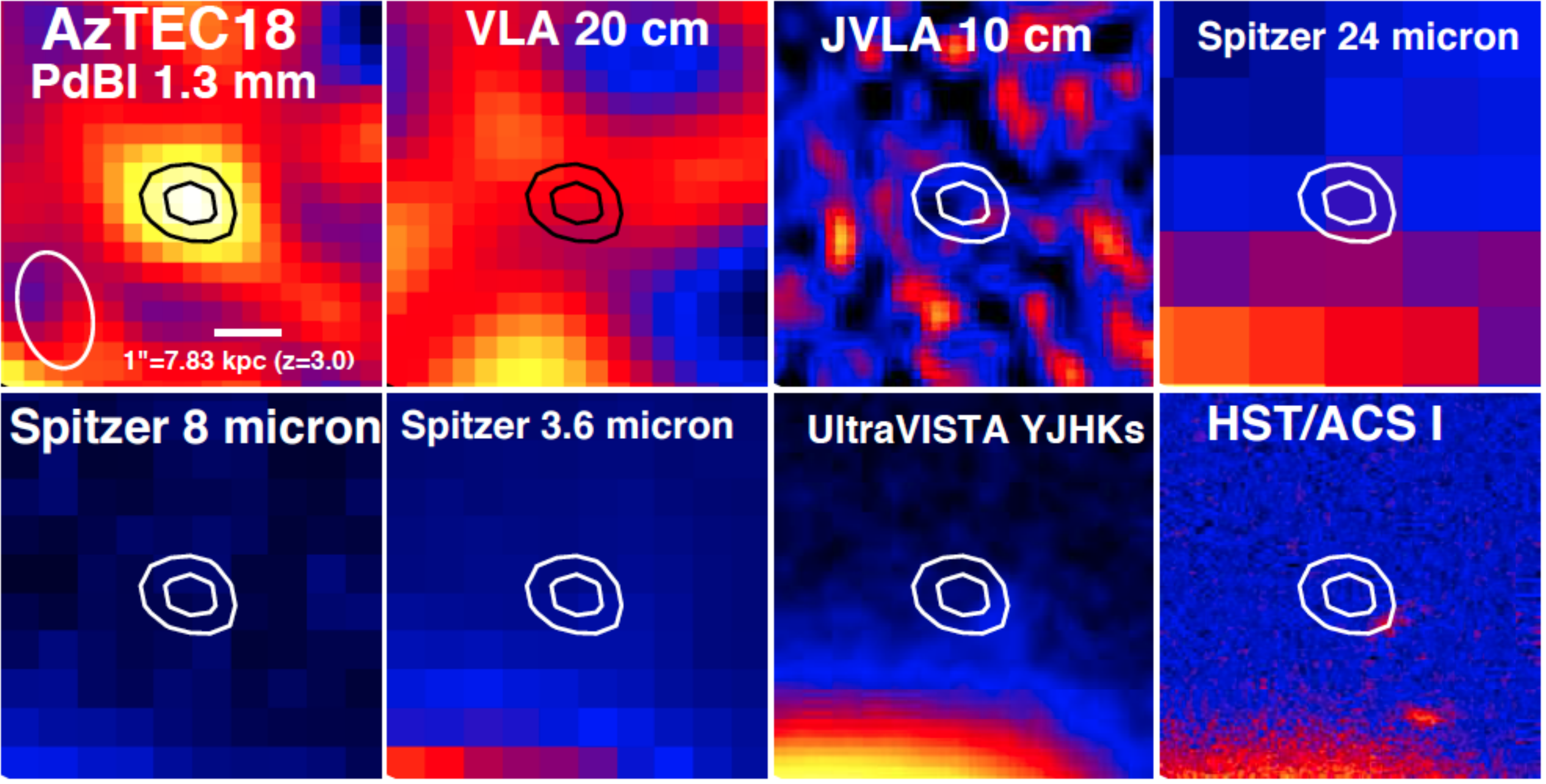}

\vspace{5mm}

\includegraphics[width=\textwidth]{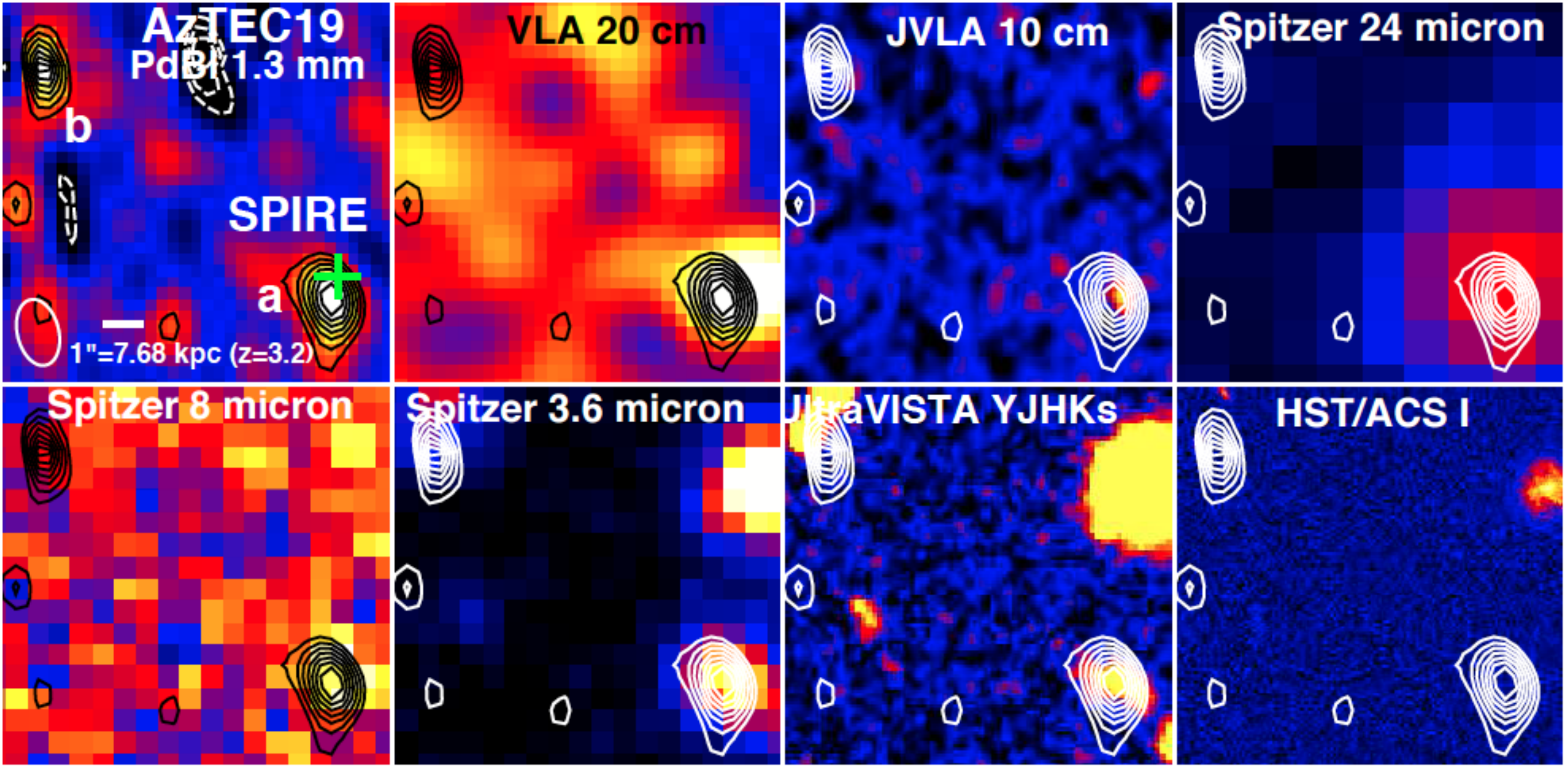}



\caption{continued.}
\label{figure:stamps}
\end{center}
\end{figure*}

\addtocounter{figure}{-1}
\begin{figure*}
\begin{center}

\includegraphics[width=\textwidth]{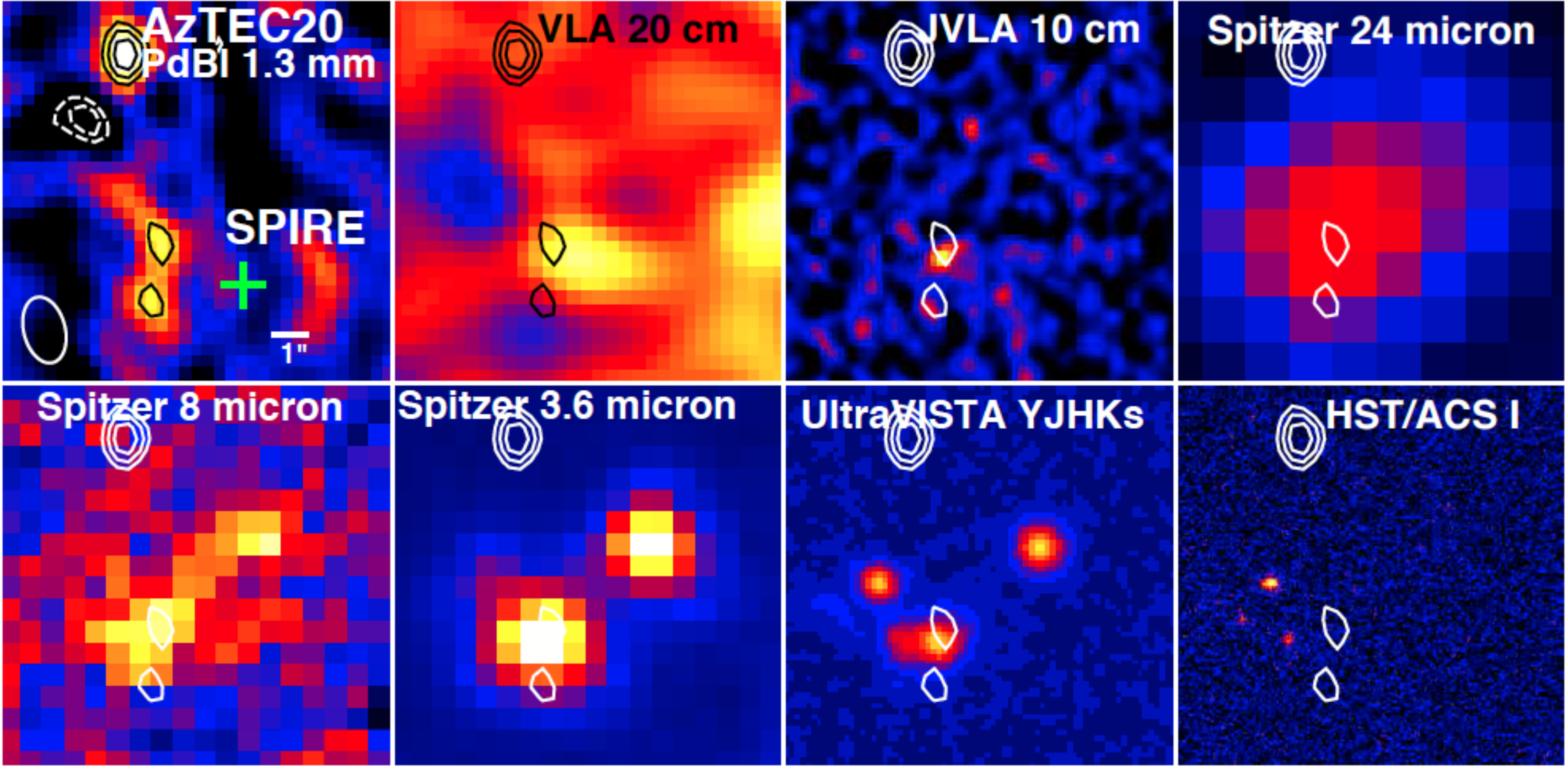}

\vspace{5mm}

\includegraphics[width=\textwidth]{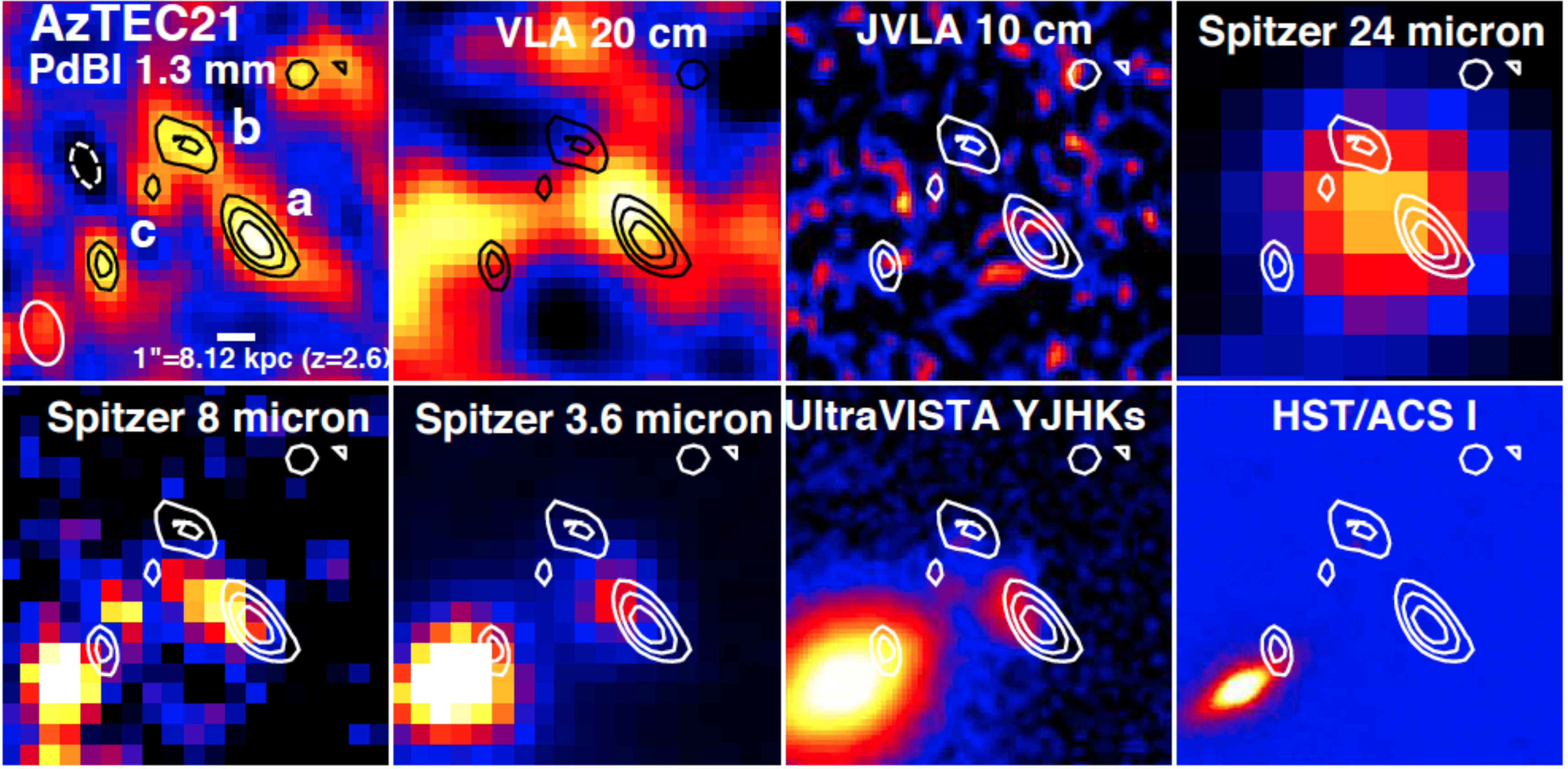}


\caption{continued.}
\label{figure:stamps}
\end{center}
\end{figure*}

\addtocounter{figure}{-1}
\begin{figure*}
\begin{center}
\includegraphics[width=\textwidth]{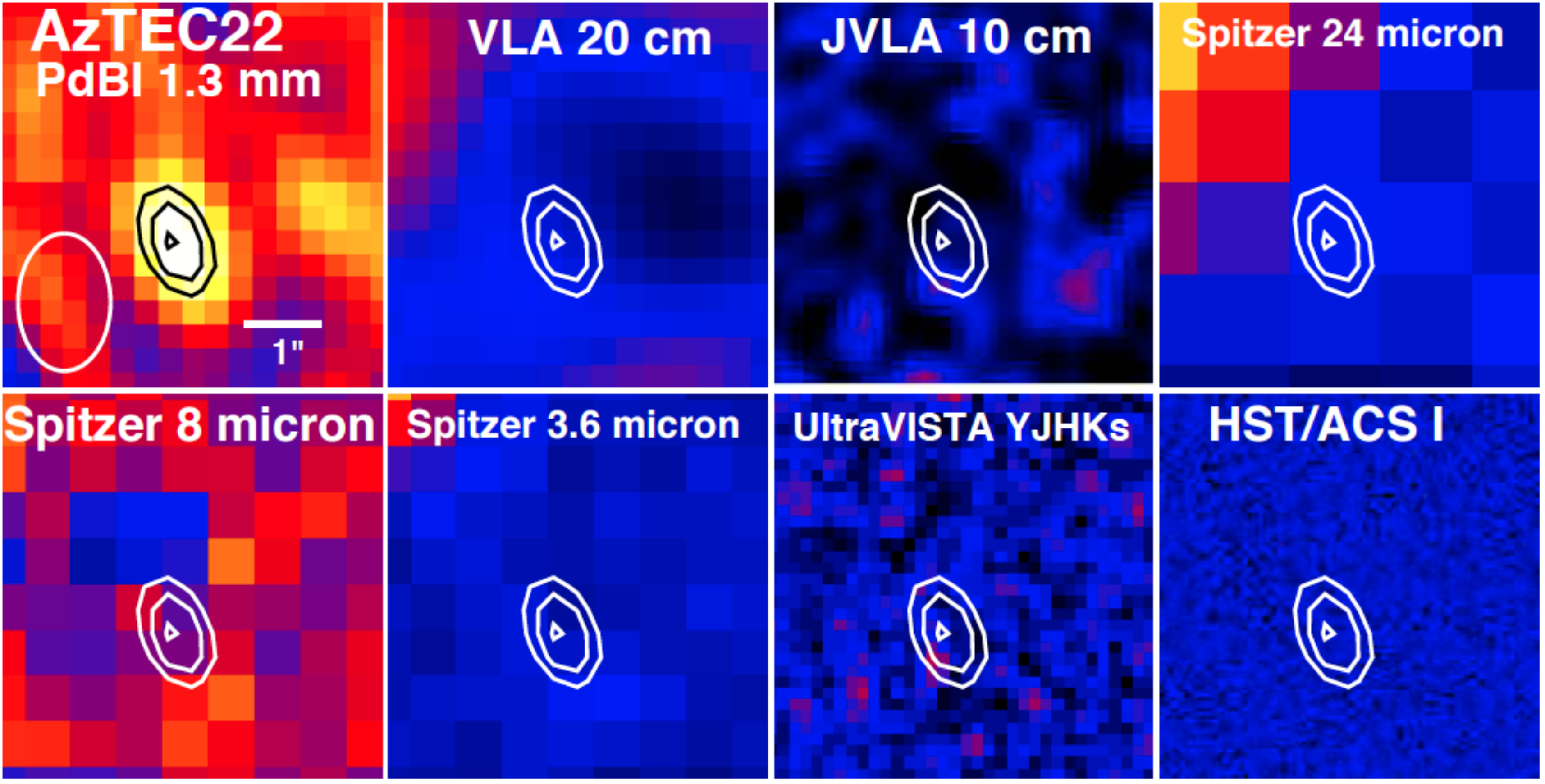}

\vspace{5mm}

\includegraphics[width=\textwidth]{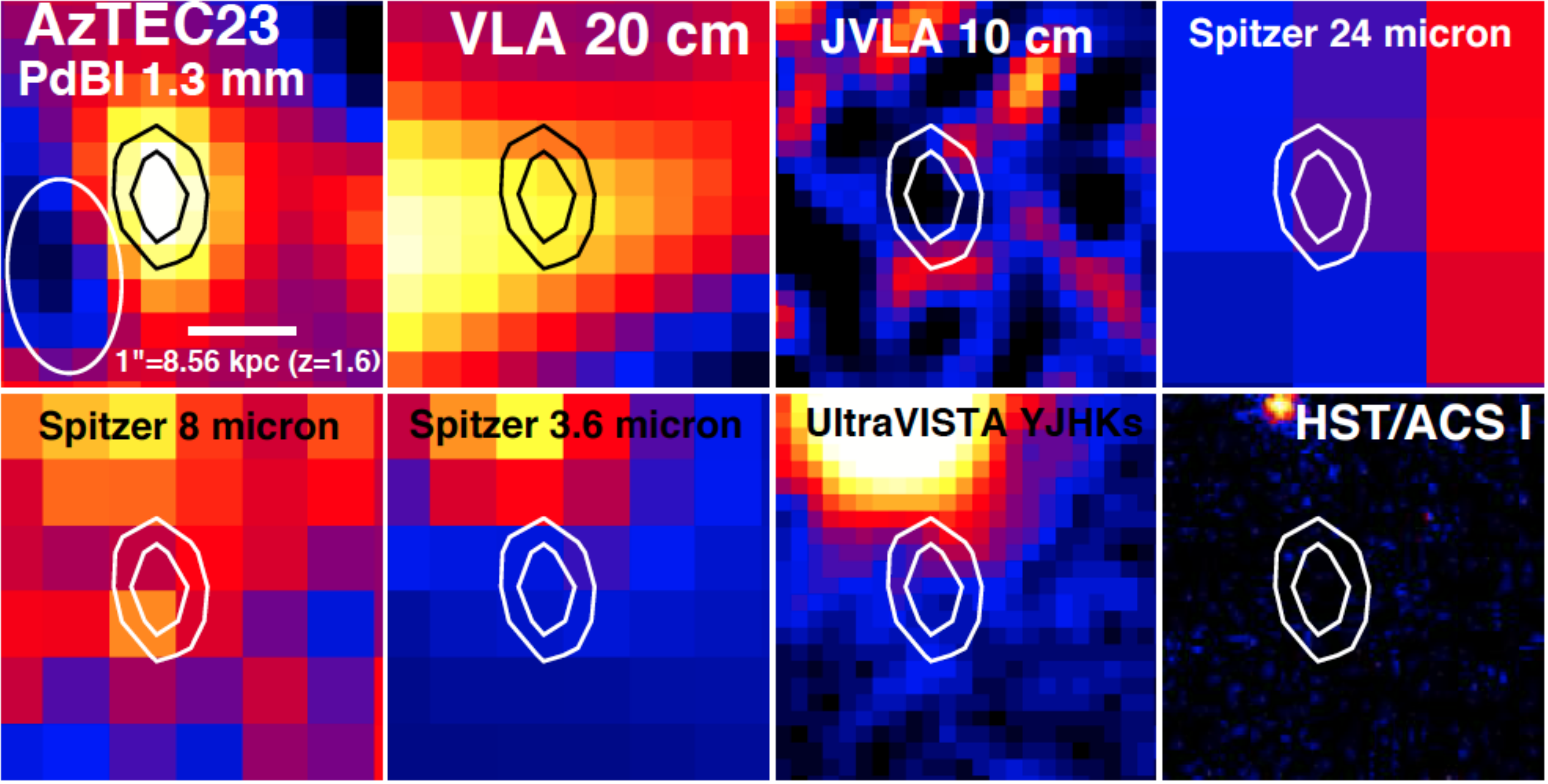}
\caption{continued.}
\label{figure:stamps}
\end{center}
\end{figure*}

\addtocounter{figure}{-1}
\begin{figure*}
\begin{center}
\includegraphics[width=\textwidth]{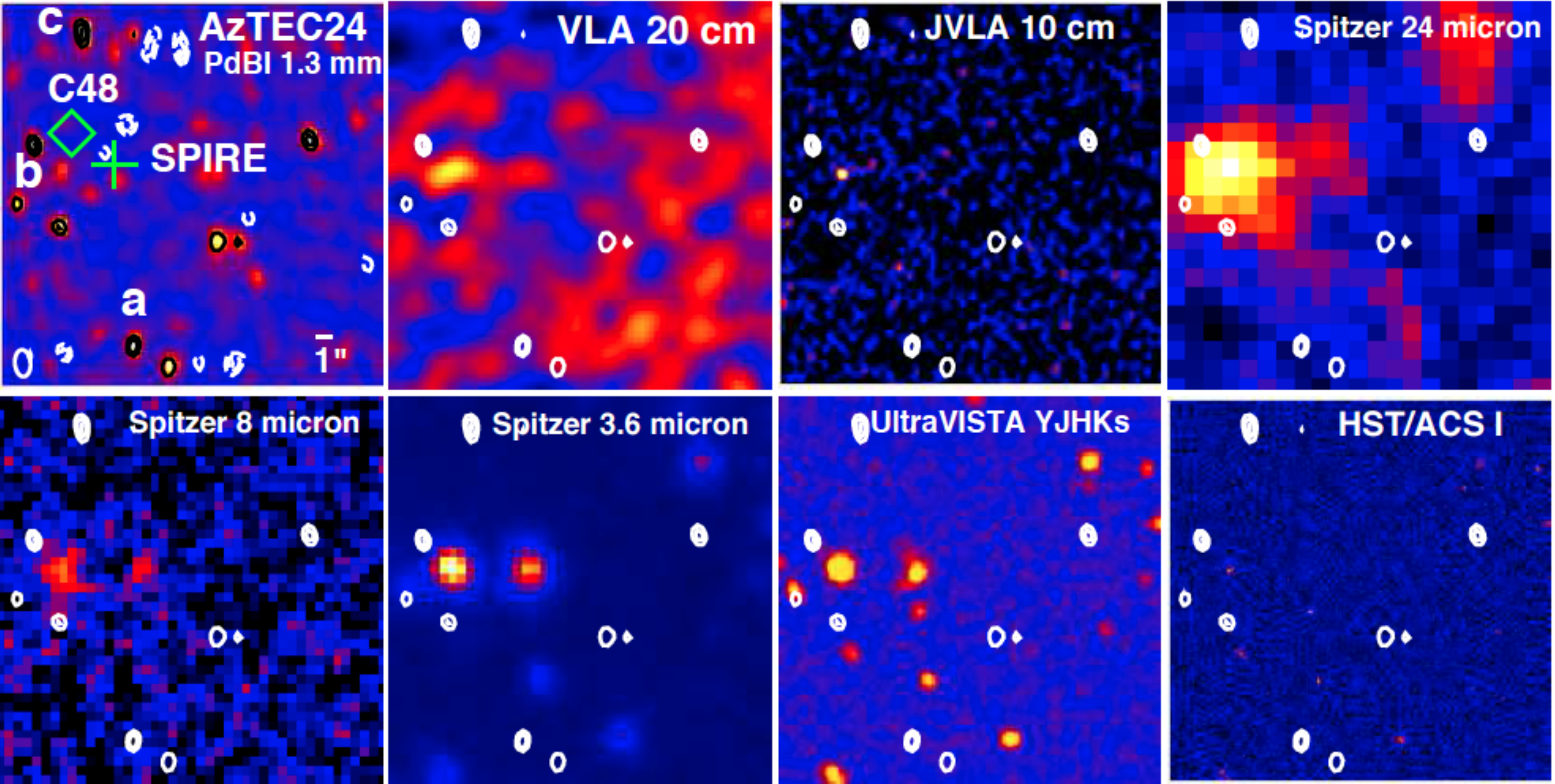}

\vspace{5mm}

\includegraphics[width=\textwidth]{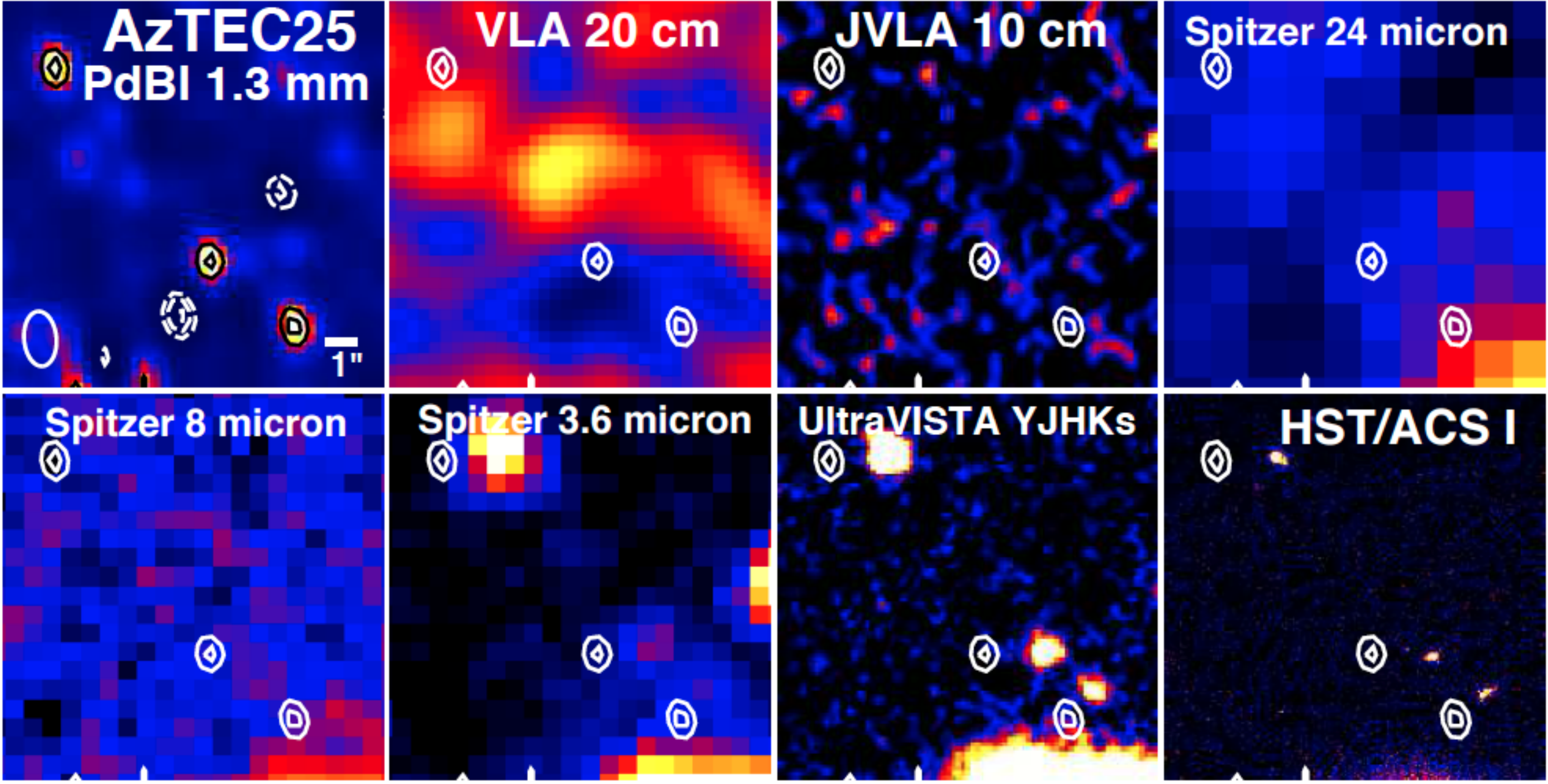}
\caption{continued.}
\label{figure:stamps}
\end{center}
\end{figure*}

\addtocounter{figure}{-1}
\begin{figure*}
\begin{center}
\includegraphics[width=\textwidth]{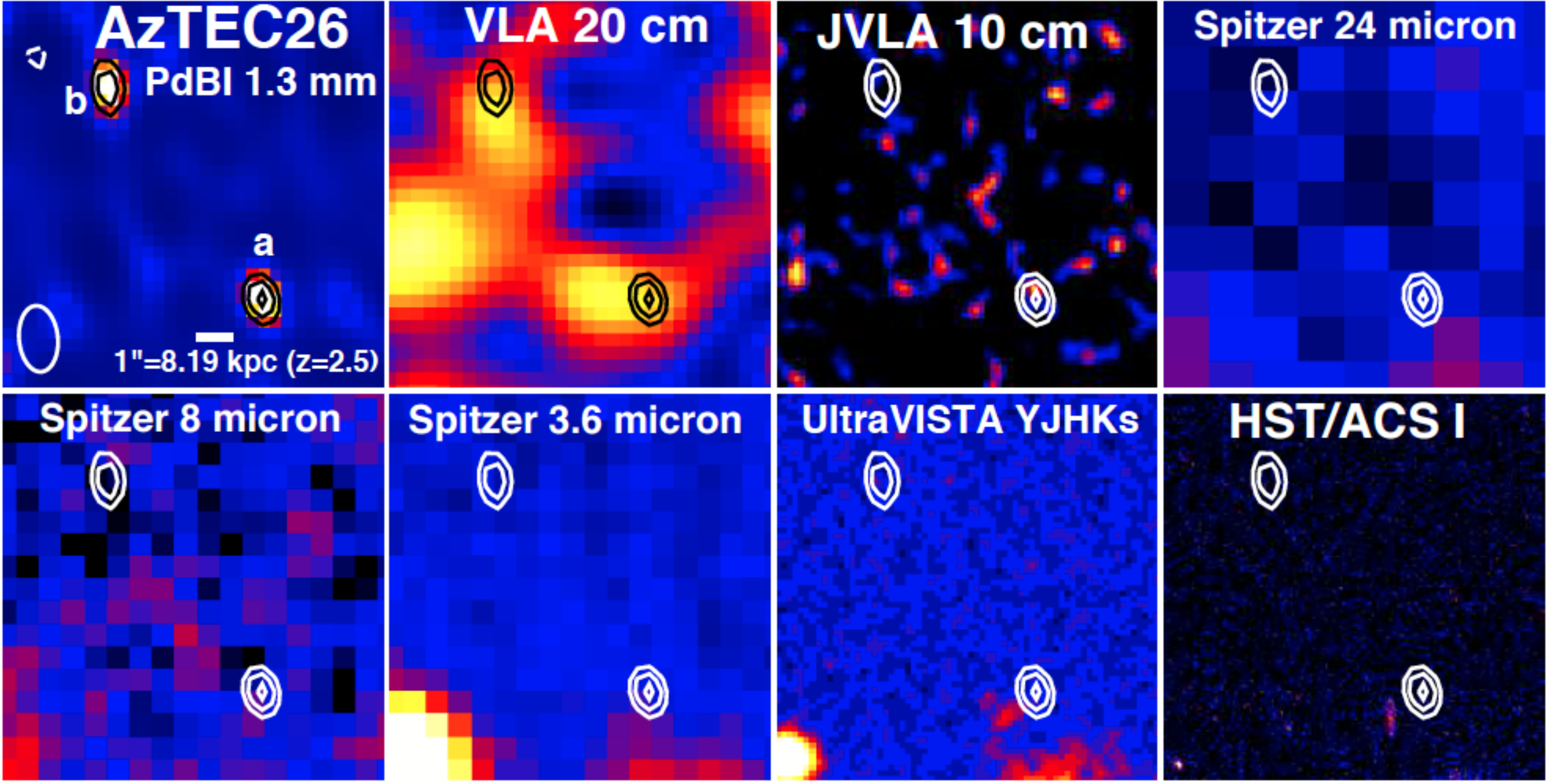}

\vspace{5mm}

\includegraphics[width=\textwidth]{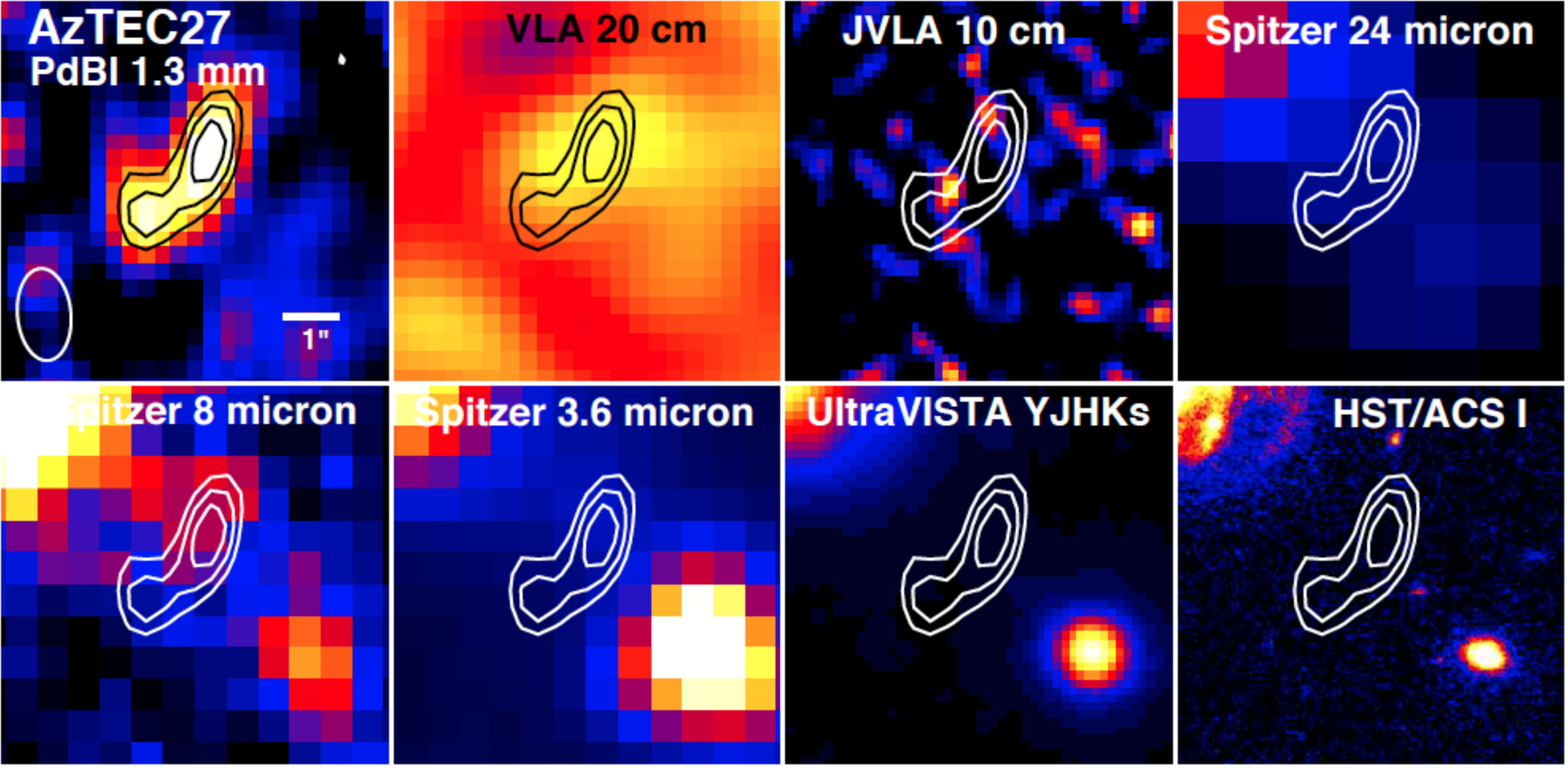}
\caption{continued.}
\label{figure:stamps}
\end{center}
\end{figure*}

\addtocounter{figure}{-1}
\begin{figure*}
\begin{center}
\includegraphics[width=\textwidth]{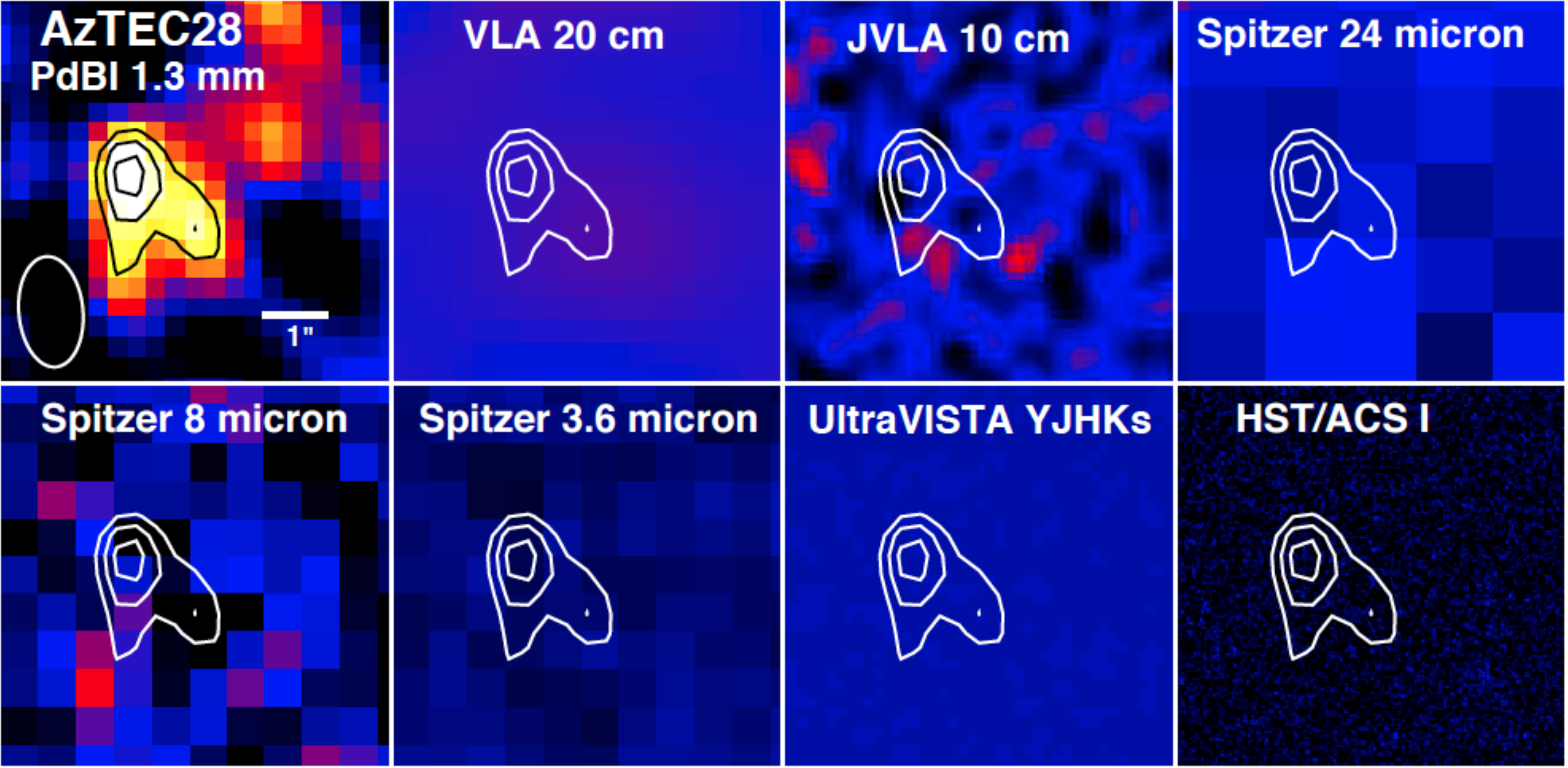}

\vspace{5mm}

\includegraphics[width=\textwidth]{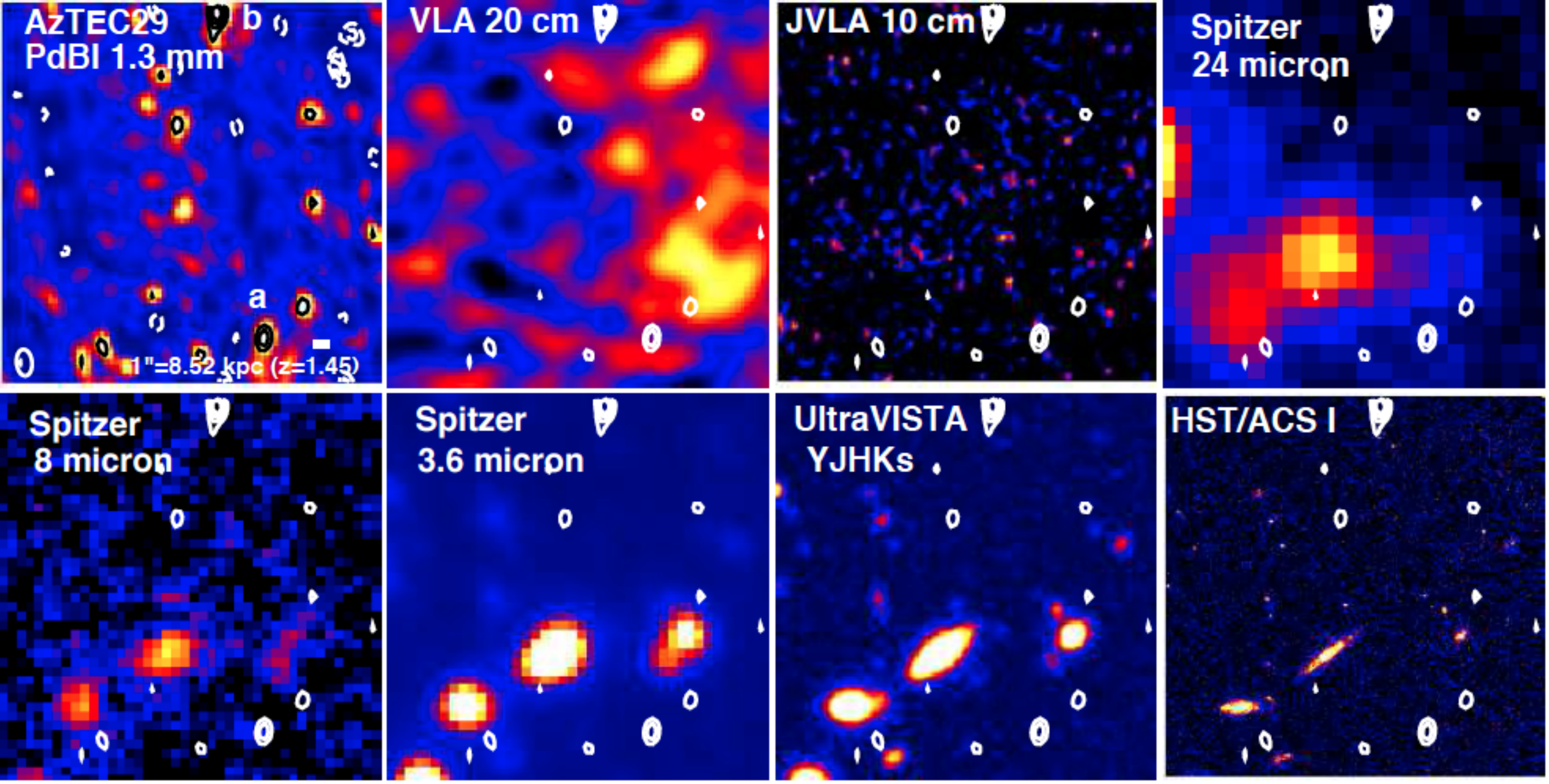}
\caption{continued.}
\label{figure:stamps}
\end{center}
\end{figure*}

\addtocounter{figure}{-1}
\begin{figure*}
\begin{center}
\includegraphics[width=\textwidth]{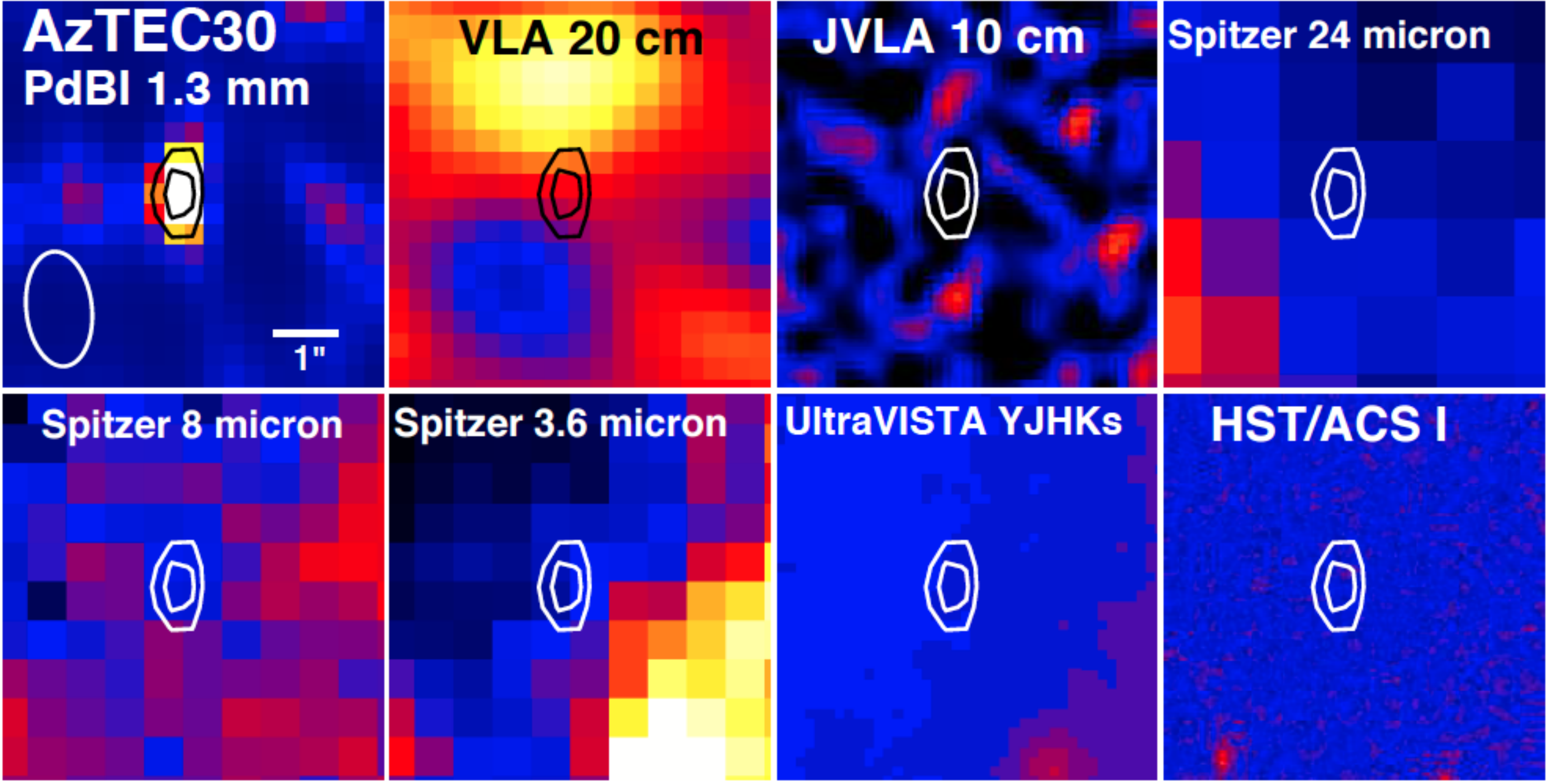}

\caption{continued.}
\label{figure:stamps}
\end{center}
\end{figure*}

\section{Updated redshifts of the 15 brightest JCMT/AzTEC-detected SMGs: AzTEC1--15}

Here we provide the reader with an overview of the redshifts of the SMA-detected SMGs AzTEC1--15.
Among these SMGs, there are eight spectroscopic redshifts reported in the literature: 
for AzTEC1, 2, 3, 5, 6, 8, 9, and 11 (see \cite{smolcic2012b}; their Tables~1 and 4 and references therein). 
Despite the partially extensive efforts and data coverage some of these redshift determinations are still uncertain. 
We discuss below the updated redshifts among AzTEC1--15 and the cases where there is some confusion about the source redshifts in the literature.

Smol{\v c}i{\'c} et al. (2011) determined a spectroscopic redshift of $4.650\pm0.005$ for AzTEC1. 
The UV--NIR photometric redshift they derived, $z_{\rm phot}=4.64^{+0.06}_{-0.08}$, was found to be very 
similar to the $z_{\rm spec}$ value, although a secondary photo-$z$ solution at $z_{\rm phot}=4.44$ was also found. 
A somewhat lower photo-$z$ of $4.26^{+0.17}_{-0.20}$ was derived by Smol{\v c}i{\'c} et al. (2012b) using 
the same method as in the present paper (Sect.~4.2). The CO spectral-line observations using the Redshift Search Receiver (RSR) on 
the Large Millimetre Telescope (LMT) performed by M.~S.~Yun et al. (in prep.) yielded a spec-$z$ value of 4.3421 for AzTEC1. 
Their SMA follow-up observations of C$^+$ emission yielded a line detection at $z_{\rm spec}=4.3415$, in very good agreement with 
the CO observations. Since it is based on interferometric observations, this last redshift is adopted in the present work. 
We note that the new spec-$z$ of AzTEC1 explains the non-detection of the CO$(5-4)$ line emission by Smol{\v c}i{\'c} et al. (2011) 
because their PdBI and Combined Array for Research in Millimetre-wave Astronomy (CARMA) observations covered the redshift ranges 
4.56--4.76 and 4.94--5.02.

The optical spectrum observed with the Deep Extragalactic Imaging Multi-Object Spectrograph (DEIMOS) on the 10~m Keck II telescope towards AzTEC2 exhibits an emission feature that can be assigned to the $[\ion{O}{II}]$ $\lambda3727$ forbidden-line doublet at $z_{\rm spec}=1.124$, and the $J=2-1$ rotational line of CO detected with CARMA suggests a similar redshift ($z_{\rm spec}=1.126$; M.~Balokovi{\'c} et al., in prep.). Following Smol{\v c}i{\'c} et al. (2012b), we adopt the value $z_{\rm spec}=1.125$ as the redshift of AzTEC2. Koprowski et al. (2014) claimed that the target position of these spectral line observations was $1\farcs4$ away from the SMA 890 $\mu$m position (\cite{younger2007}). They concluded that the SMG lies at a redshift of $3.60_{-0.18}^{+0.13}$ derived from the radio/submm flux-density ratio because the radio source is only $0\farcs4$ away from the SMA position. The redshift we derived from 
the radio/submm flux-density ratio is $z=4.28\pm0.82$. The latter difference emerges because Koprowski et al. (2014) based their calculation 
on the average $z\simeq2-3$ SMG spectral template derived by Micha{\l}owski et al. (2010), while we utilised the Carilli-Yun redshift 
indicator (\cite{carilli1999}, 2000) as described in Sect.~4.2. However, as shown in Fig.~\ref{figure:slits}, the Keck/DEIMOS slit was centred $0\farcs98$ from the SMA position, 
and the spectrum was extracted \textit{from} the SMA peak of the SMG. Moreover, 
the EW-oriented DEIMOS slit did not cover the optically visible foreground galaxy on the southern side of AzTEC2 
(UltraVISTA ID 232116, $z_{\rm phot}=0.34$; cf.~Fig.~B1 in \cite{koprowski2014}). This implies that the redshift of AzTEC2 
is close to unity instead of the higher value proposed by Koprowski et al. (2014).

The spec-$z$ of AzTEC5 was previously reported to be $z_{\rm spec}=3.971$ (\cite{smolcic2012b}). 
The Keck/DEIMOS slit position and orientation are shown in Fig.~\ref{figure:slits}. 
We note that the slit does not include emission from galaxies other than AzTEC5, but 
the corresponding DEIMOS spectrum is of poor quality. Therefore, we adopt 
the photo-$z$ of $3.05_{-0.28}^{+0.33}$ from Smol{\v c}i{\'c} et al. (2012b). 
For comparison, Koprowski et al. (2014) derived a photo-$z$ of $4.19_{-0.10}^{+0.26}$. The redshift we derived from 
the radio/submm flux-density ratio using the Carilli \& Yun (2000) formula (see Sect~4.2), $1.85\pm0.23$, 
is also lower than the value $2.90_{-0.15}^{+0.10}$ calculated by Koprowski et al. (2014).

Koprowski et al. (2014) argued that the spectroscopic redshift of AzTEC6, $z_{\rm spec}=0.802$, is uncertain because 
it is measured towards an optically visible object about $1\arcsec$ from the SMA position (\cite{younger2007}), 
and that the submm/radio flux ratio of AzTEC6 is inconsistent with a low redshift (they derived 
a value of $z=3.86_{-0.92}^{+4.91}$ from the radio/submm flux-density ratio, while we derive 
the value $z>3.52$ because AzTEC6 is \textit{not} detected at 20 cm). The photo-$z$ value derived by 
Smol{\v c}i{\'c} et al. (2012b), $z_{\rm phot}=0.82_{-0.10}^{+0.14}$, is similar to the $z_{\rm spec}$ value, while Koprowski et al. (2014) 
reported a value of $z_{\rm phot}=1.12$ for this object. The above-mentioned optically visible galaxy lies only 
$0\farcs66$ from the SMA position, and $0\farcs12$ away from the Keck/DEIMOS slit centre (see 
Fig.~\ref{figure:slits}). The above spec-$z$ value was derived from a high quality spectrum 
(flag 4; J.~S.~Kartaltepe et al., in prep.) extracted from a position that lies $0\farcs62$ from the SMA position, 
and coincides with the optical galaxy. This implies that the spectral line emission originates in this foreground 
object as suggested by Koprowski et al. (2014). 
A redshift of $z_{\rm spec}=0.802$ indeed conflicts with the non-detection of AzTEC6 at 20 cm, and 
we therefore adopt the redshift $z>3.52$. 

The DEIMOS spec-$z$ of AzTEC9 was reported to be 1.357, and its corresponding photo-$z$ was 
found to be $1.07_{-0.10}^{+0.11}$ (\cite{smolcic2012b}). However, the spec-$z$ value is based on a relatively 
weak spectrum (M.~Salvato et al., in prep.), and is therefore quite uncertain. 
Koprowski et al. (2014) reported that the above redshift values refer 
to an object as far as about $2\farcs8$ from the SMA position (\cite{younger2009}). 
Again, this is not the case, but the DEIMOS spectrum was extracted from the 
SMA position (the slit centre was offset from the SMA peak by $0\farcs45$; see Fig.~\ref{figure:slits}). 
Koprowski et al. (2014) stated that the submm/radio flux-density ratio of AzTEC9 is 
inconsistent with a redshift value close to unity [they derived $z=4.60_{-0.31}^{+0.50}$, 
while our result based on the Carilli \& Yun (2000) redshift formula is $z=2.82\pm0.76$]. 
They also derived a high photo-$z$ of $4.85_{-0.15}^{+0.50}$ for AzTEC9 
(counterpart lying $0\farcs77$ from the SMA position). 
There is a \textit{Spitzer}/IRAC source $0\farcs62$ from the SMA position, 
and the Wardlow et al. (2011) redshift formula gives a redshift of $z \simeq 2.75$, which is similar to 
the redshift we inferred from the radio/submm flux-density ratio, but 
considerably lower than the redshifts derived by Koprowski et al. (2014). In the present study, 
we adopt the photo-$z$ of $1.07_{-0.10}^{+0.11}$ from Smol{\v c}i{\'c} et al. (2012b) because the 
corresponding $\chi_{\rm tot}^2$ distribution exhibits a clear minimum at that value (see Fig.~6 in \cite{smolcic2012b}).

For AzTEC10, the photo-$z$ derived by Smol{\v c}i{\'c} et al. (2012b) is $2.79_{-1.29}^{+1.86}$, 
while Koprowski et al. (2014) determined a photo-$z$ of $5.00_{-0.50}^{+2.00}$ for the 
optical/NIR source about $1\farcs5$ from the SMA position. The most up-to-date COSMOS spectroscopic-redshift 
catalogue gives a likely (quality flag 2) DEIMOS redshift value of $z_{\rm spec}=0.547$ towards AzTEC10 
(only $0\farcs018$ offset from the SMA position; M.~Salvato et al., in prep.). As illustrated in 
Fig.~\ref{figure:slits}, the DEIMOS slit however picked up emission from a foreground galaxy at $z_{\rm phot}\simeq0.51$ 
(ID 302846 in the new UltraVISTA catalogue) that lies $0\farcs97$ NW of AzTEC10. 
Since AzTEC10 is not detected at 1.4 GHz, it appears to lie at a high redshift. 
In the present study, we adopt the photo-$z$ from Smol{\v c}i{\'c} et al. (2012b), but note that 
because of the multiple nearby counterparts of this source (three within $2\arcsec$) it is difficult to obtain accurate 
photometry for AzTEC10.

Both AzTEC13 and AzTEC14-E have neither optical nor IRAC counterparts, and we derived lower 
limits of $z>4.07$ and $z>2.95$
for their radio/submm flux-ratio based redshifts [these differ from the values $z>3.59$ and $z>3.03$ derived by 
Smol{\v c}i{\'c} et al. (2012b) because of the different assumptions that we used here]. 
These lower limits are consistent with the corresponding values of Koprowski et al. (2014), 
i.e. $z=4.70_{-1.04}^{+1.25}$ and $z=3.38_{-0.54}^{+1.00}$, respectively. The updated COSMOS spec-$z$ catalogue 
gives a high quality (flag 4) DEIMOS redshift of $z_{\rm spec}=0.471$ for a target that is $0\farcs015$ away from 
the SMA position of AzTEC13 (M.~Salvato et al., in prep.), where the spectroscopic slit centre 
was positioned $0\farcs60$ away from the SMA peak). 
This redshift is much lower than the other estimates mentioned above. 
However, as shown in Fig.~\ref{figure:slits}, there are two foreground galaxies 
lying $2\farcs22$ SE and $2\farcs45$ SW from AzTEC13 (UltraVISTA IDs 268116 and 268129 with the photo-$z$ values of 
0.49 and 0.45, respectively); these could have contaminated the spectral line measurements, although they do 
not lie within the slit boundaries. A low redshift of AzTEC13 would indeed be inconsistent with the 
radio non-detection (cf.~AzTEC6). For both AzTEC13 and AzTEC14-E, we adopt the redshifts derived from the radio/submm 
flux ratio ($z>4.07$ and $z>2.95$).

\section{Multiwavelength counterparts and redshifts of the SMGs AzTEC16--30} 

Below we describe the multiwavelength appearances of our PdBI SMGs and provide notes of their redshifts. 

\underline{\textit{AzTEC16}}. The $5\sigma$ point-like 1.3~mm emission feature near the edge of the PB FWHM appears to 
have no counterparts at other wavelengths. Altogether four negative sources were found in this field with $\vert {\rm S/N} \vert=4.0-6.2$,  
three of which lie outside the PB. Hence, AzTEC16 could be spurious despite its relatively high significance. 
We note that $4\farcs3$ west of the target field centre, there is a $\sim 3\sigma$ (35.6 $\mu$Jy~beam$^{-1}$) 
VLA 20~cm source, which also appears to be detected at \textit{Spitzer}/IRAC and 24~$\mu$m wavelengths. This source can be identified as 
the galaxy J095950.03+024416.5 from the COSMOS optical/NIR catalogue (\cite{capak2007}); 
its UltraVISTA DR1 photometric-redshift catalogue ID is 319194 ($z\simeq1.62$; \cite{ilbert2013}).
The radio non-detection of AzTEC16 suggests a lower limit of $z>2.42$ to its redshift. In Table~1 of Smol{\v c}i{\'c} et al. (2012b), 
the source called AzTEC16 at a spectroscopic redshift of 1.505 (based on high-resolution CO observations with CARMA; K.~Sheth et al., in prep.) 
corresponds to AzTEC42 in our nomenclature (cf.~\cite{scott2008}; Table~1 therein).

\underline{\textit{AzTEC17}.} The 1.3 mm source AzTEC17a (lying $4\farcs3$ SW of the pointing centre, and 
detected at a significance level of $6.2\sigma$) is clearly associated with the 20 cm source 
COSMOSVLADP$_{-}$J095939.19+023403.6 ($S_{\rm 20\, cm}=68\pm13$~$\mu$Jy), 
and a VLA 10 cm source (37 $\mu$Jy~beam$^{-1}$ or $8.2\sigma$). The source also shows 
\textit{Spitzer}/MIPS and IRAC emission. There is a \textit{Herschel} 250 $\mu$m source $3\farcs05$ east of
AzTEC17a [ID 1753 in the COSMOS SPIRE 250 $\mu$m Photo\-metry Catalogue from HerMES (\cite{oliver2012})]. 
The search for \textit{Herschel} counterparts was performed by using a search radius of $9\farcs1$, 
i.e. half the SPIRE beam FWHM at $\lambda=250$~$\mu$m. AzTEC17a has an optical-NIR counterpart about $1\farcs4$ SW of 
the 1.3~mm emission peak [ID 1475165 in the COSMOS photometry catalogue (\cite{capak2007})]. We note that the source visible 
in the Ultra\-VISTA and ACS $I$-band images, lying $1\farcs6$ NW in projection from AzTEC17a, 
is the galaxy COSMOS J095939.12+023405.1 (\cite{capak2007}), which has a photometric redshift of $z=0.793$ (the source 271694 in 
the DR1 UltraVISTA photometric-redshift catalogue; \cite{ilbert2013}). 
Cross-correlation with the COSMOS photometry catalogue yielded a candidate optical counterpart for 
AzTEC17b (ID 1475223), about $1\farcs2$ SW of the PdBI emission peak. 
However, AzTEC17b has no counterparts at UltraVISTA bands or at MIR or cm wavelengths. There are two negative sources 
($-4.5\sigma$ and $-6.2\sigma$) within $\lesssim7\arcsec$ of the phase centre. As AzTEC17a is a confirmed SMG, 
this could mean that the $4.5\sigma$ source AzTEC17b is spurious.

For AzTEC17a, the primary photo-$z$ solution is $z=0.75_{-0.12}^{+0.23}$. Because 
the source of this photo-$z$ lies about $\sim 1\farcs4$ from the PdBI position, it is questionable whether 
it is related to AzTEC17a (although it is within the synthesised beam). 
The primary photo-$z$ value is however comparable to the very secure (quality flag 3) spectroscopic redshift 
$z_{\rm spec}=0.834$ measured towards AzTEC17a ($0\farcs26$ offset from the PdBI position) with Keck/DEIMOS 
(M.~Salvato et al., in prep.). We note that for the measurements the slit was centred 
on a position $1\farcs60$ away from AzTEC17a (see Appendix~E), and that there are no 
\textit{HST}/ACS $I$-band sources within the slit boundaries. For comparison, the angular offset between 
the 1.4~GHz radio source and the PdBI detection peak is only $0\farcs22$, and the redshift derived 
from the radio/submm flux density ratio is $z=2.29\pm0.42$. A comparable redshift of $z \simeq 2.77$ was derived from  
the Wardlow et al. [2011; their Eq.~(1)] redshift estimator based on the \textit{Spitzer}/IRAC 3.6~$\mu$m and 8~$\mu$m flux densities. 
The last two values agree with a shallow ``bowl'' in the $\chi_{\rm tot}^2$ distribution at $z\simeq2.7$, 
but the $z_{\rm spec}$ value is adopted in the present work. For AzTEC17b we derive a photo-$z$ of
$4.14_{-1.73}^{+0.87}$. 
There is a dip in the $\chi_{\rm tot}^2$ distribution also at $z\simeq0.4$. However, the 
1.4~GHz non-detection towards AzTEC17b results in a lower limit to its redshift of $z>2.49$, which is consistent with 
the above photo-$z$ value of $4.14_{-1.73}^{+0.87}$.

\underline{\textit{AzTEC18}.} The 1.3 mm $4.5\sigma$ source seen towards AzTEC18 has
counterparts at optical and NIR wavelengths ($0\farcs82$ from the PdBI position), and is therefore considered a potential SMG. 
This is further supported by the fact that only one negative source, being of $-4.2\sigma$ significance, was detected in this field.

AzTEC18 has a photo-$z$ solution of $z_{\rm phot}=3.00_{-0.17}^{+0.19}$, which is consistent 
with the redshift derived from the radio/submm flux-density ratio of $z>2.20$.

\underline{\textit{AzTEC19}.} AzTEC19a lies $3\farcs1$ NE of the 
SCUBA-2 source SMMJ100028.6+023201 (or 450.00 or 850.07) identified by Casey et al. (2013). This angular 
offset is within the JCMT/SCUBA-2 beam size (FWHM) of $\sim 7 \arcsec$ at 450~$\mu$m. 
With the deboosted flux densities of $S_{\rm 450\, \mu m}=37.54\pm6.58$~mJy and 
$S_{\rm 850\, \mu m}=9.21\pm1.45$~mJy, this was the 
strongest 450~$\mu$m source found by Casey et al. (2013) in the COSMOS field. 
These authors identified two possible optical counterparts to SMMJ100028.6+023201 (see their Table~6), 
but our higher-resolution observations show that only one of them -- lying $1\farcs03$ west of 
the PdBI peak position -- can be taken as a candidate counterpart 
(the other source lies $2\farcs65$ SW of the PdBI peak). 
AzTEC19a was also detected by \textit{Herschel}. In the COSMOS SPIRE 250 $\mu$m Photometry Catalogue, the
source ID is 2277 ($0\farcs43$ offset). AzTEC19a is associated with both a 20 and 10 cm radio\textbf{-}continuum source. 
In the VLA Deep Catalogue, the corresponding source has the name COSMOSVLADP$_{-}$J100028.70+023203.7 
($S_{\rm 20\, cm}=78\pm12$~$\mu$Jy). The 10 cm peak flux density is $S_{\rm 10\, cm}=28.8$~$\mu$Jy~beam$^{-1}$, 
making it a $6.4\sigma$ detection. The source is also associated with \textit{Spitzer} IR emission. 
The $9.7\sigma$ 1.3 mm source AzTEC19b lying at the border of the PdBI PB has some 
\textit{Spitzer}/IRAC 3.6 $\mu$m and lower wavelength emission 
just north of it. This emission can be associated with the galaxy COSMOS J100029.24+023211.5 (\cite{capak2007}), which 
has a photo-$z$ of about 1.27 (source 262768; \cite{ilbert2013}). As it lies $1\farcs67$ NW of the 1.3 
mm peak, it is probably unrelated to AzTEC19b. However, there is also a NIR source within 
the $3\sigma$ contour of 1.3~mm emission, about $0\farcs5$ from the mm peak. In the COSMOS 
ACS $I$-band photometry catalogue (\cite{leauthaud2007}), the ID of this source is 1486, 
while in the DR1 Ultra\-VISTA catalogue its ID is 262766, and its reported photo-$z$ value is about 1.30 (\cite{ilbert2013}). 
Three negative sources ($-4.8\sigma$, $-6.6\sigma$, $-9.5\sigma$) were detected in the AzTEC19 field. However, only one of them 
($-6.6\sigma$) lies within the PB FWHM ($9\farcs9$ from the phase centre), while the remaining two lie outside the PB 
($13\arcsec-13\farcs8$ away from the phase centre). Because AzTEC19a is confirmed, and AzTEC19b 
has a high 1.3 mm detection S/N of 9.7 and is associated with multiwavelength emission, we are not expecting to have any spurious sources in 
this field.

The optical/IR counterpart of AzTEC19a (CFHT $i^*-K=1.49$), located only $0\farcs2$ 
from the PdBI position, has a photo-$z$ solution of $z=3.20_{-0.45}^{+0.18}$. 
There is a spec-$z$ value of 1.048 measured for a source only $0\farcs195$ 
from the PdBI position of AzTEC19a with the VLT Visible Multi-Object Spectrograph (VIMOS) in the zCOSMOS project. 
However, the corresponding quality flag is 1.1, meaning that the $z_{\rm spec}$ value is 
insecure ($<25\%$ reliability). For the 1.4~GHz source -- situated $0\farcs53$ from the PdBI position -- 
we derived a radio/submm-based redshift of $4.22\pm0.91$, which is comparable within the errors 
with our photo-$z$ solution. For comparison, Casey et al. (2013) derived a photometric redshift of 
$z=2.86_{-0.26}^{+0.21}$ for their SMG source SMMJ100028.6+023201, which is associated with AzTEC19a. 
A comparable value of $z\sim2.3$ can be derived from the \textit{Spitzer} 
IRAC/MIPS flux densities [\cite{pope2006}; their Eq.~(2)], and the value $z \simeq 2.90$ is obtained when using the 
IRAC 3.6~$\mu$m and 8~$\mu$m flux densities (\cite{wardlow2011}). In the present paper, we adopt 
our photo-$z$ solution of $z=3.20_{-0.45}^{+0.18}$ for AzTEC19a. For AzTEC19b, the photo-$z$ solution of $z_{\rm phot}=1.11\pm0.10$ 
is adopted as the redshift of the aforementioned SMG, while the 1.4~GHz non-detection suggests a very high radio/submm-based redshift 
of $z>6.57$, where no SMGs have been discovered to date. There is, however, a degeneracy between the 
dust temperature of the source and its redshift (both affecting the source SED), and the radio dimness of 
AzTEC19b could in principle be the result of a low dust temperature (e.g. \cite{blain2002}; \cite{kovacs2006}).

\underline{\textit{AzTEC20}.} Interestingly, the $6\sigma$ source about $5\farcs8$ north of 
the phase centre shows no emission at other wavelengths. The $\sim 3\sigma$ features of 
1.3~mm emission seen near the phase centre appear instead to be associated with a source seen at several different wavelengths 
from 20~cm to NIR. The Ultra\-VISTA catalogue ID of the latter source is 306331 ($z \simeq1.98$; \cite{ilbert2013}).
The 20~cm source has a peak flux density of 42.3~$\mu$Jy~beam$^{-1}$, hence has a S/N ratio of about 
3.5. We note that the VLA Deep catalogue contains sources down to $4\sigma$ or about 
48~$\mu$Jy~beam$^{-1}$ (\cite{schinnerer2010}). About $5\farcs7$ west of this source, 
a slightly stronger 20~cm source candidate (45~$\mu$Jy~beam$^{-1}$ or $\sim 3.8\sigma$) can be seen. 
The 10~cm source near the PdBI phase-tracking centre is a $\sim 6.2\sigma$ detection 
(28.1~$\mu$Jy~beam$^{-1}$). The \textit{Spitzer}/MIPS 24~$\mu$m emission is quite extended, but is 
clearly resolved into two sources in the IRAC 3.6~$\mu$m image and yet more sources in 
the UltraVISTA NIR images. The \textit{Herschel}/HerMES/SPIRE 250~$\mu$m catalogue (\cite{oliver2012}) 
contains a source (ID 3076) near the phase centre, $7\farcs2$ from our PdBI source. There is one negative source of $-4.8\sigma$ 
significance in this field, located on the SE side of our 1.3 mm source. Although the positive source is more significant ($6\sigma$) compared 
to the negative feature, and it fulfils our detection criterion of ${\rm S/N}>4.5$, it has no optical-to-IR counterparts that could confidently 
confirm that it is real. The lack of radio emission from AzTEC20 yields a lower redshift limit of $z>2.35$.

\underline{\textit{AzTEC21}.} The south-western clump of the detected filamentary structure, AzTEC21a, 
is associated with a 20 cm source of peak flux density of 63 $\mu$Jy~beam$^{-1}$ ($\sim 3.9\sigma$). 
This also coincides with the position of a \textit{Spitzer} IR source. AzTEC21b is probably part of the 
same structure (see below). The bright galaxy lying $1\farcs4$ SE of AzTEC21c
is COSMOS J100002.93+024639.9 ($V=20.681$; \cite{capak2007}). The ID of this galaxy in the 
DR1 UltraVISTA catalogue is 327783, and its photo-$z$ is about 0.34 (\cite{ilbert2013}). We note that the source 
ID in the zCOSMOS catalogue is 846495, but its spectroscopic redshift measurement could not be attempted 
(confidence class 0; \cite{lilly2007}, 2009).

For AzTEC21a, the photo-$z$ solution is $z=2.60_{-0.17}^{+0.18}$ 
(for a source with the CFHT colour $i^*-K=2.39$). The optical/NIR source lies only $0\farcs13$ away (NE) from 
the PdBI source according to the previous COSMOS/UltraVISTA catalogue (ID 1688587), but in the most recent UltraVISTA-TERAPIX 
DR the nearest source (ID 328878) lies $1\farcs1$ NE, 
i.e. 8.5 times further away from our source, making the reliability of the proposed counterpart questionable. 
The 1.4~GHz non-detection (only $3.9\sigma$) suggests a high redshift of $z>3.45$. 
On the other hand, the \textit{Spitzer} photometric redshift is estimated to be about 1.5 (\cite{pope2006}), 
and the value $z \simeq 2$ is derived using the Wardlow et al. (2011) IRAC flux-density method. 
The last value is comparable to our photo-$z$ of $z=2.60_{-0.17}^{+0.18}$.
For AzTEC21b, we derived a redshift of $ z_{\rm phot}=2.80_{-0.16}^{+0.14}$ 
(source $0\farcs45$ from the PdBI peak; CFHT/$i^*-K=1.74$), while the radio/submm based value is $z>2.47$, consistent with our photo-$z$ 
solution. The most recent UltraVISTA DR does not contain a nearby ($<2\farcs4$) counterpart to AzTEC21b. 
We note that the overall 1.3 mm emission morphology (Fig.~\ref{figure:stamps}) could indicate a relation 
between AzTEC21a and 21b, and this is further supported by their comparable photo-$z$ values of $z=2.60_{-0.17}^{+0.18}$ and 
$z=2.80_{-0.16}^{+0.14}$ (i.e. their redshifts could be identical). For AzTEC21c, we derived a lower limit of $z>1.93$ from the upper limit to 
the 1.4 GHz flux density. 

\underline{\textit{AzTEC22}.} The candidate point-like $5.1\sigma$ PdBI source at the southern edge 
of the field has no counterparts at other wavelengths. The detection of two negative features with $\vert {\rm S/N} \vert=5.7$ and 5.9 in this field 
provides a hint that the 1.3 mm detection is spurious. A submm source was detected with \textit{Herschel} 
towards our phase centre (250 $\mu$m ID 5470), about $8\arcsec$ north from the above mentioned 
PdBI feature. Moreover, about $2\farcs5$ from our pointing centre, there is the 20 cm source COSMOSVLADPP$_{-}$J095950.57+022827.5 
($S_{\rm 20\, cm}=124\pm12$~$\mu$Jy), which is also seen in the VLA 10~cm image (58.6~$\mu$Jy~beam$^{-1}$ or $\sim 13\sigma$). 
The \textit{Spitzer} IR images show a ``double source'', and a stronger 24 $\mu$m emitter is associated with 
the above-mentioned radio-continuum source J095950.57+022827.5. A trace of 20 cm emission (43.8~$\mu$Jy~beam$^{-1}$ peak surface brightness) can also be seen 
towards the position of the weaker 24 $\mu$m source. The two \textit{Spitzer} sources can be seen 
in the UltraVISTA NIR images: the NW one has the ID 244762 ($z_{\rm phot}\simeq1.8-1.9$), 
while the SE source is 244405 at $z_{\rm phot}\simeq2$ (\cite{ilbert2013}). For our PdBI detection the lack of a radio counterpart suggests a redshift of $z>3.0$.

\underline{\textit{AzTEC23}.} The 1.3 mm feature seen towards this source ($5\farcs4$ NW of the AzTEC centroid) has counterparts at optical-NIR 
wavelengths ($1\farcs24$ from the PdBI position). Cross-correlation with the \textit{Herschel}/SPIRE 250~$\mu$m catalogue 
shows the presence of a source (ID 2659) about $3\arcsec$ NW from the 1.3 mm feature.
A visual inspection of the VLA 20~cm image reveals an EW-oriented, elongated emission feature, 
whose western emission peak ($2\farcs48$ from the 1.3 mm source) has a peak surface brightness of 
43.5 $\mu$Jy~beam$^{-1}$ and the eastern peak has the surface brightness of 39.8 $\mu$Jy~beam$^{-1}$. 

We derived a primary photo-$z$ of $z=1.60_{-0.50}^{+0.28}$ with a secondary solution at $z\simeq 4.3$, while the radio non-detection 
implies the lower limit $z>2.06$. We adopt the redshift $z=1.60_{-0.50}^{+0.28}$ but note that the $\chi_{\rm tot}^2$ distribution of 
AzTEC23 is quite complex.

\underline{\textit{AzTEC24}}. This source is called AzTEC/C48 in the ASTE/AzTEC 1.1 mm catalogue of Aretxaga et al. (2011). 
The ASTE/AzTEC peak position lies $8\farcs8$ NE of the JCMT/AzTEC centroid. We have found three candidate PdBI sources 
of $4.9-5.1\sigma$ significance (outside/at the border of the PB), but none of them have counterparts at other wavelengths. 
Altogether five negative sources with $\vert {\rm S/N} \vert=4.3-5.9$ were detected in the field, and two of them lie outside the 
PB FWHM. Hence, the identified positive 1.3 mm sources might be spurious and should be treated with caution. 
About $5\farcs5$ SW from AzTEC24b there is a \textit{Herschel} submm source (250 $\mu$m ID 4991).
Moreover, $2\farcs5$ SW from AzTEC24b there is the VLA 20 cm source COSMOSVLADP$_{-}$J100039.28+023845.3 ($S_{\rm 20\, cm}=63\pm13$~$\mu$Jy). 
Aretxaga et al. (2011) associated this source with the radio counterpart of their 
source AzTEC/C48 ($2\farcs7$ separation). In the VLA 10 cm image, a source with a peak flux density of 
28.5~$\mu$Jy~beam$^{-1}$ ($6.3\sigma$) can be seen at the 20~cm source position. 
The cm radio-continuum source is also associated with \textit{Spitzer} IR emission. 
When referring to the COSMOS catalogue of Capak et al. (2007), the source can be identified as the galaxy COSMOS 
J100039.29+023845.4, and in the UltraVISTA catalogue the source ID is 293896 
($z_{\rm phot}\simeq 2.1$; \cite{ilbert2013}). For the three radio non-detected components AzTEC24a, 24b, and 24c we derived lower redshift limits of $z>2.35$, $z>2.28$, $z>3.17$, respectively. 

\underline{\textit{AzTEC25}.} In this field, none of the 1.3 mm point-like features fulfilled our detection criteria. 
We note that $1\farcs37$ of the $4.2\sigma$ feature (the most south-western feature shown in Fig.~\ref{figure:stamps}) 
there is the source 1925434 from the COSMOS+UltraVISTA catalogue, but its CFHT/$i^*$-band magnitude ($24.27\pm0.49$ mag) with 
respect to the CFHT/$K$-band and UltraVISTA $K_{\rm s}$-band magnitudes ($23.56\pm0.20$ mag and $24.30\pm0.23$ mag, respectively) suggests a very blue 
colour index, hence it is unlikely related to the 1.3 mm feature.\footnote{If the dust column density in the SMG is high, it is expected to appear reddened. 
However, if the spatial distribution of dust is inhomogeneous, the unobscured star formation could cause the galaxy to appear blue. 
Moreover, the viewing angle can contribute to the observed effective colour of the galaxy (see e.g. \cite{wang2013}).} 
There are also two negative sources ($-4.3\sigma$ and $5.0\sigma$) in the field, 
inside the PB FWHM. In the VLA 20 cm map, a source candidate with a peak surface brightness of 58.7 $\mu$Jy~beam$^{-1}$ (only $\sim3.3\sigma$) 
is visible close to our pointing centre ($1\arcsec$ away). However, no shorter-wavelength emission can be seen towards this 
source.

\underline{\textit{AzTEC26}.} Two 1.3~mm point sources were detected around the phase centre of this source, 
and the source we called AzTEC26a has an optical-to-NIR counterpart ($0\farcs94$ from the 1.3~mm peak). 
The weaker, $4.8\sigma$ source AzTEC26b has no multiwavelength counterparts. The VLA 20~cm image detects some 
emission near both sources, but the S/N ratio of the radio detection is less than 2. 

For AzTEC26a, we derived a photo-$z$ value of $z=2.50_{-0.14}^{+0.24}$. 
Although the Subaru/$i^+$-CFHT/$K$ colour of this source, 1.06, is relatively blue, the radio/submm-based 
redshift of $z>1.87$ supports the aforementioned $z_{\rm phot}$ solution. The radio non-detection of AzTEC26b 
gives a lower redshift limit of $z>1.79$. 

\underline{\textit{AzTEC27}.} There is a hint of 20~cm emission associated with the elongated 1.3~mm source, although its
20~cm peak surface brightness is only 32.1~$\mu$Jy~beam$^{-1}$ or about $2.5\sigma$. No other wavelength counterparts 
are detected. At the projected distance of $5\farcs2$ NE of the 1.3~mm peak position, there is the bright ($V=20.359$) 
galaxy COSMOS J100039.47+024055.5 (\cite{capak2007}). The Ultra\-VISTA catalogue gives a photo-$z$ of about 0.25 for 
this galaxy (ID 303584; \cite{ilbert2013}). 
About $2\farcs9$ SW of the 1.3~mm peak position, there is another galaxy, namely COSMOS J100039.07+024050.2 (\cite{capak2007}), whose
photometric redshift is about unity (source 303782 in the Ultra\-VISTA catalogue; \cite{ilbert2013}). 
As described in Appendix~D, AzTEC27 appears to be subject to gravitational lensing by the foreground galaxies J100039.47 
and J100039.07, and our lens model suggests that their combined lens effect amplifies the $\lambda_{\rm obs}=1.3$~mm flux density 
by a factor of $\sim2$. At the distance of $4\farcs5$ north from AzTEC27, there is a \textit{Herschel}/HerMES 250~$\mu$m source 
(ID 1500; \cite{oliver2012}). Based on its radio dimness, we derived a redshift of $z>4.17$ for AzTEC27, which makes this source potentially 
the highest-redshift SMG among AzTEC16--30.

\underline{\textit{AzTEC28}.} This clearly detected 1.3 mm source, $3\farcs3$ SE of the phase centre, 
has no counterparts at other wavelengths shown in Fig.~\ref{figure:stamps}. 
Casey et al. (2013) detected AzTEC28 with SCUBA-2 (their source SMMJ100004.5+023042 or 450.20), with the 450~$\mu$m 
peak lying about 
$5\farcs2$ NW of the PdBI peak position. The deboosted flux density at 450 $\mu$m was reported to be $19.11\pm5.91$~mJy. 
The optical counterpart (at $z_{\rm phot}=0.76_{-0.03}^{+0.04}$) reported by Casey et al. (2013) lies $4\farcs67$ north of 
the PdBI 1.3 mm peak, hence is unrelated to the SMG. We also note that the ASTE-detected source AzTEC/C150 from 
Aretxaga et al. (2011) lies $8\farcs44$ NE of our PdBI 1.3 mm source -- still within the $34\arcsec$ beam of ASTE/AzTEC at 1.1~mm. 
About $6\farcs6$ NE of AzTEC28, there is the \textit{Herschel} 250~$\mu$m source 4388 from the HerMES survey (\cite{oliver2012}).

\underline{\textit{AzTEC29}.} There are no clear signatures of PdBI 1.3~mm emission inside the PB.
The source candidate AzTEC29a ($4.7\sigma$) lies at the border of the PB and the strong source AzTEC29b 
($7.3\sigma$) at the northern edge of the map, is outside the PB. The latter could be associated ($0\farcs76$ offset) 
with the source 1685295 from the COSMOS+UltraVISTA catalogue. Two negative sources of $-4.3\sigma$ and $-5.4\sigma$ were detected 
in the field outside the PB FWHM. Hence, AzTEC29a, which does not show up at other wavelengths, could be spurious.
The VLA 20~cm image of the source region detects emission in its 
south-western corner (peak surface brightness of 50.5~$\mu$Jy~beam$^{-1}$ or $\sim3.4\sigma$). 
The \textit{Spitzer}/MIPS 24~$\mu$m emission near the field centre can be associated with the galaxy COSMOS 
J100026.79+023749.4 (\cite{capak2007}) or 289240 in the UltraVISTA catalogue ($z_{\rm phot}\simeq0.58$; \cite{ilbert2013}).

For AzTEC29a, we derived a radio/submm-based redshift of $z>2.96$. 
The photo-$z$ of $1.45_{-0.38}^{+0.79}$ derived for AzTEC29b (Subaru/$i^+$-CFHT/$K=2.67$) is much lower than 
the unrealistically high value $z>7.25$ derived from the radio/submm flux-density ratio, hence the photo-$z$ value is adopted.

\underline{\textit{AzTEC30}.} A $4.6\sigma$ candidate point source with no counterparts is detected in this field. 
There is a \textit{Spitzer}/MIPS 24~$\mu$m source in the NW part of the field, $\sim4\farcs5$ SE of our tentative source, that is
also visible at the IRAC wavelengths but has no catalogue identification. The radio non-detection of AzTEC30 gives a redshift of $z>2.51$.

Finally, we note that none of our sources are detected in X-rays, implying that none host 
a prominent AGN. Cross-correlation with the \textit{Chandra}-COSMOS Bright Source Catalogue 
v2.1 and COSMOS \textit{XMM} Point-like Source Catalogue v2.0 revealed that the nearest 
X-source to any of our SMGs is XMMU J100002.8+024635, lying $9\farcs96$ south of the PdBI phase centre 
towards AzTEC21. However, it is possible that some of the studied SMGs host an extremely Compton-thick AGN 
(i.e. with obscuration due to high column densities of dust) that remains undetected in the existing X-ray images.

\section{Gravitational lens modelling of AzTEC27} 

\begin{figure}[!h]
\centering
\resizebox{\hsize}{!}{\includegraphics{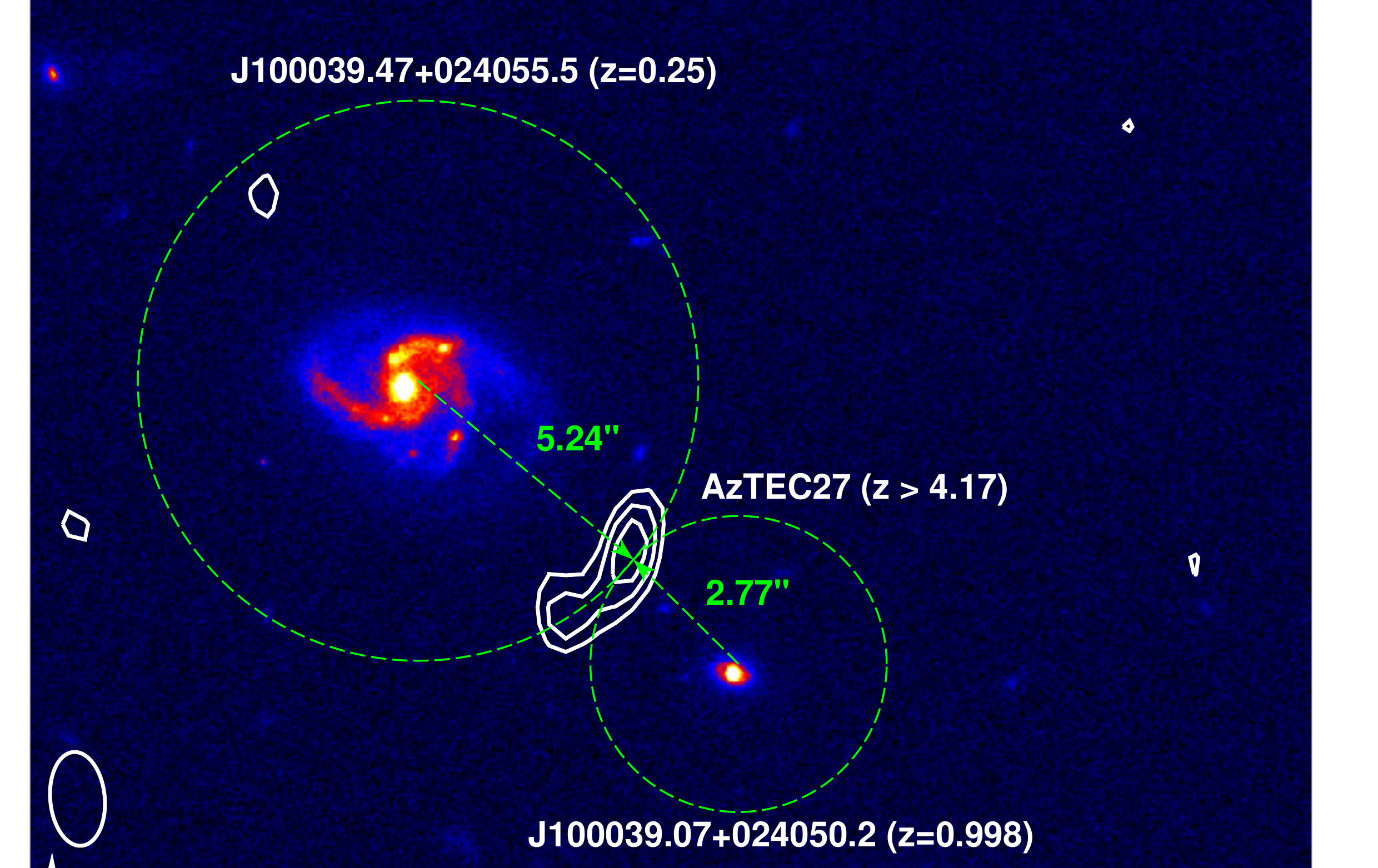}}
\caption{\textit{HST}/ACS $I$-band ($\lambda_{\rm central}=8\,333~\AA$) image towards AzTEC27 (linear colour scale), 
overlaid with white contours showing the PdBI 1.3~mm emission. The positive contour levels are as in Fig.~\ref{figure:pdbi}. 
The green dashed circles of radius $5\farcs24$ (20.31 physical kpc at $z=0.25$) and $2\farcs77$ 
(22.26 physical kpc at $z=0.998$) are centred on the galaxies COSMOS J100039.47+024055.5 and J100039.07+024050.2, 
respectively. The synthesised PdBI beam ($1\farcs76 \times 1\farcs02$, P.A. $5\fdg19$) is shown in the bottom left corner.}
\label{figure:lens}
\end{figure}

There are two foreground galaxies seen in projection close to the SMG AzTEC27 (see Fig.~\ref{figure:lens}).  
The north-eastern galaxy, at $(\Delta \alpha,\, \Delta \delta)=(+4\farcs02,\, +3\farcs36)$ 
from the 1.3 mm peak position of AzTEC27, is COSMOS J100039.47+024055.5. The photometric 
redshift and stellar mass of J100039.47 are $z_{\rm phot}\simeq0.25$ and $\log(M_{\star}/{\rm M_{\sun}})=10.084$ 
(\cite{ilbert2013}). On the south-western side, at $(\Delta \alpha,\, \Delta \delta)=(-1\farcs98,\, -1\farcs94)$ 
from AzTEC27, the foreground galaxy is COSMOS J100039.07+024050.2 at $z_{\rm phot}=0.998$ with a stellar mass 
of $\log(M_{\star}/{\rm M_{\sun}})=10.713$. AzTEC27 is potentially subject to gravitational lensing by 
these two intervening galaxies, and therefore to better understand its intrinsic physical properties requires a lens model.

To estimate the strength of the lensing effect for AzTEC27, we carried out a gravitational lens modelling using the publicly 
available {\tt Python} software called {\tt uvmcmcfit}\footnote{{\tt https://github.com/sbussmann/uvmcmcfit}} that will 
be described in detail by R.~S.~Bussmann et al. (in prep.). The code is a modified version of the one used in the papers 
by Bussmann et al. (2012, 2013) and uses the visibilities to determine the goodness of fit. To sample the posterior probability 
density function of our model parameters, we used the MCMC sampling code {\tt emcee} (\cite{foreman2013}). 

The lensed background source in the model is assumed to be an elliptical Gaussian source, described by the following parameters: 
the total intrinsic (unlensed) flux density ($S_{\rm in}$), the effective radius ($R_{\rm eff}=\sqrt{a\times b}$, 
where $a$ and $b$ are the semi-major and semi-minor axes), the projected angular offset from the model image centroid, 
the axial ratio ($b/a$), and the position angle (P.A. measured E of N). The lens is assumed to be a singular isothermal ellipsoid 
(SIE), parameterised by the angular Einstein radius ($\theta_{\rm E}$), the angular offset from the model image centroid, 
the axial ratio, and the P.A. The magnification factor is then computed as $\mu=S_{\rm out}/S_{\rm in}$, where $S_{\rm out}$ is the 
source's total lensed flux density in the best-fit model. 

The modelling was performed assuming three different scenarios: \textit{i)} J100039.07 at $z_{\rm phot}=0.998$ is acting as 
a lens; \textit{ii)} J100039.47 at $z_{\rm phot}\simeq0.25$ is responsible for lensing; and \textit{iii)} both the above 
intervening galaxies act as lenses. The magnification factor in these three cases was found to be $\mu=1.36\pm0.11$, $1.17\pm0.04$, 
and $2.04\pm0.16$, respectively. Both the galaxies J100039.07 and J100039.47 are therefore causing
only a weak lensing effect, the former one, having a $\sim4.3$ times higher stellar mass and lying at a higher redshift (i.e. closer 
to AzTEC27) than the latter, being the slightly more stronger lens. The weak lensing is consistent with the fact that we see only a 
single image of AzTEC27 in the PdBI 1.3 mm map. In the present paper we assume the two-lens system and adopt the 
value $\mu=2.04\pm0.16$ for AzTEC27.   

\section{The DEIMOS spectrograph slit parameters}

\begin{figure*}
\begin{center}
\includegraphics[width=0.3\textwidth]{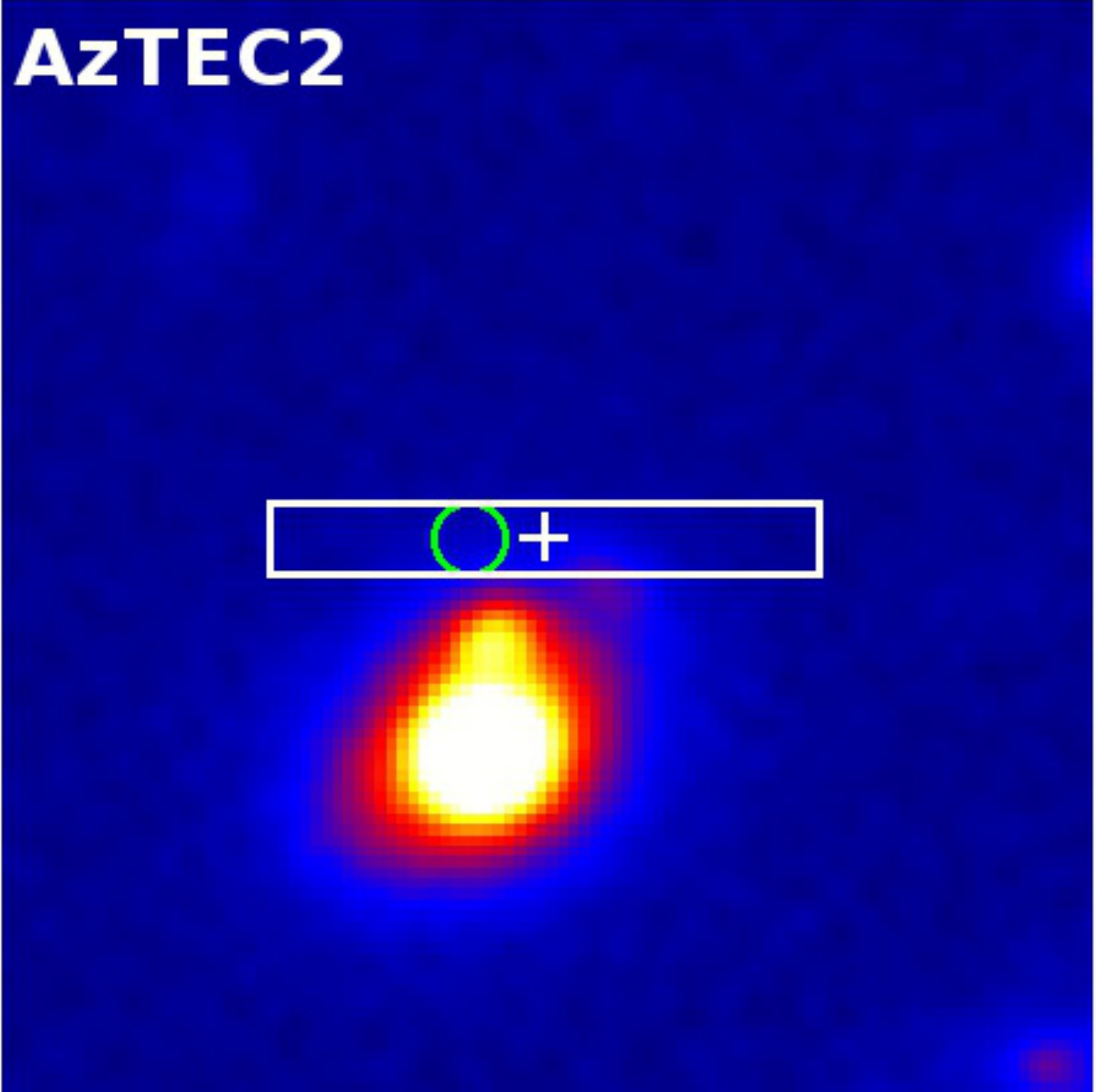}
\includegraphics[width=0.3\textwidth]{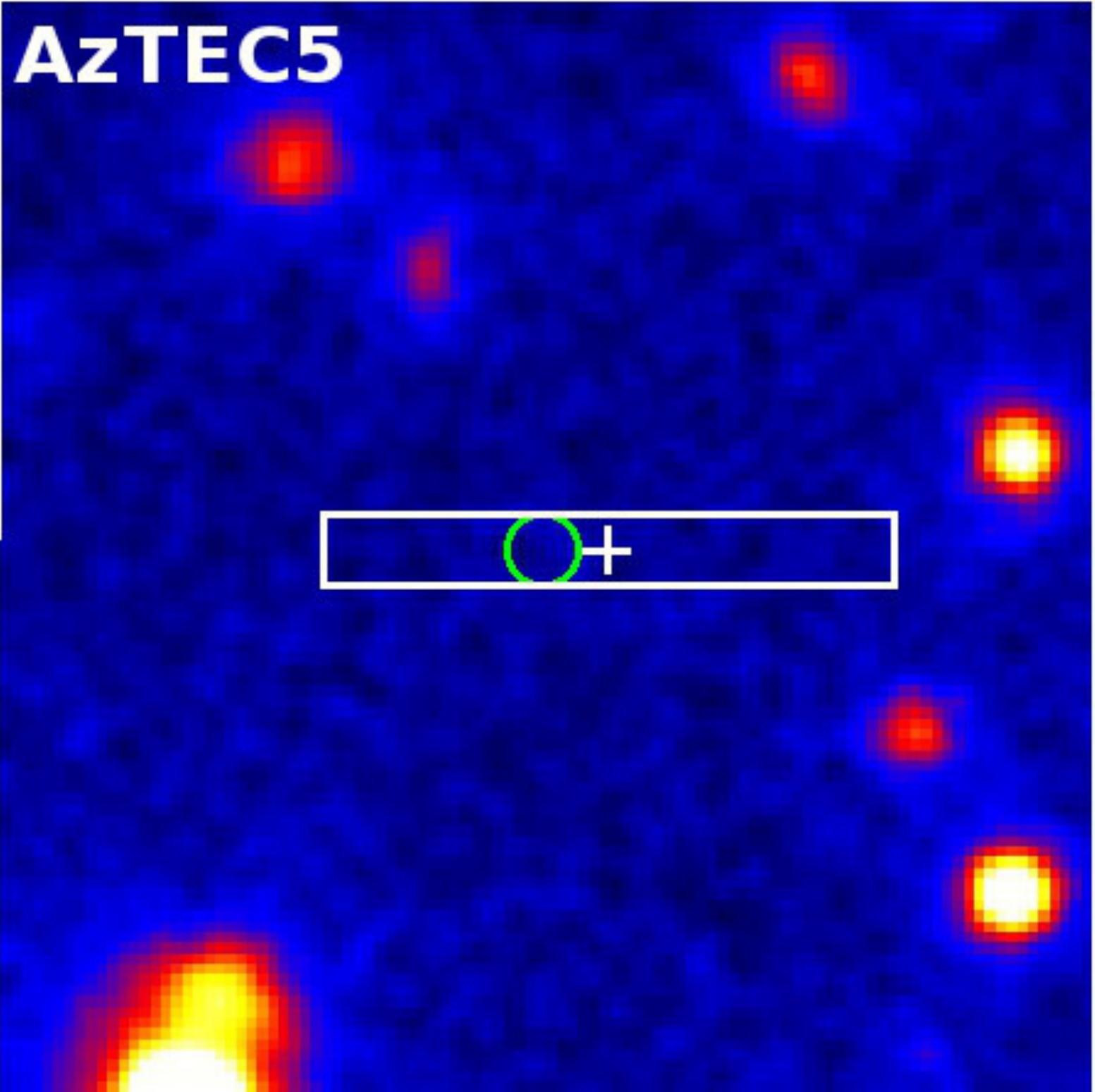}
\includegraphics[width=0.3\textwidth]{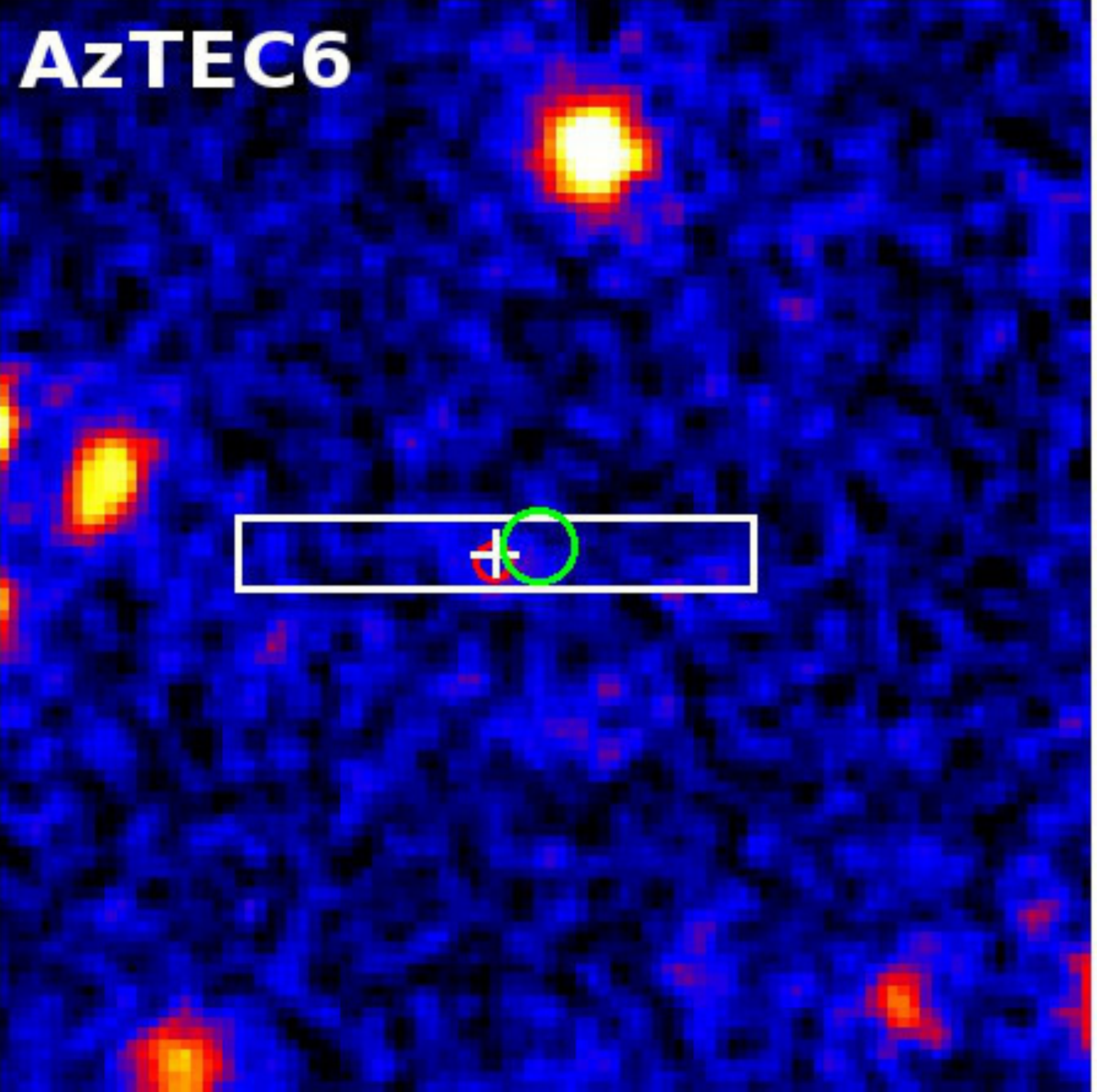}
\includegraphics[width=0.3\textwidth]{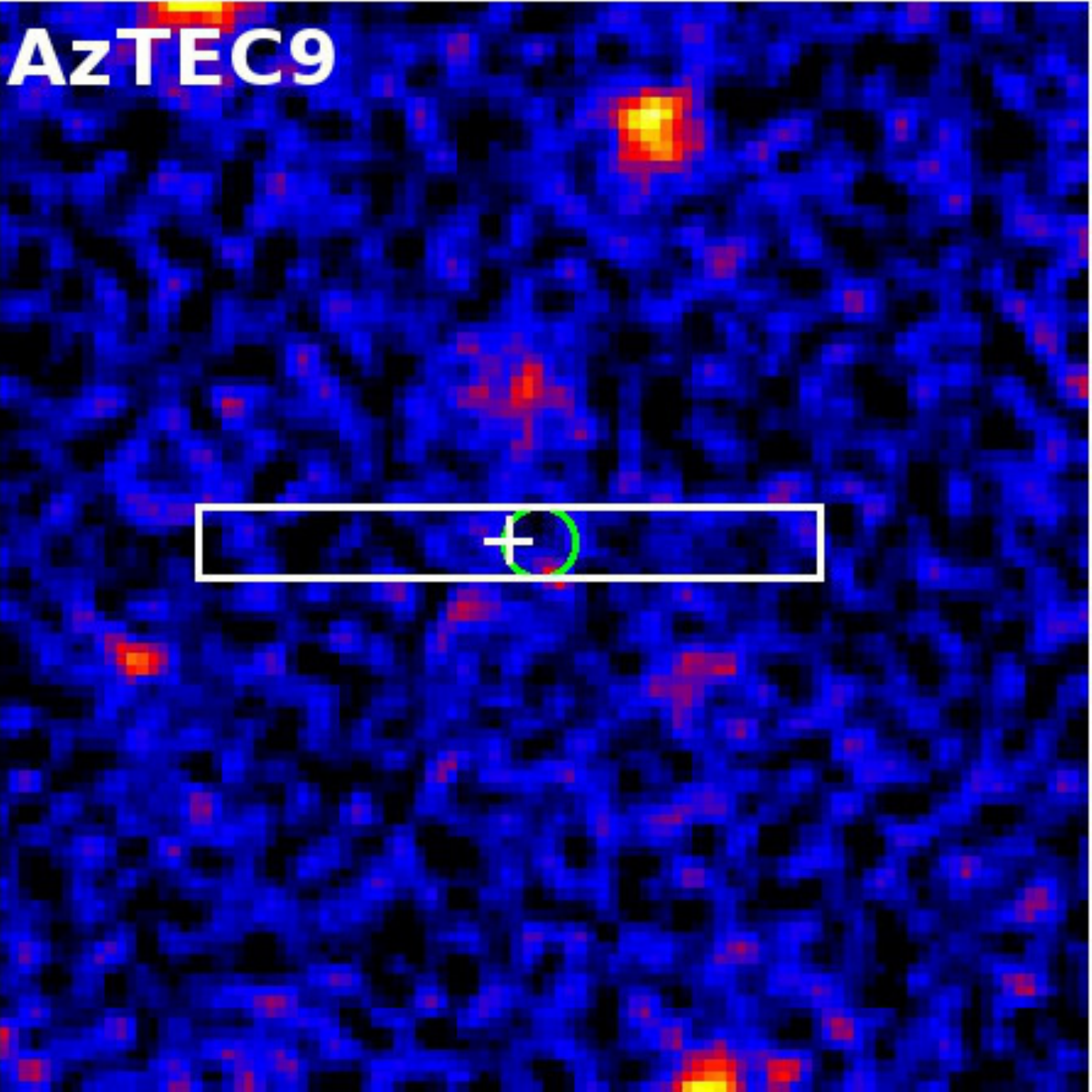}
\includegraphics[width=0.3\textwidth]{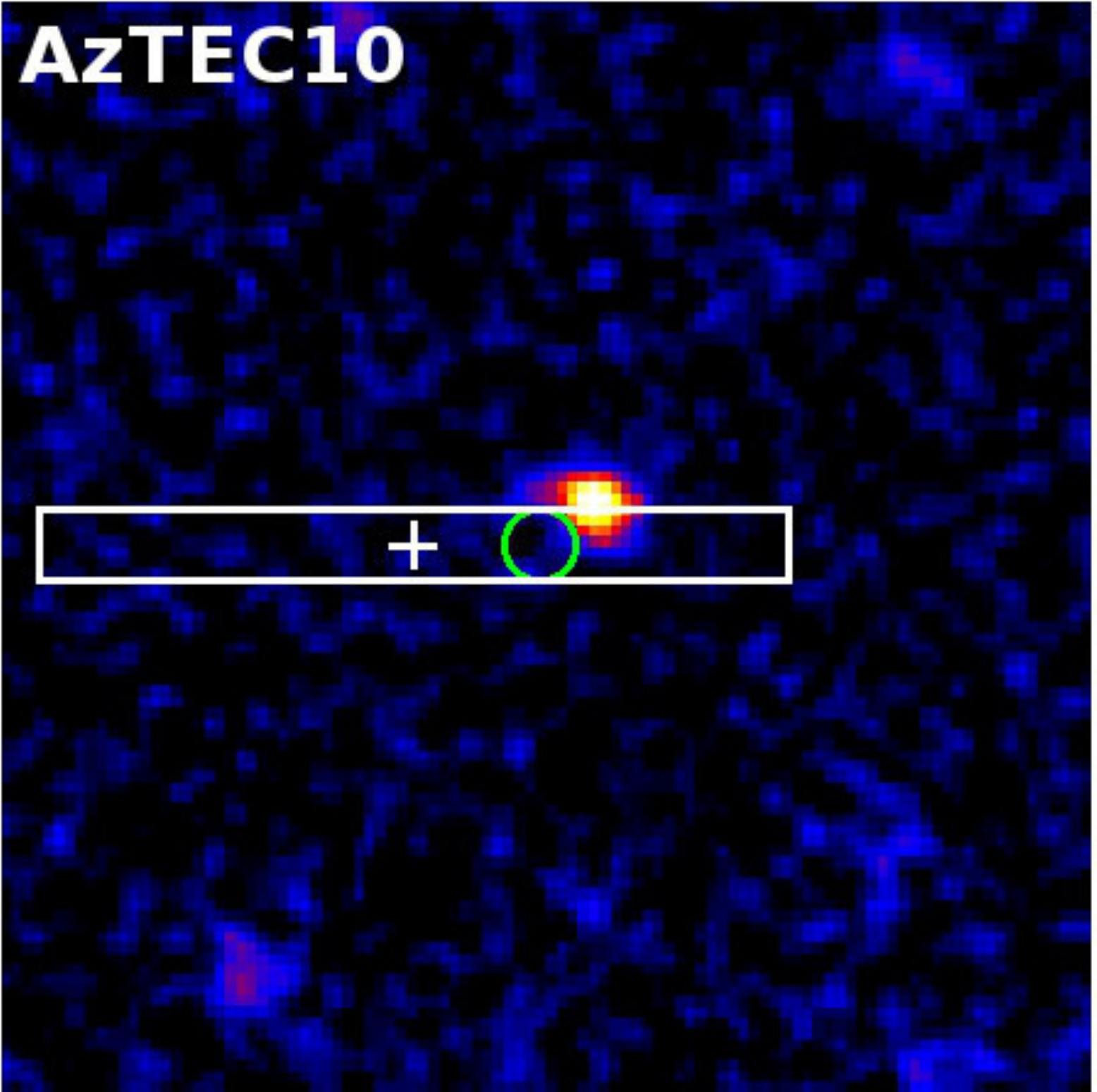}
\includegraphics[width=0.3\textwidth]{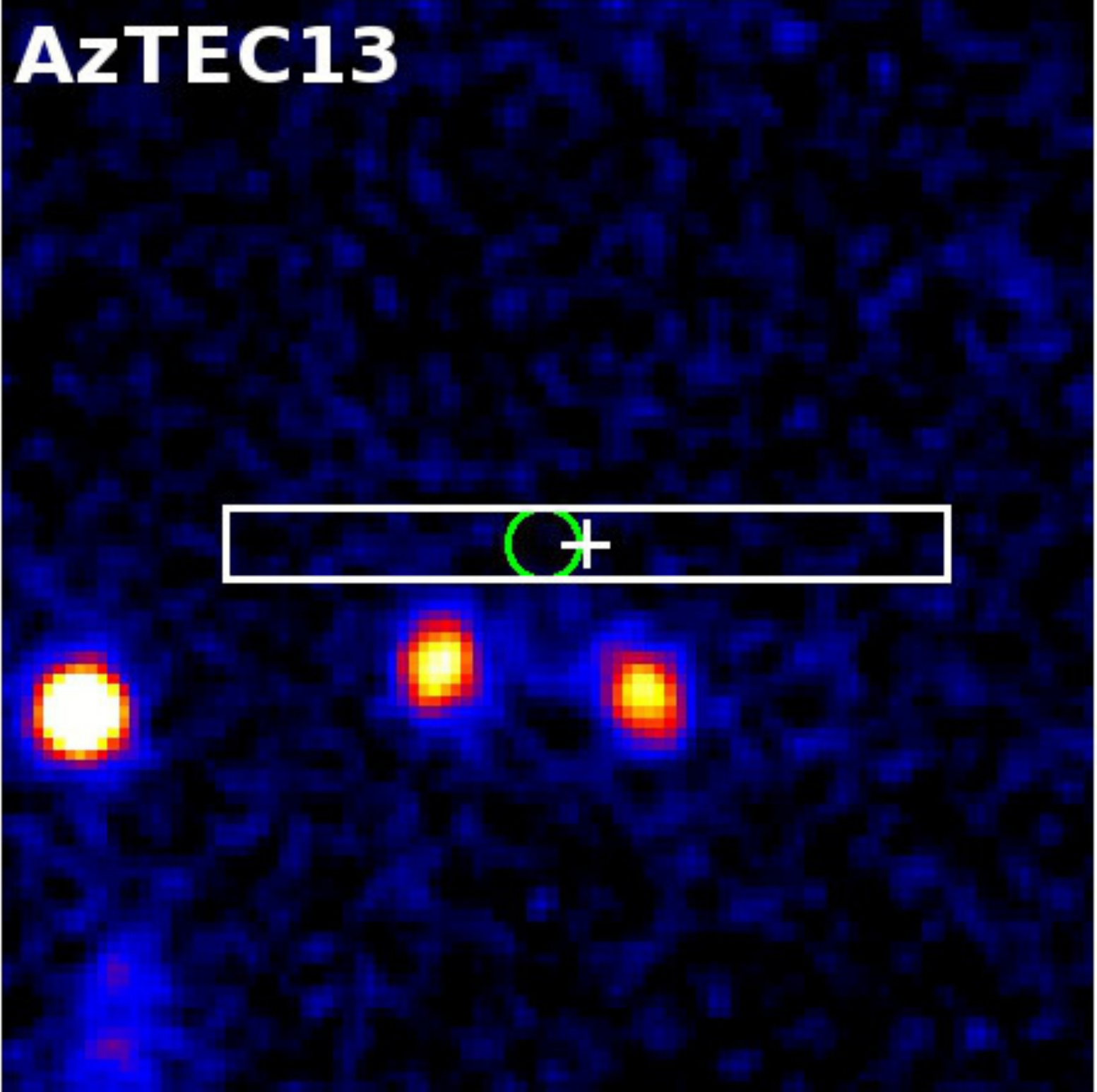}
\includegraphics[width=0.3\textwidth]{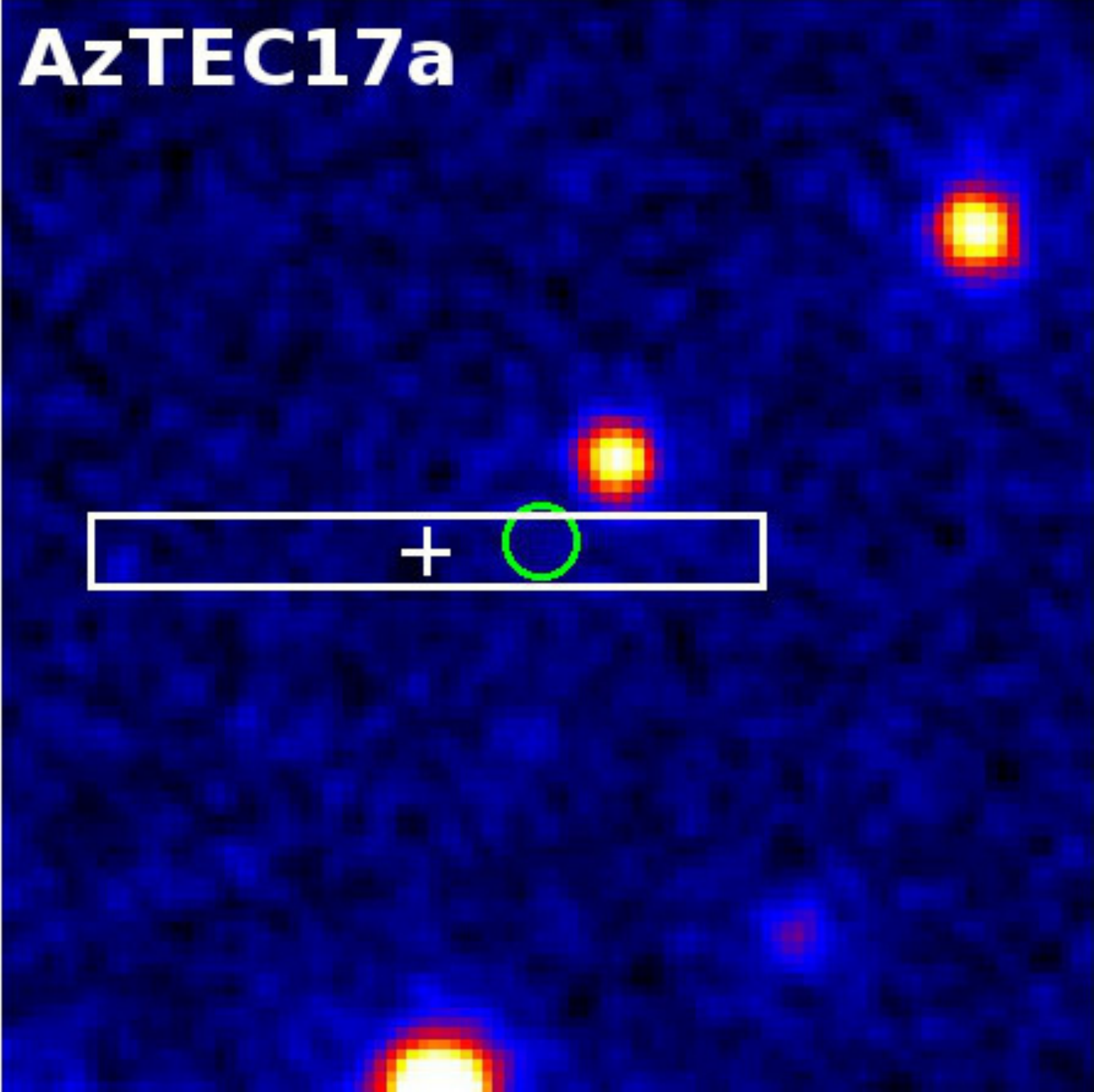}
\caption{UltraVISTA $Y$-band ($\lambda_{\rm eff}=1.02$ $\mu$m) images towards AzTEC2, 5, 6, 9, 10, 13, 
and 17a shown in linear scale. All the images are $15\arcsec$ on a side. The white rectangles indicate the DEIMOS 
slit positions, sizes, and orientations (see Table~\ref{table:slit}). 
We note that all the slits were aligned horizontally along the east-west direction. 
The central position of the slit is marked with a white plus sign, while the green circle of radius $0\farcs5$ 
shows the SMA 890~$\mu$m peak position (\cite{younger2007}, 2009) except in the case of AzTEC17a where it marks 
the PdBI 1.3~mm peak position. The red circle in the AzTEC6 panel represents the optical galaxy discussed by 
Koprowski et al. (2014); see Appendix~B for details.}
\label{figure:slits}
\end{center}
\end{figure*}

In Table~\ref{table:slit} we list the central coordinates, sizes (length and width), and position angles of 
the slits used for the Keck/DEIMOS spectral line observations towards AzTEC2, 5, 6, 9, 10, 13, and 17a. 
The DEIMOS slits are also illustrated in Fig.~\ref{figure:slits} where we show the UltraVISTA $Y$-band 
NIR images towards the above sources.

\begin{table*}
\renewcommand{\footnoterule}{}
\caption{Slit parameters of the Keck~II/DEIMOS spectral line observations.}
\begin{minipage}{2\columnwidth}
\centering
\label{table:slit}
\begin{tabular}{c c c c c c c}
\hline\hline 
Source & $\alpha_{2000.0}$ & $\delta_{2000.0}$ & Length & Width & P.A. & Offset\tablefootmark{a}\\
       & [h:m:s] & [$\degr$:$\arcmin$:$\arcsec$] & [$\arcsec$] & [$\arcsec$] & [$\degr$] & [$\arcsec$]\\
\hline
AzTEC2 & 10 00 07.98 & +02 26 12.2 & 7.7 & 1.0 & 90 & 0.98 \\
AzTEC5 & 10 00 19.69 & +02 32 04.4 & 8.0 & 1.0 & 90 & 0.90 \\
AzTEC6 & 10 00 06.54 & +02 38 37.6 & 7.2 & 1.0 & 90 & 0.61 \\
AzTEC9 & 09 59 57.28 & +02 27 30.6 & 8.7 & 1.0 & 90 & 0.45 \\
AzTEC10 & 09 59 30.88 &	+02 40 33.9 & 10.5 & 1.0 & 90 & 1.80 \\
AzTEC13 & 09 59 37.01 &	+02 33 20.0 & 10.1 & 1.0 & 90 & 0.60\\
AzTEC17a & 09 59 39.30 & +02 34 03.7 & 9.4 & 1.0 & 90 & 1.60\\
\hline 
\end{tabular} 
\tablefoot{\tablefoottext{a}{Angular offset between the SMG position and the slit centre 
(cf.~Fig.~\ref{figure:slits}).}}
\end{minipage} 
\end{table*}

\end{document}